\documentclass[aps,showpacs,eps,preprint]{revtex4}
\usepackage{graphicx}
\usepackage[font=small]{caption}
\usepackage{colordvi}
\usepackage{amssymb}
\usepackage{amsmath}
\usepackage{bbold}
\usepackage{epsf}
\usepackage{mathrsfs} 
\usepackage{float}
\usepackage{subfig}
\begin{document}
\def \beq{\begin{equation}}
\def \eeq{\end{equation}}
\def \bea{\begin{eqnarray}}
\def \eea{\end{eqnarray}}
\def \bem{\begin{displaymath}}
\def \eem{\end{displaymath}}
\def \P{\Psi}
\def \Pd{|\Psi(\boldsymbol{r})|}
\def \Pds{|\Psi^{\ast}(\boldsymbol{r})|}
\def \Po{\overline{\Psi}}
\def \bs{\boldsymbol}
\def \bl{\bar{\boldsymbol{l}}}		
\title{Fock space exploration by angle resolved transmission through quantum diffraction grating of cold atoms in an optical lattice} 
\author{Adhip Agarwala, Madhurima Nath, Jasleen Lugani, K. Thyagarajan and Sankalpa Ghosh }
\affiliation{Department of Physics, Indian Institute of Technology Delhi, New Delhi-110016, India}
\begin{abstract}	Light transmission or diffraction from different quantum phases of cold atoms in an optical lattice has recently come up as a useful tool to probe such ultra 
cold atomic systems. The periodic nature of the optical lattice potential closely resembles the structure of a diffraction grating in real space, but loaded with a strongly correlated quantum many body state which interacts with the incident electromagnetic wave, a feature 
that controls the nature of the light transmission or dispersion through such quantum medium. In this paper we show that as one varies the relative angle between the cavity mode and the optical lattice, the peak of the transmission spectrum through such cavity also changes reflecting the  statistical distribution of the atoms in the illuminated sites.
Consequently the angle resolved transmission spectrum
of such quantum diffraction grating can  provide a plethora of information about  
the Fock space structure of the many body quantum state of ultra cold atoms in such an optical cavity that can be explored in current state of the art experiments. 

\end{abstract}
\date{\today}
\maketitle
\section{Introduction}
Ultra cold atomic condensates loaded in an optical lattice \cite{Jaksch, RMP} provide a unique opportunity to study the properties of an ideal quantum many body system. After the first successful experiment in this field \cite{Greiner1} where a quantum phase transition from a Mott Insulator(MI) to Superfluid(SF) phase was observed, extensive theoretical as well 
as experimental study in this direction took place. The field continues to be a frontier research area of atomic and molecular physics, quantum condensed matter systems as well as quantum optics simultaneously, and holds promise for application in fields like quantum metrology, quantum computation and quantum information processing \cite{Cirac,Raimond}.

The relevance of the field of ultra cold atomic condensates to quantum optics 
was suggested much earlier when it was pointed out that the refractive index of a degenerate Bose gas gives a strong indication of quantum statistical effects \cite {Morice} and the interaction between quantized modes of light and such ultra-cold atomic quantum many body system is going to lead to a new type of quantum optics \cite{Meystre}. 
Subsequently,
it was pointed out that the optical transmission spectrum of a Fabry-Perot cavity loaded with ultra cold atomic BEC in an optical lattice can clearly distinguish between a SF and MI state \cite{mekhov1} and may be used as an alternative way of detecting such phase transition without directly perturbing the cold atomic ensemble through absorption spectroscopy.
A successful culmination of some of these theoretical predictions happened with the recent experimental success  
of realization of a strongly coupled atom-photon system where an ultra cold atomic system is placed inside an ultra-high finesse optical cavity\cite{Bren1, Colom1}, such that a photon in a given quantum state can interact with a large collection of atoms in same quantum mechanical state and thereby enhancing the atom-photon coupling strength. Superradiant Rayleigh scattering from 
ultracold atoms  in a ring cavity, which can be either Bose Einstein condensed or in the thermal phase was also observed experimentally \cite{Zimmerman1}.  
 As an aftermath, a host of interesting phenomena such as cavity optomechanics \cite{Bren2}, observation of optical bistability 
and Kerr nonlinearity \cite{Gupta} has been experimentally achieved with such systems. It may be also mentioned in this context that Bragg diffraction pattern from cold atoms in 
three dimensional optical lattice  \cite{Miyake}  and from quasi-two dimensional 
Mott Insulator, but without any cavity, was also recently observed experimentally \cite{Weit}.

The theoretical progress in understanding such atom-photon systems involving ultra cold atomic condensates is also impressive. A series of work by the Innsbruck group \cite{mekhov1, mekhov2, mekhov3, mekhov4, mekhov5, mekhov6, mekhov7, mekhov8} clearly pointed out how the optical properties of the cavity reveals the quantum statistics of these many body systems. In another set of work, cavity induced bistability in the MI to SF transition either due to strong cavity-atom coupling \cite{Larson} or due to the change in the boundary condition of the cavity \cite{Chen1} has been studied and its relation to cavity quantum optomechanics \cite{Aranya,Chen2} has also been explored. The self organization of atoms in a multimode cavity due to atom-photon interaction leading to the formation of exotic quantum phases and phase transition \cite{Gopal1, Gopal2, Vidal} is another major development in this direction. The recent observation of Dicke quantum phase transition through which a transition to a supersolid phase was achieved \cite{Bren3} through such self organization is an important experimental landmark in this direction.

The physics of ultracold atoms loaded inside an ultrahigh-finesse Fabry-Perot cavity can be analyzed from two different, but highly correlated perspectives. For example, 
ultra cold atomic ensemble loaded in such optical lattices with short range interaction can exist in two different types of quantum phases, MI and SF. The former is a definite state in the Fock space with well defined number of particles at each lattice site and lacks phase coherence between the atomic wavefunctions at different sites. The latter is a superposition of various Fock space states and has phase coherence. A phase transition between these phases takes place as the lattice depth varies. The statistical distribution of number of atoms in  lattice sites that characterizes these many body states, consequently influences the transmitted or diffracted electromagnetic wave  through atom-light interaction and thereby changes the dielectric response of such a cavity in the same way as the change of material leads to the change in refractive index. 

From another perspective, the periodic optical lattice potential forms a grating like structure in the real space, but now each slit of the grating contains ultra cold atoms in their quantum many-body state, that interacts with the light quanta of the electromagnetic field through the dipole interaction.
Such a system has been dubbed as a quantum-diffraction grating in the literature \cite{mekhov1}. It is well known that 
any quantum mechanical scattering process 
leads to the diffraction effect and thus such effect is ubiquitous in various quantum systems. As early as in 1977 in 
a review article by Frahn \cite{Frahn} an overview of such wide range of quantum mechanical diffraction process was presented in a common theoretical framework by comparing them with classical optical diffraction. Though some element of such quantum diffraction is also present in atom-photon system under consideration, it is unique in the sense that here  electromagnetic wave is getting diffracted by a quantum phase of matter wave loaded inside a cavity. 
A classical description of such diffraction of electromagnetic wave by a single atom or an atomic ensemble placed inside a cavity was also discussed in detail in ref. \cite{chapter}. 

It has 
been pointed out that diffraction properties of  scattered  ultra cold fermionic atoms \cite{Meystre2} by light  is
 strongly dependent on the mode of quantization of the eletromagnetic wave that scatters such fermions. 
Whereas in the current set of the problems one is concerned with the properties 
of the scattered light from ultra cold atoms placed in a cavity, a similar question on the dependence on the mode of quantization can be asked. 
In the limit of very large cavity atom detuning also, the features of such quantum diffraction for ultra cold bosonic atoms and its departure from the classical behavior has been studied \cite{mekhov2}.  
The results from these earlier studies indicate that a detailed analysis of the diffraction properties of such quantum diffraction grating has the potential to characterize the many-body quantum states of ultra cold atoms in more detail. 
Since the relevant experimental system is already available, such a study is even more encouraging. 

 In the current paper we carry out an analysis of the diffraction property of such quantum diffraction grating in detail.  
The most significant result from our analysis is that whereas  the transmission spectrum from the cavity at a given angle  between the cavity mode and the one dimensional optical lattice 
can detect the  MI and SF phase\cite {mekhov1}, 
the variation of the transmission spectrum as a function of this angle contains information about the Fock space structure of the quantum-many body state of the ultra cold atoms 
in either of the MI and SF phase. This is due to the fact that the dispersion shift or the frequency shift in the cavity mode corresponding to a 
transmission peak at a given angle is contributed by a set of Fock states that corresponds to a certain number of atoms in the illuminated sites. We analyze 
this feature in detail  by 
considering when a single cavity mode or two modes in two different cavities but with same frequency is excited. Some  comments on further generalization of this scheme is also mentioned.

The plan of the paper is as follows. In the next section \ref{model} we begin with a brief review of the formalism that is used to calculate the transmission from such a cavity loaded with ultra cold atoms in an optical lattice. Subsequently in section \ref{method}, we consider cases of  single  as well as two standing wave cavity modes for Fabry Perot cavities loaded with such ultra cold atoms 
and show 
how the transmitted intensity through such cavity can be calculated in these two cases.  As pointed out, the particular emphasis in this work will be on how the transmitted intensity changes as one varies the relative angle between the optical lattice and the cavity modes for both the MI and SF phases. An analysis of these results and their comparison against various classical diffraction patterns provides us a sound understanding of this quantum diffraction phenomena. In the next section \ref{TW} we extend the results to a ring shaped cavity and will show how cavity quantization procedure changes the transmitted intensity. We conclude the paper  after mentioning the relevance of our results to current experimental situations. 
\section{Model}\label{model}
\begin{figure}[H]
\begin{center}
\includegraphics[width=9cm, height=7cm]{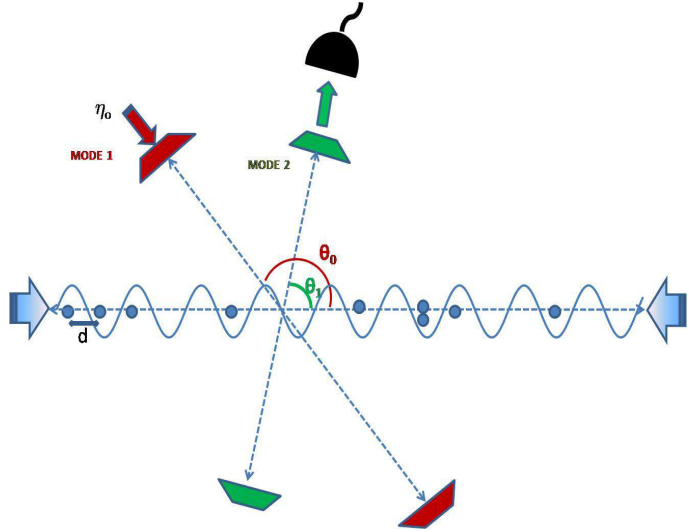}
\caption{(Color Online) Schematic diagram for cold atoms in an optical lattice loaded in a cavity. 
The optical lattice is created from two counter propagating laser beams and has a site spacing of length $d$. The two standing wave cavity modes, MODE 1 and MODE 2 are at angles $\theta_{0}$ and $\theta_{1}$ with the axis of the optical lattice respectively. The MODE 1 is being pumped by a pump laser with amplitude $\eta_{0}$ while MODE 2 is not being pumped and is used to collect the scattered photons by a detector. In the single mode case, the detector is also on MODE 1 and has not been shown in this figure.}
\label{S1}
\end{center}
\end{figure}

The physical system we describe is  depicted in Fig. \ref{S1} and consists of $N$ identical two-level bosonic atoms placed in an optical lattice of $M$ sites inside a Fabry-Perot cavity. $K$ sites among these are illuminated by cavity modes, pumped into the system by external lasers. We shall consider both the cases, in which single cavity mode, and two cavity modes will be excited.
These modes can be composed of either standing waves(SW) or travelling waves(TW). The setup given in Fig. \ref{S1} can realize standing waves; whereas later we shall discuss the corresponding setup for travelling wave solutions. Such a system was studied in \cite{mekhov1,mekhov2,mekhov3,mekhov4,mekhov5,mekhov6,mekhov7,mekhov8} and for a detailed treatment, the reader may refer to \cite{mekhov2}. Here we describe this theoretical framework briefly.

 The above mentioned system can be  theoretically modeled  as a collection of $N$ two level atoms, that are approximated as linear dipoles, to account for 
their interaction with the quantized electric field of the cavity modes. To describe the system through an effective Hamiltonian one then uses well known  
rotating wave approximation in which all fast oscillating terms in the Hamiltonian are neglected.  The excited state  of the two level atoms is then adiabatically eliminated assuming that
the cavity modes are largely off resonant to energy difference between the atomic levels. Thus in the resultant system all the 
atoms are in their ground states.
The effective Hamiltonian arrived in this way is given by   
\bea 
H &  =  & H_{f} + J_{0}^{cl} \hat{N} + J^{cl} \hat{B} +
 \hbar g^{2} \sum_{l,m} \frac{\hat{a}_{l}^{\dag} \hat{a}_{m}}{\Delta_{ma}} \Bigg( \sum_{j=1}^{K} J_{j,j}^{lm} \hat{n}_{j} \Bigg) \nonumber \\
&  & \mbox{}  + \hbar g^{2} \sum_{l,m} \frac{\hat{a}_{l}^{\dag} \hat{a}_{m}}{\Delta_{ma}} \Bigg( \sum_{<j,k>}^{K} J_{j,k}^{lm}\hat{ b}_{j}^{\dag}\hat{ b}_{k} 
\Bigg) + \frac{U}{2} \sum_{j=1}^{M} \hat{n}_{j} (\hat{n}_{j} -1)   \nonumber \\
  &    & \label{Ham}  \eea 
Here 
\beq
H_{f} = \sum_{l} \hbar \omega_{l} \hat{a}_{l}^{\dag} \hat{a}_{l} - i\hbar \sum_{l} (\eta_{l}^{*}(t) \hat{a}_{l} - \eta_{l}(t) \hat{a}_{l}^{\dag}) \nonumber\\
\eeq 
 where first term denotes the free field Hamiltonian and the second depicts the interaction of classical pump field with cavity mode, $\hat{a}_{l}$ is the annihilation operators of light modes with the frequencies $\omega_{l}$, wave vectors $ {\bf{k}}_{l}$, and mode functions $u_{l}({\bf{r}})$. $\eta_{l}(t) = \eta_{0} e^{-i\omega_{p}t}$ is the time dependent amplitude of the external pump laser of frequency $\omega_{p}$ that populates the cavity mode.

Here $J_{j,k}^{cl}$ correspond to the matrix element of the atomic Hamiltonian in the site localized Wannier basis, $ w({\bf{r}} - {\bf{r}}_{j}) $, namely 
\beq
J_{j,k}^{cl} = \int d{\bf{r}}w({\bf{r}} - {\bf{r}}_{j})  H_{a} w({\bf{r}} - {\bf{r}}_{k}) 
\eeq 
where $H_{a} =  -\frac{\hbar^{2} \nabla^{2}}{2m_{a}} + V_{cl}({\bf{r}})$ is the Hamiltonian of a free atom of mass $m_{a}$ in an optical lattice potential $V_{cl}({\bf{r}})$. Therefore, $J_{0}^{cl}$ = $J_{j,j}^{cl}$, and $J^{cl}$ =$J_{j,j\pm 1}^{cl}$  are respectively the onsite energy and the hopping amplitude of proto-type Bose Hubbard model given in \cite {Jaksch}.
 At the atomic site $j$, $\hat{b}_{j}$ is the annihilation operator,  and $ \hat{n}_{j} = \hat{b}_{j}^{\dag} \hat{b}_{j} $ is the corresponding atom number operator, $\hat{N} = \sum_{j=1}^{M} \hat{n}_{j} $ denotes the total atom number and $ \hat{B} = \sum_{j=1}^{M} \hat{b}_{j}^{\dag}\hat{ b}_{j+1} $. 

The coefficients $J_{j,k}^{lm}$  is similar to $J_{j,k}^{cl}$, but  now generated from interaction between atoms and quantized cavity modes and is given by ,	
\beq
J_{j,k}^{lm} = \int d{\bf{r}}w({\bf{r}} - {\bf{r}}_{j})  u_{l}^{*}({\bf{r}}) u_{m}({\bf{r}}) w({\bf{r}} - {\bf{r}}_{k}) 
\eeq
$\Delta_{la}$  = $\omega_{l}$ - $\omega_{a}$ denotes the cavity atom detunings where $\omega_{a}$ is the frequency corresponding to the energy level separation of the two-level atom and $ g $ is the atom-light coupling constant. Thus the fourth and fifth term in (\ref{Ham}) respectively contribute to the onsite energy and hopping amplitude due to the interaction between atoms and quantized cavity modes. 

In the last term, $ U = \frac{4\pi a_{s} \hbar^{2}}{m_{a}} \int d{\bf{r}} |w({\bf{r}})|^{4}$ , where $a_{s}$ denotes the $s$-wave scattering length and gives the onsite interaction energy. For a sufficiently deep optical lattice potential  $V_{cl}({\bf{r}})$, the overlap between Wannier functions can be neglected. In this limit  $J^{cl} = 0$ and $J_{j,k}^{lm} = 0$ for $j\neq k$. Such Wannier functions can be well approximated as delta functions centered at lattice sites ${\bf{r}}_{j}$ and consequently  $J_{j,j}^{lm} = u_{l}^{*} ({\bf{r}}_{j})u_{m}({\bf{r}}_{j})$.

 The above Hamiltonian in (\ref{Ham}) describes the zero temperature quantum phase diagram of ultra cold bosonic atoms loaded in an optical lattice placed inside a optical cavity. 
This is because their many body quantum mechanical ground state can exist in various quantum phases \cite{Jaksch,Greiner1} as a function of  parameters like $U$  and $J$. 
 In the subsequent analysis in this work, the physical system that diffracts the photons is therefore 
 a novel type of quantum diffraction grating not only because the diffracting medium corresponds to a quantum phase of ultra cold atoms, but also due to the fact that it is embedded in a optical lattice/grating like structure in real space which in turn affects the nature of such quantum phase. As we shall point out, one particular way of understanding the nature of such quantum diffraction and differentiate from classical diffraction or any other quantum diffraction \cite{Frahn} is to study it as a function of the relative angle between the cavity mode and direction of the optical lattice in which the cold atoms are loaded. Quantum diffraction of electromagnetic wave  by such ultra cold atomic condensate inside a Fabry-Perot cavity, but without loading them in an optical lattice (other than the one dynamically generated due to cavity-atom coupling), was already experimentally studied in ref \cite{Bren2} in the context of cavity quantum optomechanics. Thus the physical system under consideration is very much realizable experimentally. We start our discussion by briefly outlining the relevant theoretical framework 
to understand such quantum diffraction following ref \cite{mekhov2}. 
\section{Methodology}\label{method}
From the atom-photon Hamiltonian (\ref{Ham}), the Heisenberg equation of motion of the photon annihilation operator $\hat{a}_{l}$ is given by 
\beq
\dot{\hat{a}}_{l} = -i\omega_{l}\hat{a}_{l} - i \delta_{l}\hat{D}_{ll} \hat{a}_{l} - i\sum_{m}\delta_{m}\hat{D}_{lm}\hat{a}_{m} - \kappa \hat{a}_{l} + \eta_{l}(t) \label{heisen}
\eeq 
with $\hat{D}_{lm} \equiv \sum_{j=1}^{K} {u_{l}}^{*}({\bf{r}}_{j})u_{m}({\bf{r}}_{j})\hat{n}_{j}$, where $ l\neq m$ and $\delta_{l} = g^{2}/ \Delta_{la} $. $ \kappa $ is the cavity relaxation rate introduced phenomologically. The first, fourth and the fifth terms on the right hand side correspond to property of light transmission through an empty cavity. The second and third terms give the information about the atom-light interaction in the cold atomic condensates. As we have already mentioned the above equation (\ref{heisen}) is valid in the limit of deep optical lattice where the wannier functions are approximated as delta functions. 
\subsection{Single Mode}\label{smode}
First we shall consider the case when a single cavity mode is excited. 
From the stationary solution for one mode case, namely $\dot{\hat{a}}_{l}=0$ 
we obtain the expression for the corresponding photon number operator as,  
\beq \hat{a}_{0}^{\dag} \hat{a}_{0} = \frac{|\eta_{0}|^{2}}{(\Delta_{p}  - \delta_{0}\hat{D}_{00})^{2} + \kappa^{2}}
\eeq
Here $\Delta_{p} = \omega_{0p} - \omega_{0}$ is the probe-cavity detuning and $\hat{a}_{l} = \hat{a}_{0}$. In this case $\hat{a}_{m} = 0$ and $\hat{D}_{l,m}= \hat{D}_{00}$.
The single mode transmission through the cavity is calculated by taking the expectation value of the above expression in given many-body atomic ground state. As expected such an expression is similar to the standard Breit Wigner form. 
However $\hat{D}_{00}$ is in terms of the Fock space operators acting on the atomic ensemble, revealing the statistical properties of the quantum matter of ultra cold atoms.

Now in the denominator of the above expression the shift in frequency is determined by the eigenvalue of the operator $\hat{D}_{00}$, which is dependent on both the atomic configuration, $i.e.,$ the number of illuminated atoms and the mode functions. For plane standing waves the mode function, $u({\bf{r}}_{j})_{SW}= cos({\bf{k.r}}_{j} +\phi)$ where $\phi$ is constant phase factor which has been set to zero and ${\bf r}_{j}$ denotes the position vector of the $j$th site on the optical lattice. Here we consider a one dimensional optical lattice with site spacing $d$. For a cavity mode of wavelength $\lambda$ incident at an angle $\theta$ with the optical lattice, $u({\bf{r}}_{j})_{SW}= cos( {{\frac{2\pi}{\lambda}jd cos \theta}})$. We assume the cavity mode wavelength to be $2d$ and thus, the mode function is $ u({\bf{r}}_{j})_{SW}= cos(j\pi cos\theta)$, where $j \in I$. For such standing waves, the factor $\hat{D}_{00}$ becomes,
\beq
\hat{D}_{00} = \sum_{j} u_{l}^{*} u_{l} \hat{n_{j}} =  \sum_{j=1:K} cos^{2}(j\pi cos\theta) \hat{n}_{j} \label{D00}
\eeq
where $K$ are the number of illuminated sites. 
This shows that shift in the cavity resonant frequency is dependent on the relative angle of the cavity mode with the optical lattice. To simplify the analysis, here it has been assumed that while changing this angle, the light beam waist is modified in a way that we always illuminate only fixed $K$ sites. However, as explained below, a few sites fall at intensity minima of the cavity mode, thus changing the effective number of illuminated sites.

For example in Fig. \ref{M13} we show the cases when light with wavelength(=2$d$) is incident at $\theta=0^{\circ}$ and $\theta =60^{\circ}$. When the angle $\theta=0^{\circ}$, all the atoms are at the points of maximum intensity or the anti-nodes of the cavity mode wavelength. Thus all the atoms are illuminated. When $\theta$ changes to $60^{\circ}$, the projected wavelength along the optical lattice direction changes and a few atoms which were at the maxima points are now placed at the points of minimum(or zero) intensity or nodes. Thus now only the alternate sites are illuminated as can be seen in Fig. \ref{M13}. Therefore, the effective number of illuminated sites in the lattice at $\theta=60^{\circ}$ reduces to half its value at $\theta=0^{\circ}$. Hence the dispersive shift varies with the change in the relative angle of the cavity mode and the optical lattice.
\subsubsection{Mott Insulator} \label{SMI}
\begin{figure}[H]
\centering
\subfloat[Part 1][]{\includegraphics[width=15cm, height=7cm]{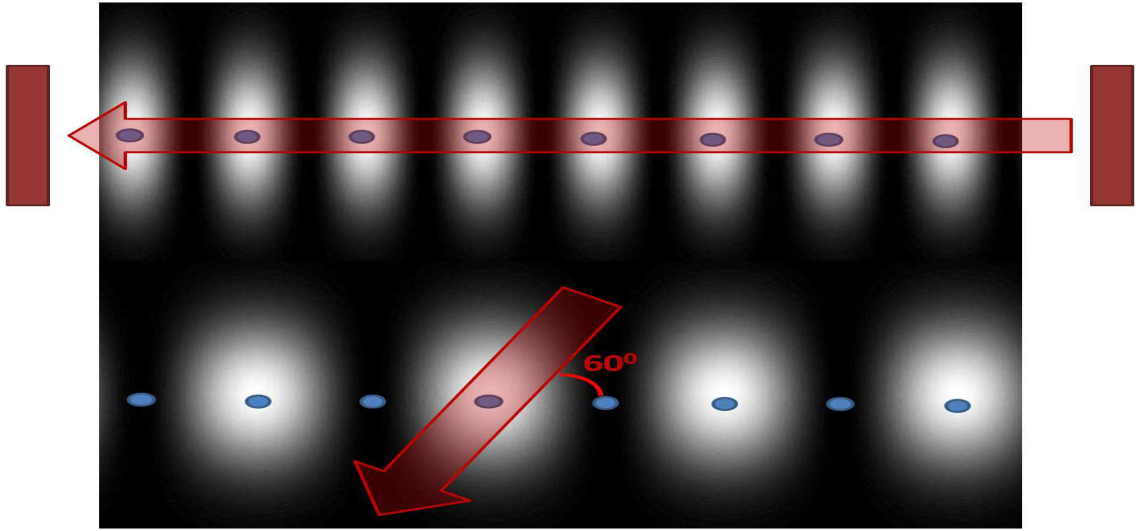} \label{M13}} \\
\subfloat[Part 2][]{\includegraphics[width=15cm, height=8cm]{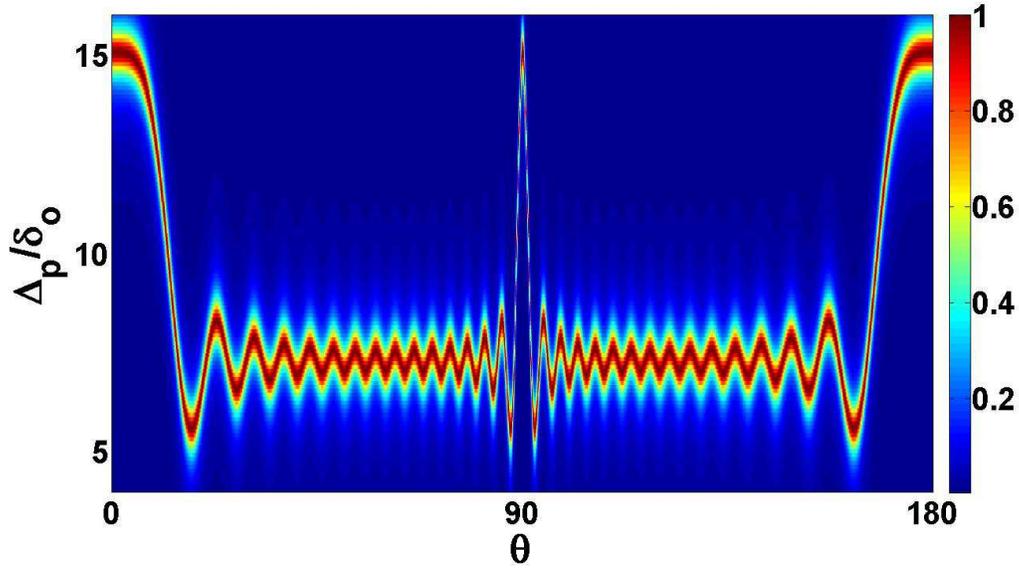}\label{M11}}
\caption{(Color Online) (a)The top part of this schematic shows the way atoms are present at the intensity maxima regions of the illuminating cavity mode when the optical lattice is illuminated at $0^{\circ}$. However, changing the angle by $60^{\circ}$ results in the decrease of the number of illuminated sites to half. (b)Variation of the photon number (\ref{MI1}) (color axis) with detuning $\frac{\Delta_{p}}{\delta_{0}}$ and $\theta$(in degrees) for N=M=30, K=15, $\kappa$ =0.5$\delta_{0}$, when the atoms are in a  MI state and are illuminated by a single standing wave cavity mode.}
\end{figure}

We shall first consider the case when the ground state of the atomic ensemble is a MI, ie., a single state in the Fock space. 
\beq| \Psi \rangle = |n,n,n,....,n\rangle   \label{mott} \eeq with $n = \frac{N}{M}$. This state is also an eigenstate of the operator $\hat{D}_{00}$ with eigenvalue $F(\theta, K) n$ 
where 
\beq  F( \theta , K) = \frac{1}{2} [  K + \frac{sin(K\pi cos\theta) }{sin(\pi cos\theta)} cos((K +1)\pi cos\theta) ]  \label
{Ftheta}  \eeq
which has been calculated using (\ref{D00}).
The corresponding transmission spectrum will be proportional to the photon number, which is given by 
\beq \langle \Psi | \hat{a}_{0}^{\dag} \hat{a}_{0} | \Psi \rangle = \frac{|\eta_{0}|^{2}}{(\Delta_{p}  - \delta_{0}F(\theta,K) n)^{2} + \kappa^{2}} \label{MI1} \eeq 
This has been plotted in Fig.\ref{M11} with the angle $\theta$ and detuning $\Delta_{p}/\delta_{0}$.

Let us first point out that from the left and the right side, the intensity plot is strikingly similar to the real space intensity variation in classical light wave diffraction from a straight edge \cite{Born,Ghatak} even though intensity variation in these two cases are function of completely different set of physical variables. We shall here briefly explain this apparent similarity 
inspite of these differences. Here we have plotted the variation of the photon number as a function of the angle $\theta$ and the cavity detuning. Therefore the plot is not an intensity plot in the real space. At each value of $\theta$, we obtain a maximum intensity at that value of the dispersion shift which corresponds to the number of atoms illuminated in the lattice at that angle. As the number of illuminated atoms changes with the change in the relative angle $\theta$ so the dispersive shift takes different values depending on the atomic arrangement. 
Nevertheless, the similarity stems from the fact that the factor $\hat{D}_{00}$ which is written in the following form,
\beq
\hat{D}_{00} = \sum_{j=1:K} \frac{1}{2} \big[1 + cos(4\pi j cos^{2} \frac{\theta}{2})\big]\hat{n}_{j} \label{fid}
\eeq
mathematically has a similar form of the Fresnel integral, encountered in the intensity profile for a straight edge diffraction pattern. There the intensity is a function of $C(\tau)$ given by
\beq C(\tau) = \int_{0}^{\tau} cos(\pi x^{2}/2) dx  \label{fi} \eeq 
where $\tau$ is dependent on the geometry of the system including the distance from screen. An increase in $\tau$ implies, the evaluation of intensity at a point farther from the straight edge. The oscillating behavior of the intensity can be attributed to the functional dependence of Eq.(\ref{fi}). However, it is to be noted, that Eq.(\ref{fid}) involves a summation over the illuminated sites K, which is constant and the variation is plotted with respect to $\theta$, which is the angle made by the cavity mode with lattice. This summation can be understood in the context of the lattice being discrete. But for a fixed K, as we change $\theta$, we are effectively changing the number of illuminated sites, as mentioned earlier, and hence the nature of the plot seems similar. 

Since the dispersion shift is an indicator of the refractive index of the medium, the above result suggests that the refractive index of a given quantum 
phase is dependent not only on the site distribution of the atomic number, but also on the angle between the propagation direction and the optical lattice. 
This is a unique feature of this system. It may be recalled that in well known optical phenomenon like Raman Nath scattering due to diffraction through a medium with periodically modulated 
refractive index \cite{Raman} or in Brillouin scattering in non-linear medium \cite{Brill}, there is also a frequency shift due to dispersion of the transmitted electromagnetic wave through the medium.
However the mechanism of the dispersion shift as a function of angle between the cavity mode and optical lattice as explained 
in the preceding discussion is fundamentally different from these cases. 
\subsubsection{Superfluid}
Next we consider the case when the atoms are in SF phase. 
The SF wave function in the Fock space basis can be written as superposition state, namely 
 \beq	| \Psi \rangle = \frac{1}{M^{N/2}} \sum_{\langle n_{j} \rangle} \sqrt{\frac{N!}{n_{1}!n_{2}!...n_{M}!}} | n_{1}, n_{2}, .., n_{M}\rangle	\eeq 
where  $ n_{j}$ denotes the number of atoms at the $j$th site while $\langle n_{j} \rangle$ denotes a set of $n_{j}$ for a particular Fock state.
Unlike the MI case, here $\hat{D}_{00}$ acts on a  superposition of Fock states each of which is 
an eigenstate of this operator. Each such Fock state carries a different set of $| n_{1} ,n_{2}, .., n_{M}\rangle$. 
Hence,
\beq \langle \Psi | \hat{a}_{0}^{\dag} \hat{a}_{0} | \Psi \rangle = \frac{1}{M^{N}} \sum_{\langle n_{j} \rangle}\frac{N!}{n_{1}!n_{2}!...n_{M}!} \frac{|\eta_{0}|^{2}}{{(\Delta_{p}  - \delta_{0}F_s(\theta,K, n_{j}))}^{2} + \kappa^{2}} \\ \label{sf} \eeq 
where,  $F_s(\theta,K, n_{j})$ is the eigenvalue of the operator $\hat{D}_{00}$ acting on a particular Fock state. Here the $F_{s}$ functions are generalization of the $F$ function described 
in (\ref{Ftheta}), for the case of SF phase in which the number of particles in each site is different as,
\beq F_s( \theta, K, n_{j}) =\sum_{j=1:K} cos^{2}(j\pi cos\theta) {n_{j}} \label{D00sf} \eeq 
where $j$ is the site index and $n_{j}$ is the occupancy of site $j$.  It can be easily checked when $n_{j}=n$ for all $j$ that corresponds to the MI state in (\ref{mott}), $F_{s}=nF$. 

The decomposition of the many body states in Fock space basis is now mapped in the frequency shifts of the cavity mode. The probabilistic weight factor is mapped in the intensity of the peak at that particular value of dispersion shift. At a particular angle of incidence, the singular peak of MI now breaks into multiple peaks with varied peak strengths and dispersion shifts. Each particular peak corresponds to a particular group of Fock states which have the same value of $F_{s}(\theta,K, n_{j})$ as defined in Eq.($\ref{D00sf}$). Now as the angle $\theta$ is being changed, the effective number of illuminated sites change. This changes the value of $F_{s}(\theta,K, n_{j})$ as well as the set of Fock states which yield the same value of $F_{s}(\theta,K, n_{j})$.

\begin{figure}[H]
\centering
\subfloat[Part 1][]{\includegraphics[width=18cm, height=9cm]{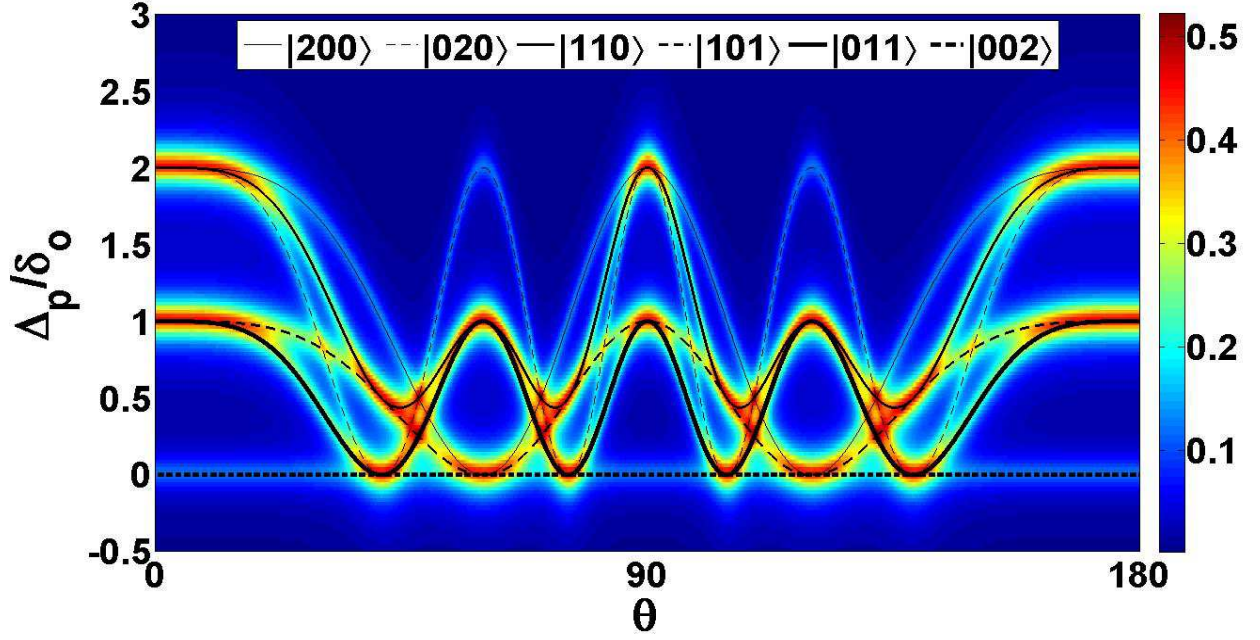}} \\ 
\subfloat[Part 2][]{\includegraphics[width=6cm, height=5cm]{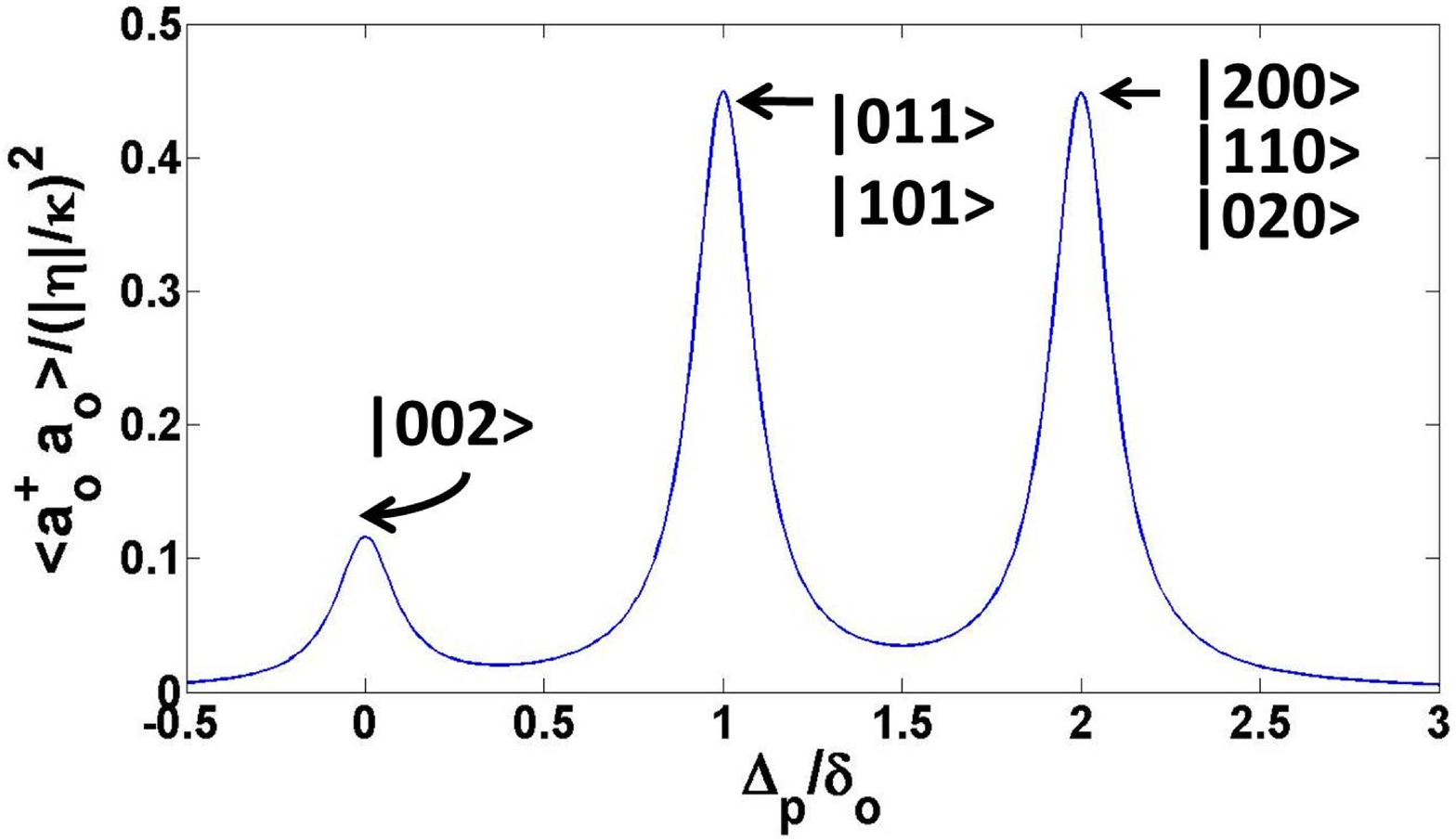}}
\subfloat[Part 3][]{\includegraphics[width=6cm, height=5cm]{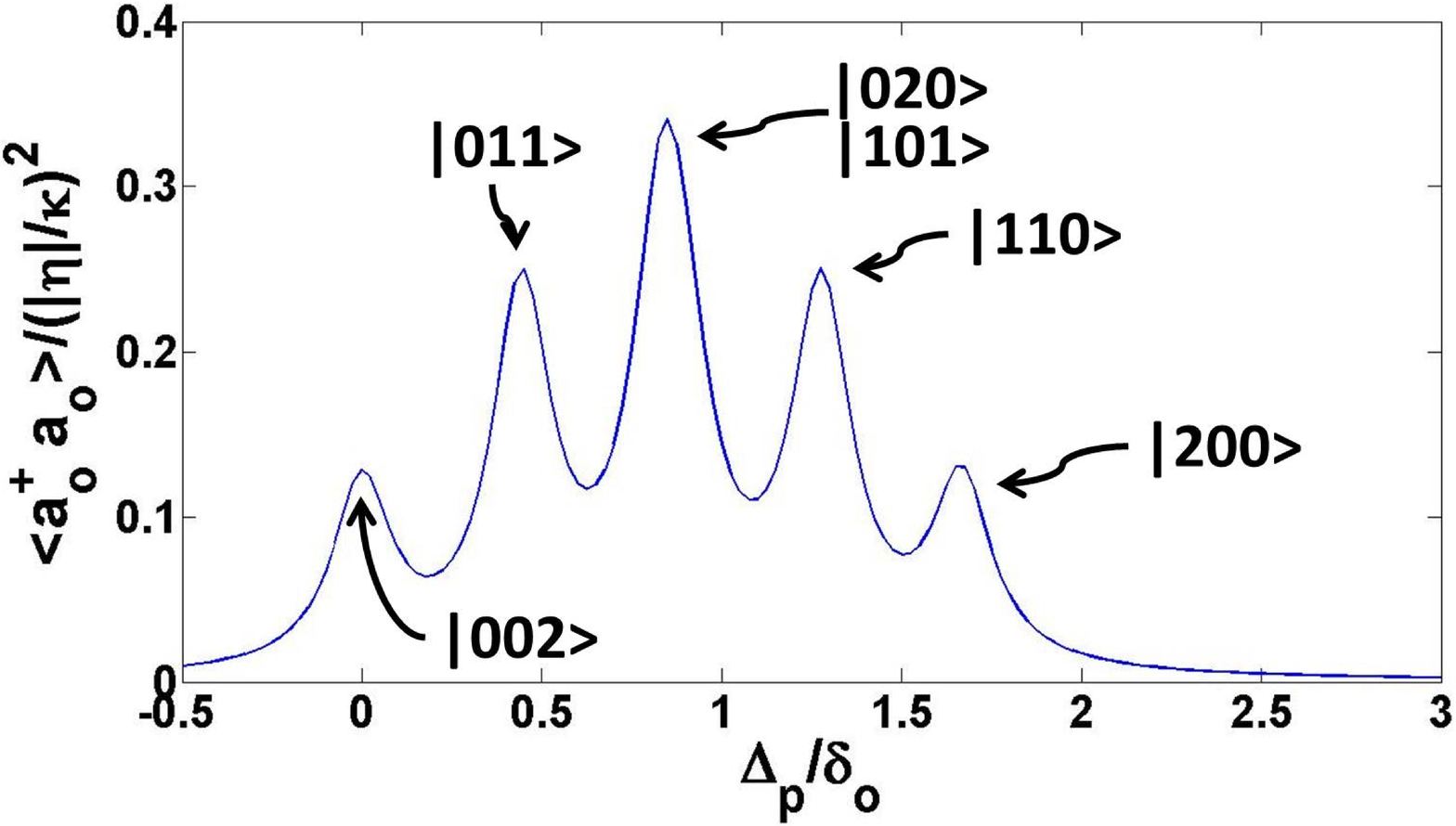}}
\subfloat[Part 4][]{ \includegraphics[width=6cm, height=5cm]{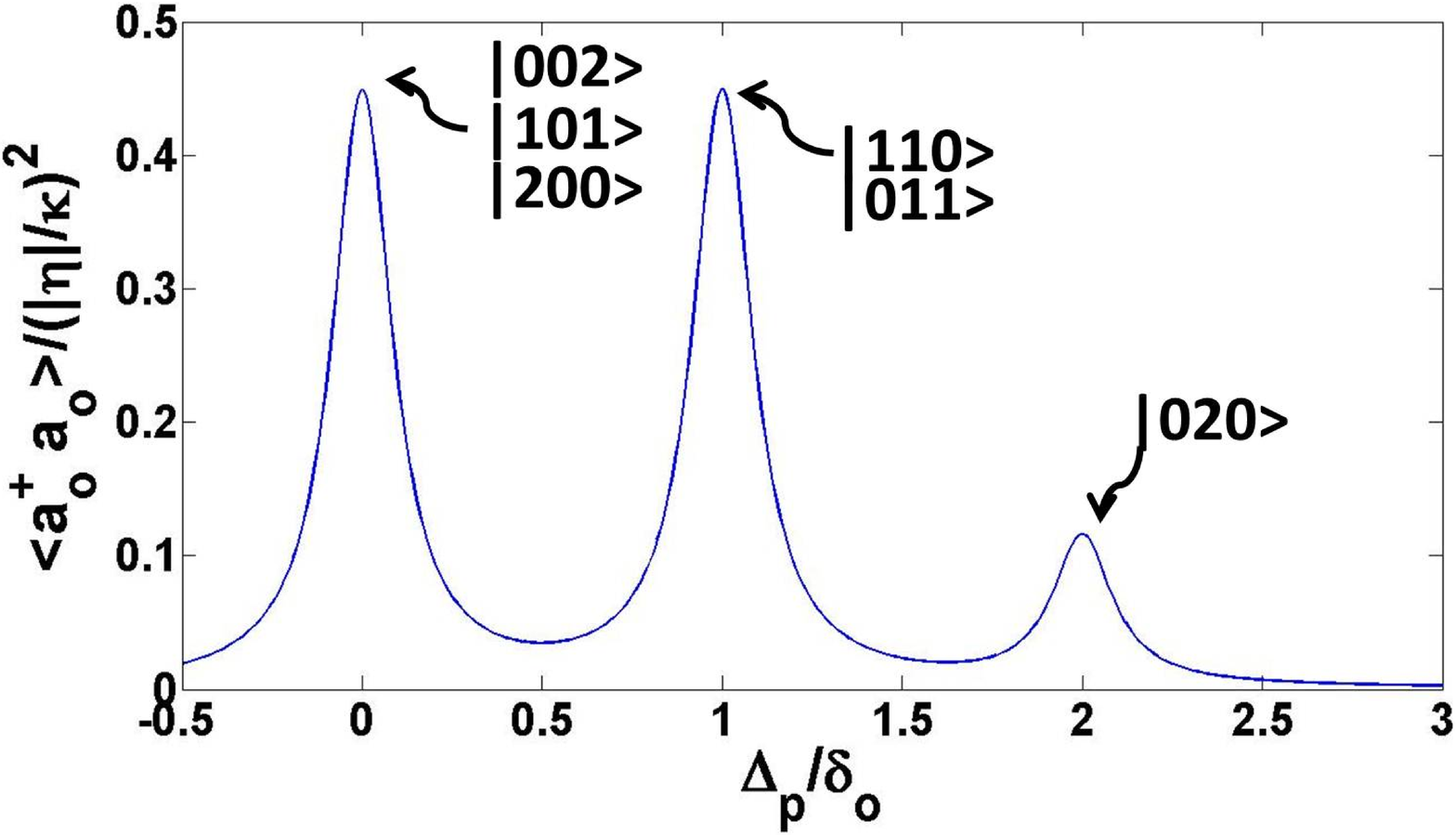} }

\caption{(Color Online) Plot (a) shows the variation of the photon number (\ref{sf}) (color axis) with detuning $\frac{\Delta_{p}}{\delta_{0}}$ and $\theta$(in degrees) for N=K=2, M=3, when the atoms are in SF state and are illuminated by a single standing wave cavity mode.This is superposed with the variation of how individual Fock states(black lines) corresponding to different peaks change with change in $\theta$. Plots (b)-(d) are the two dimensional plots for photon number with respect to $\Delta_p / \delta_0$ for $\theta = 0^\circ,30^\circ$ and $60^\circ$. (b)when $\theta = 0^{\circ}$ the six Fock states corresponding to this model system divides into groups of 1,2,3 Fock states ( see text). The corresponding Fock states for each peak is mentioned beside the respective peak. (c) when $\theta = 30^{\circ}$ we observe five peaks corresponding to various different Fock states . (d) At $\theta = 60^{\circ}$ we again observe three peaks but unlike (b), now  $|020\rangle$ shows a separate shift. In all the above cases $\kappa =0.1 \delta_{0}$.}
\label{SF00}
\end{figure}

In Fig.\ref{M13}, we showed  how variation in the $\theta$ changes the effective number of illuminated sites.  In Fig. \ref{SF00}(a)  we show how this variation in angle leads to separation of fock states, when the ground state of the ultracold condensate is superfluid.
This can be understood clearly by taking a case where the number of Fock states involved is small.
The corresponding states are  few body correlated states  that are  few body analogues of a superfluid state, where the number of Fock states involved is thermodynamically large.

In Fig. \ref{SF00}(a)-(d) we consider a case 
when 2 atoms are placed in 3 sites among which the first 2 sites are illuminated through a cavity mode. Let us first analyze, the condition at $\theta = 0^{\circ}$ when all K sites get illuminated. And the coefficient of $n_{i}$ in $\hat{D}_{00}$ defined in (\ref{D00}) i.e. $cos^2(m\pi cos(\theta))$ is identically $1$. Thus at this value,  the eigenvalue of $\hat{D_{00}}$ is simply the number of atoms in the illuminated sites. If only a part of optical lattice is illuminated ie. $K<M$, then at any point of time, there can be $q$ atoms (such that $q \le N$) in the illuminated sites. The eigenvalue of $\hat{D}_{00}$, namely $F_{s}(\theta,K, n_{j})$ for a particular value of $q$ will also be $q$. Hence states yielding the same amount of shift would be $|2,0,0\rangle$ , $|0,2,0\rangle$ and $|1,1,0\rangle$  giving  $q=2$. On the other hand the states in which only a single atom is present in the illuminated sites, such as $|1,0,1\rangle$ or $|0,1,1\rangle$ corresponds to $F_{s}(\theta, K, n_{j})=1$. However, $|0,0,2\rangle$ does not show any shift. Hence, the Fock states get distributed into groups having $3,2,1$ states respectively, each group having a different value of the dispersion shift. This has been demonstrated in Fig. \ref{SF00} (b). 

\begin{figure}[H]
\centering
\subfloat[Part 1][]{\includegraphics[width=16cm, height=6cm]{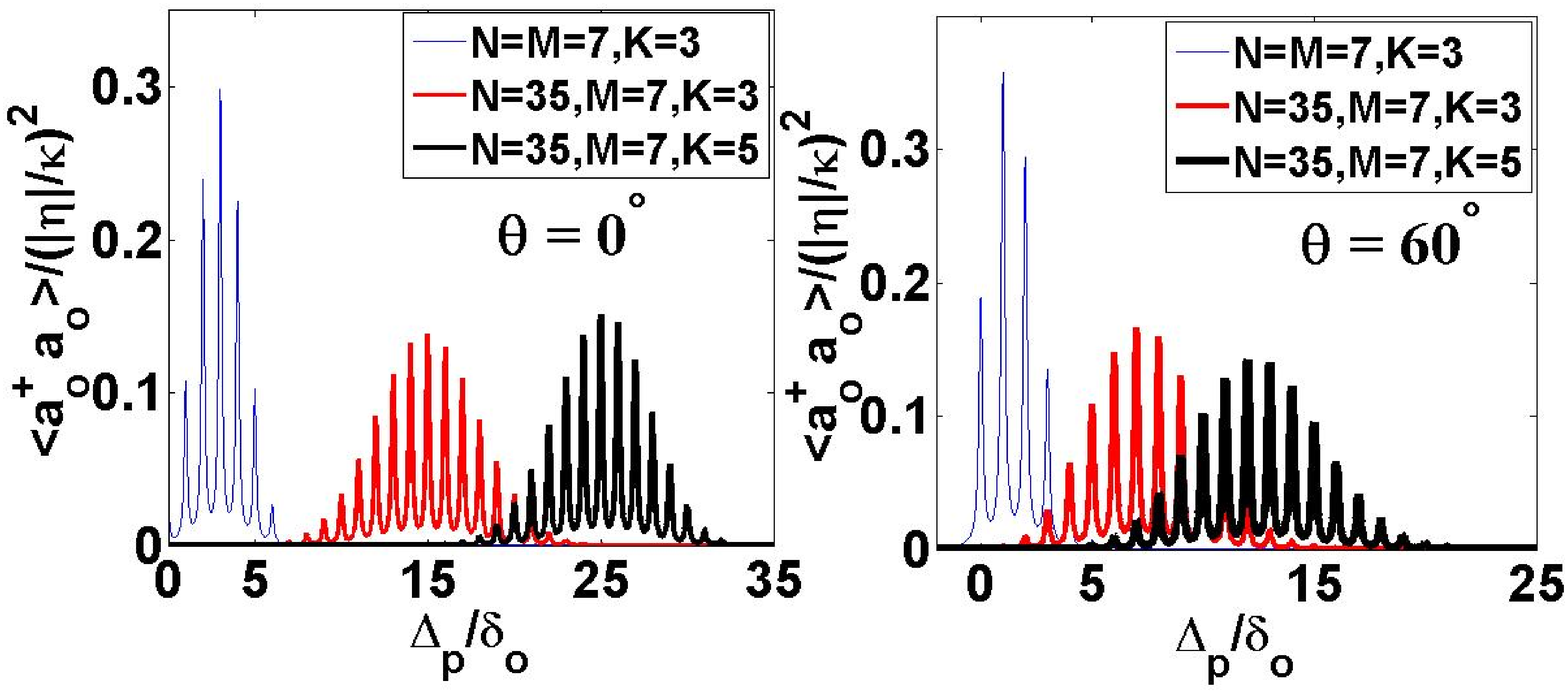}\label{SFLN1}}\\
\subfloat[Part 2][]{\includegraphics[width=12cm, height=6cm]{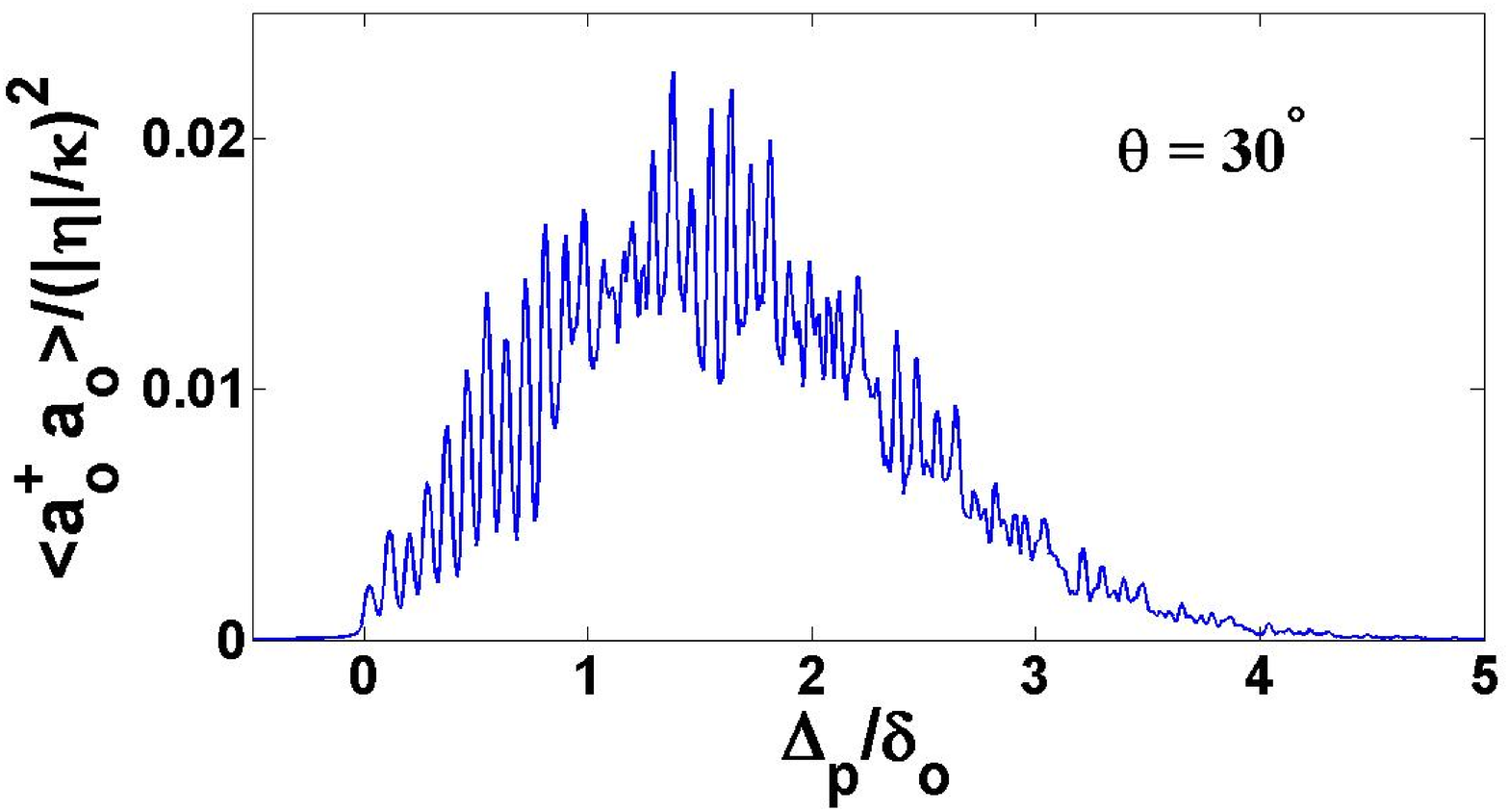}\label{SFLN3}}
\caption{(Color Online) Variation of the photon number (\ref{sf}) with detuning $\frac{\Delta_{p}}{\delta_{0}}$. (a) The left plot is  for $\theta = 0^{\circ}$(in degrees)  and the right  plot is for $\theta = 60^{\circ}$(in degrees) for N=M=7, K=3 (blue), N=35,M=7, K=3(red) and N=35,M=7 and K=5(black) when the atoms are in SF state and are illuminated by a single standing wave cavity mode.
Here $\kappa =0.1\delta_{0}$.
In (b) $\theta = 30^{\circ}$(in degrees) and N=M=8, K=5 . Here $\kappa= 0.01\delta_{0}$.}
\label{SF11}
\end{figure}

Therefore, all the Fock states corresponding to those $q$ atoms, which includes  various permutations of $q$ identical atoms in $K$ sites, will map to one single lorentzian in terms of photon number with $\Delta_p/\delta_0$. Therefore in this case all $K$ sites are equivalent to each other. The height of this peak is given by the probability corresponding to those $q$ atoms in $K$ sites. $q$ changes by $\Delta q$ which is always an integer with  minimum value of $=1$. Thus the distance between two adjacent lorentzians can only be $1$ for $\theta=0^{\circ}$. It is to be noted, this is independent of the total number of atoms, or the number of sites. This can be seen in the left plot of Fig.\ref{SFLN1}. Here for larger values of $N$ , $M$ or $K$, the peak separation remains unity.

Now as the angle between the cavity mode and lattice is varied, the effective number of illuminated sites change and so changes the set of Fock states that has same $F_{s}(\theta, K, n_{j})$.
However the above feature of equidistant peaks of the lorentzians discussed for $\theta=0^{\circ}$  is also observed for $\theta= 60^{\circ}$. At this angle  sites are  either completely illuminated or not at all illuminated. Here again if $q'$ atoms can be considered to be present in the illuminated alternates sites,  the corresponding shift will be $q'$ in terms on $\Delta_p/\delta_0$. All permutations of atoms in these alternate sites will contribute to same peak thus showing that all alternate sites become equivalent to each other. Therefore the shift between two adjacent lorentzians will be unity as minimum value of $\Delta q'$ = $1$. For the case of $2$ atoms in three sites  demonstrated in Fig. \ref{SF00} (d) it is just the central site which is illuminated. Now under these circumstances, state $|0,2,0\rangle$ show a distinctively separate shift from the states $|2,0,0\rangle$ and $|1,1,0\rangle$. It may be recalled that at $\theta=0^{\circ}$ all these states had the same shift. Moreover, now the states $|0,1,1\rangle$ and $|1,1,0\rangle$ will have the same shift since only one atom gets illuminated in this case.  As pointed out earlier, the shift between successive lorentzians will remain $1$ for large number of atoms and sites as demonstrated in  the right plot of Fig. \ref{SFLN1}. 

 For the other angles such that $ 0 < \theta < \frac{\pi}{2}$, sites get partially illuminated and the above equivalence among all the illuminated sites changes. Consequently the separation between 
the two successive lorentzians will also differ from $1$. A particular case of interest is 
if $\theta$ is such that $cos(\theta)$ is an irrational number (for eg., at $\theta=30^{\circ}$), such that no two sites can be completely equivalent. Consequently we see in Fig. \ref{SF00} (c) that all $6$ Fock states have different shift. However the shift corresponding to the state $|0,2,0 \rangle$ and $|1,0,1 \rangle$ are very close to each other and thus are not resolved in the plot.  In Fig. \ref{SFLN3} we have plotted the corresponding case of $\theta=30^{\circ}$ for a somewhat larger system, namely for $N=8, M=8, K=5$ which corresponds to a larger number of Fock states. 
  To increase the resolutions between the adjacent peaks we also choose $\kappa =0.01\delta_{0}$ that controls the width of each lorentzian. Nevertheless, some of these peaks correspond to more than one Fock state, where the difference  between the peaks of such Fock states cannot be resolved in the current plot.
As one can see, compared to $\theta=0^{\circ}, 60^{\circ}$ for the comparable values 
of $N,M,K$ plotted in Fig. \ref{SFLN1} and \ref{SFLN3}, the transmitted intensity has many more peaks.

\begin{figure}[H]
\centering
\subfloat[Part 1][]{\includegraphics[width=9cm, height=6.5cm]{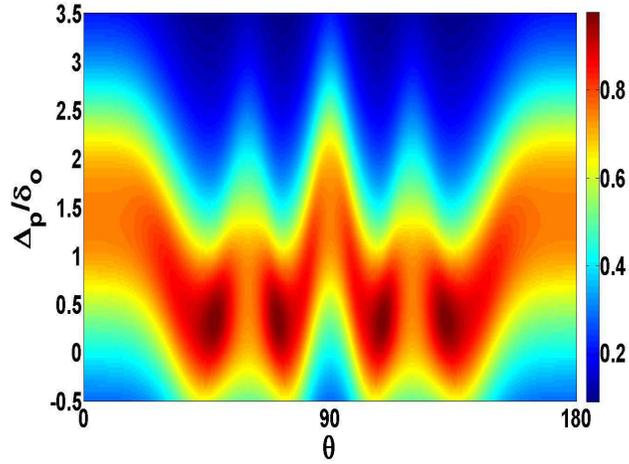}} 
\caption{(Color Online) Variation of the photon number (\ref{sf}) (color axis) with detuning $\frac{\Delta_{p}}{\delta_{0}}$ and $\theta$(in degrees) for N=K=2, M=3, when the atoms are in SF state and are illuminated by a single standing wave cavity mode for $\kappa$ =$\delta_{0}$, the spectrum is blurred.}
\label{SF11kappa}
\end{figure}

Thus the variation of the shift in the cavity frequency as a function of $\theta$ contains the information about the Fock states. 
This is demonstrated in Fig.{\ref{SF00}} (a) for a smaller system. As pointed out with the help of Fig. \ref{SFLN1} and \ref{SFLN3} some of these features should be detectable in relatively larger 
systems as well. 
The transmitted intensity from the cavity mode will be proportional to the photon number in that cavity mode. The color plot depicts this photon number whereas the black lines (explained in the legend of Fig.\ref{SF00}(a)
 correspond to the $F_{s}(\theta,K,n_{j})\delta_{0}$ for each Fock state as a function of $\theta$. As can be inferred wherever there is a maximum overlap of $F_{s}(\theta,K,n_{j})$ corresponding to different Fock states, the same location corresponds to the intensity peak. However an increase in the cavity decay rate $\kappa$ shows that the individual peaks cannot be resolved as shown in Fig.\ref{SF11kappa}. 

\subsection{Double mode}\label{dmode}
We shall now consider the case where two cavity modes are excited. The corresponding photonic annihilation operators are given by $\hat{a}_{0}$ and $\hat{a}_{1}$. Following \cite{mekhov1}  we also assume that the probe is injected only into $ \hat{a}_{0} $, and hence, $ \eta_{1} = 0$. Also both have the same frequencies, $i.e.,$ $\omega_{0} = \omega_{1}$ and are oriented at angles $ \theta_{0}$ and $\theta_{1}$ with respect to the optical lattice.
From Eq.(\ref{heisen}), $\dot{\hat{a}}_{1}=0$ yields  
\beq \hat{a}_{1} =  \frac{i\eta_{0} \delta_{1} \hat{D}_{10} e^{-i\Delta_{p}t}}{
([\Delta_{p} - (\hat{\omega}_{m} + \hat{\Omega}_{m})] + i \kappa)([\Delta_{p} - (\hat{\omega}_{m} - \hat{\Omega}_{m})] + i \kappa)}
\label{splitting} \eeq 
where 
\bea \hat{\omega}_{m} & = & \frac{\delta_{1}}{2}(\hat{D}_{11} + \hat{D}_{00})   \nonumber \\
\hat{\Omega}_{m} & = & \sqrt{\frac{\delta_{1}^{2} ({\hat{D}_{11} - \hat{D}_{00}})^{2}}{4} +   \delta_{1}^{2} \hat{D}_{10}^{\dag} \hat{D}_{10}} \label{omega} \eea 
are now operators acting on the Fock space. 
This leads to 

\beq \hat{a}_{1}^{\dag} \hat{a}_{1} = \frac{\delta_{1}^{2} \hat{D}_{10}^{\dag} \hat{D}_{10} |\eta_{0}|^{2}} {([\Delta_{p} - (\hat{\omega}_{m} + \hat{\Omega}_{m})]^2 +  \kappa^2)([\Delta_{p} - (\hat{\omega}_{m} - \hat{\Omega}_{m})]^2 +  \kappa^2)} \label{tmode} \eeq 
Here, $ \hat{D}_{01} = \hat{D}_{10}^{\dag} $ and the expectation value of $\hat{a}_{1}^{\dag} \hat{a}_{1}$ gives the photon number in this mode.

The above problem is equivalent to two linearized coupled harmonic oscillators which show mode splitting\cite{mekhov1}. Briefly two such harmonic oscillators with natural frequencies $\omega_{1} $ and $\omega_{2} $ coupled to each other by a perturbation $\zeta$, is described by the following set of coupled equations.
\bea
\frac{dx_{1}}{dt} &=& - i\omega_{1} x_{1} + \zeta x_{2}  \nonumber \\
\frac{dx_{2}}{dt} &=& - i\omega_{2} x_{2} + \zeta x_{1} \nonumber
\eea
 The normal modes of such a system are  
\beq
\omega = \frac{\omega_{1} + \omega_{2}}{2} \pm \sqrt{(\frac{\omega_{1} -  \omega_{2}}{2})^2 + {\zeta}^2} \label{harmonic}
\eeq
 
In the current problem, the shifted frequencies are given by the eigenvalues of
\beq \hat{\omega}_{m}\pm \hat{\Omega}_{m} \label{modes} \eeq
acting on a particular state of the system.
	                                                                                                                                                                                                                                                                                                                                                                                                                                                                                                                                                                                                                                                                                                                                                 A particular case of interest will be when 
$\theta_{0} = \theta_{1}$, this implies $ \hat{D}_{00} = \hat{D}_{11} = \hat{D}_{10}= \hat{D}$. Then the photon number $\hat{a}_{1}^{\dag} \hat{a}_{1}$ is
\beq
\hat{a}_{1}^{\dag}\hat{a}_{1} = \frac{\delta_{1}^{2}\hat{D}^{\dag}\hat{D}|\eta_{0}|^{2}}{[(\Delta_{p} - 2\hat{\Omega}_{m})^2 + \kappa^{2}][\Delta_{p}^{2} + \kappa^{2}]} \label{tteq}
\eeq
Thus one of the normal modes is  independent of the atomic dispersion, however the other mode disperses by twice the value for a single mode. 
We shall now consider the case when the cavity modes are Standing Waves. The many body ground state shall be considered as either a MI or SF phase.
\subsubsection{Mott Insulator}
Again we shall first calculate the two mode transmission spectrum when the cavity contains atomic ensemble in a MI state given in (\ref{mott}).
The operators $\hat{D}_{00}$ and $\hat{D}_{11}$ are given by (\ref{D00}) with eigenvalues  $F(\theta_{0},K)n$ and  $F(\theta_{1},K)n$. Also for such a MI state the operator $\hat{D}_{10}$ is given by
\beq
 \sum_{j=1:K} cos(j\pi cos\theta_{0})cos(j\pi cos\theta_{1}) \hat{n_{j}}
\eeq  
When this operator $\hat{D}_{10}$ acts on (\ref{mott}) its eigenvalue is given by $F(\theta_{0},\theta_{1},K) n$, where 
\bea
F(\theta_{0},\theta_{1},K) & = & \frac{1}{2} \Bigg[ \bigg(\frac{sin(K\pi \frac{cos\theta_{0} + cos\theta_{1}}{2}) }{sin(\pi \frac{cos\theta_{0} + cos\theta_{1}}{2})} cos((K +1) \pi \frac{cos\theta_{0} + cos\theta_{1}}{2}) \bigg)  \nonumber  \\
& & \mbox{} + 
 \bigg( \frac{sin(K\pi \frac{cos\theta_{0} - cos\theta_{1}}{2}) }{sin(\pi \frac{cos\theta_{0} - cos\theta_{1}}{2})} cos((K +1)\pi \frac{cos\theta_{0} - cos\theta_{1}}{2})\bigg)\Bigg]  \nonumber  \eea

 The photon number is hence given by 
\beq\langle \Psi_{MI} | \hat{a}_{1}^{\dag} \hat{a}_{1} |\Psi_{MI} \rangle = \frac{\delta_{1}^{2} |\eta_{0}|^{2} [F(\theta_{0},\theta_{1},K) n]^{2}} {([\Delta_{p} - (f + \mathcal{F})]^2 +  \kappa^2)([\Delta_{p} - (f - \mathcal{F})]^2 +  \kappa^2)} \label{inMI}
\eeq
where, 
\bea 
f &=&\langle \Psi_{MI}|\omega_{m}|\Psi_{ MI} \rangle= \frac{\delta_{1}}{2} (F(\theta_{0},K)+F(\theta_{1},K)) n\\ \nonumber \\
\mathcal{F} &= &\langle \Psi_{MI}|\Omega_{m}|\Psi_{ MI} \rangle = n\sqrt{\frac{\delta_{1}^{2}({ F(\theta_{1},K) - F(\theta_{0},K) )}^{2}}{4} + \delta_{1}^{2}[F(\theta_{0},\theta_{1},K)]^2} \label{modesplitting}
\eea
are the eigenvalues of the operators $\hat{\omega}_m$ and $\hat{\Omega}_m$ acting on MI state respectively. The normal modes are hence given by $f \pm \mathcal{F}$, and therefore the amount of mode splitting is given by $2\mathcal{F}$.
\begin{figure}[H]
\centering
\subfloat[Part 1][]{\includegraphics[width=9cm, height=6.5cm]{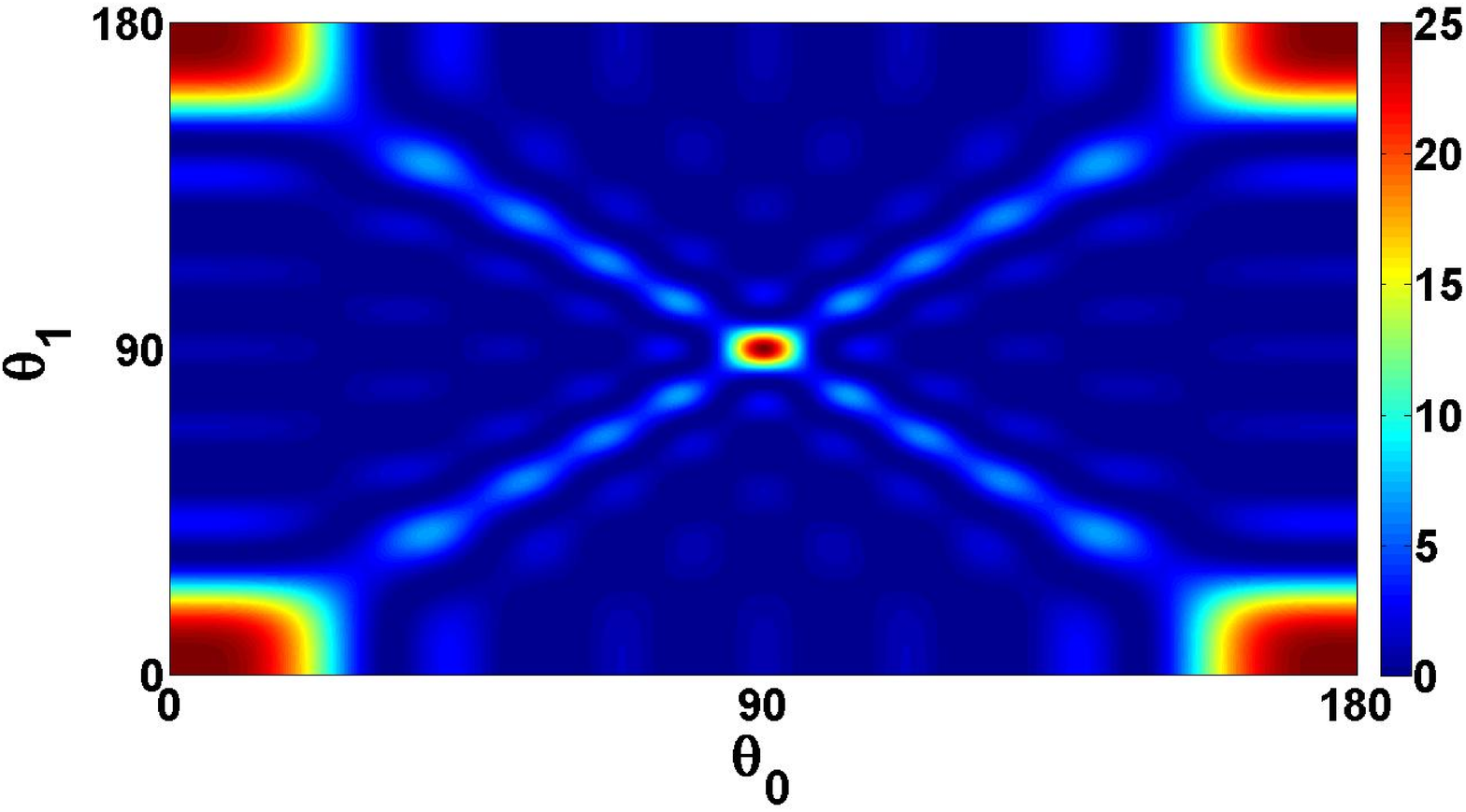}}
\subfloat[Part 2][]{\includegraphics[width=9cm, height=6.5cm]{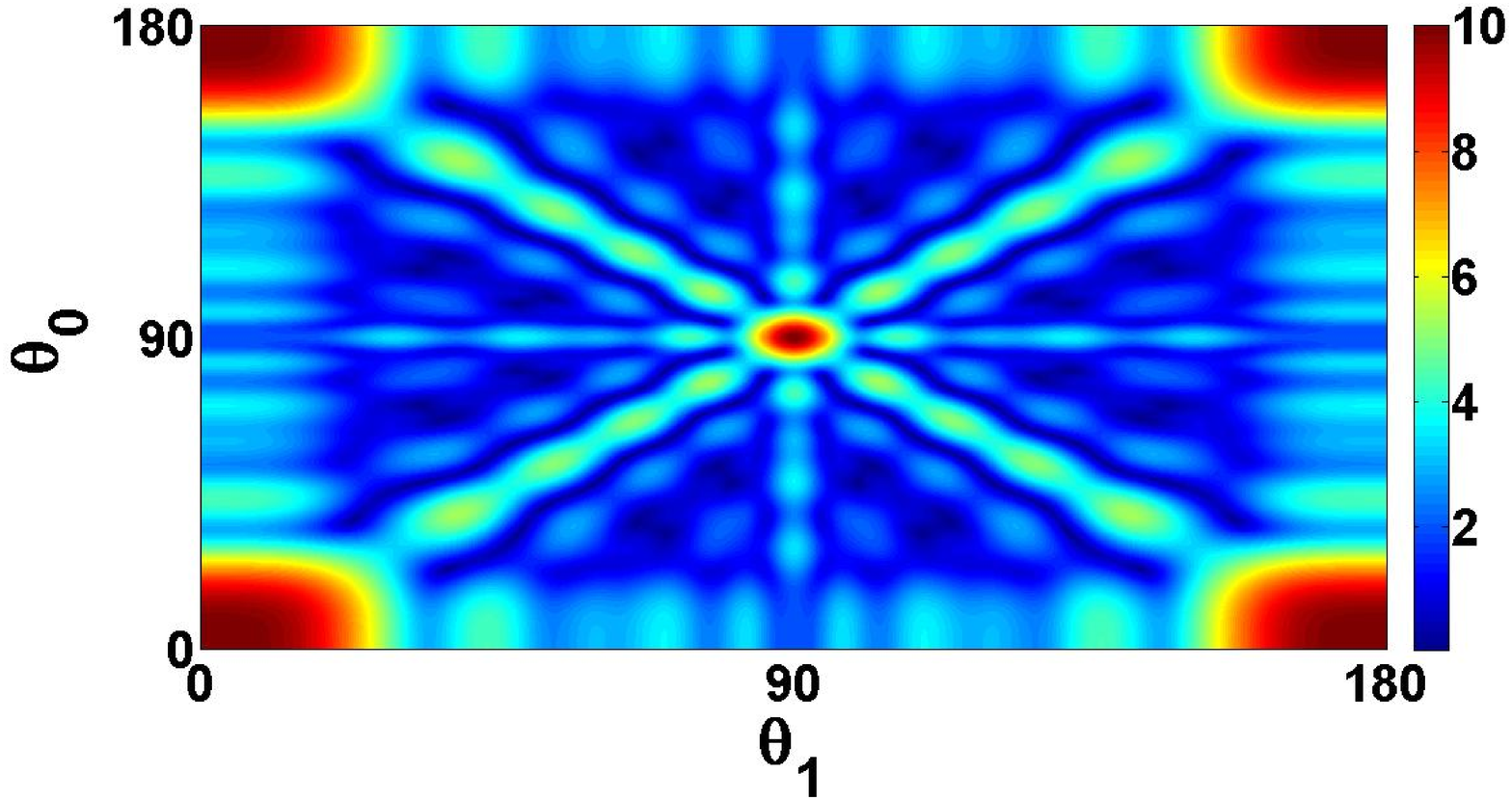}}
\caption{(Color Online)(a) shows the variation of $|F(\theta_{0},\theta_{1},K)|^{2}$(color axis) with the angles $\theta_{0}$ and $\theta_{1}$. (b) shows the mode splitting $2\mathcal{F}$ given in Eq.(\ref{modesplitting}) variation(color axis) in units of $\delta_{1}$ with angles $\theta_{0}$ and $\theta_{1}$,  N=5, K=5, M= 5. In both cases, $\theta_{0}$ and $\theta_{1}$ varies from $0^{\circ}$ to $180^{\circ}$ and they correspond to MI phase.}
\label{TT1}
\end{figure}
The photon number and the mode splitting, are given by the expressions (\ref{inMI}) and (\ref{modesplitting}) respectively. Both are dependent on the value $|F(\theta_{0},\theta_{1},K)|^2$ and are thus related to each other. Fig. \ref{TT1} (a) shows the plot of the $|F(\theta_{0},\theta_{1},K)|^2$ for $n=1$ MI state and  Fig. \ref{TT1} (b) shows the mode splitting at specific values of $\theta_{0}$ and  $\theta_{1}$ and thus very clearly demonstrates their inter-dependence.

\begin{figure}[H]
\centering
\subfloat[Part 1][]{\includegraphics[width=9cm, height=6.5cm]{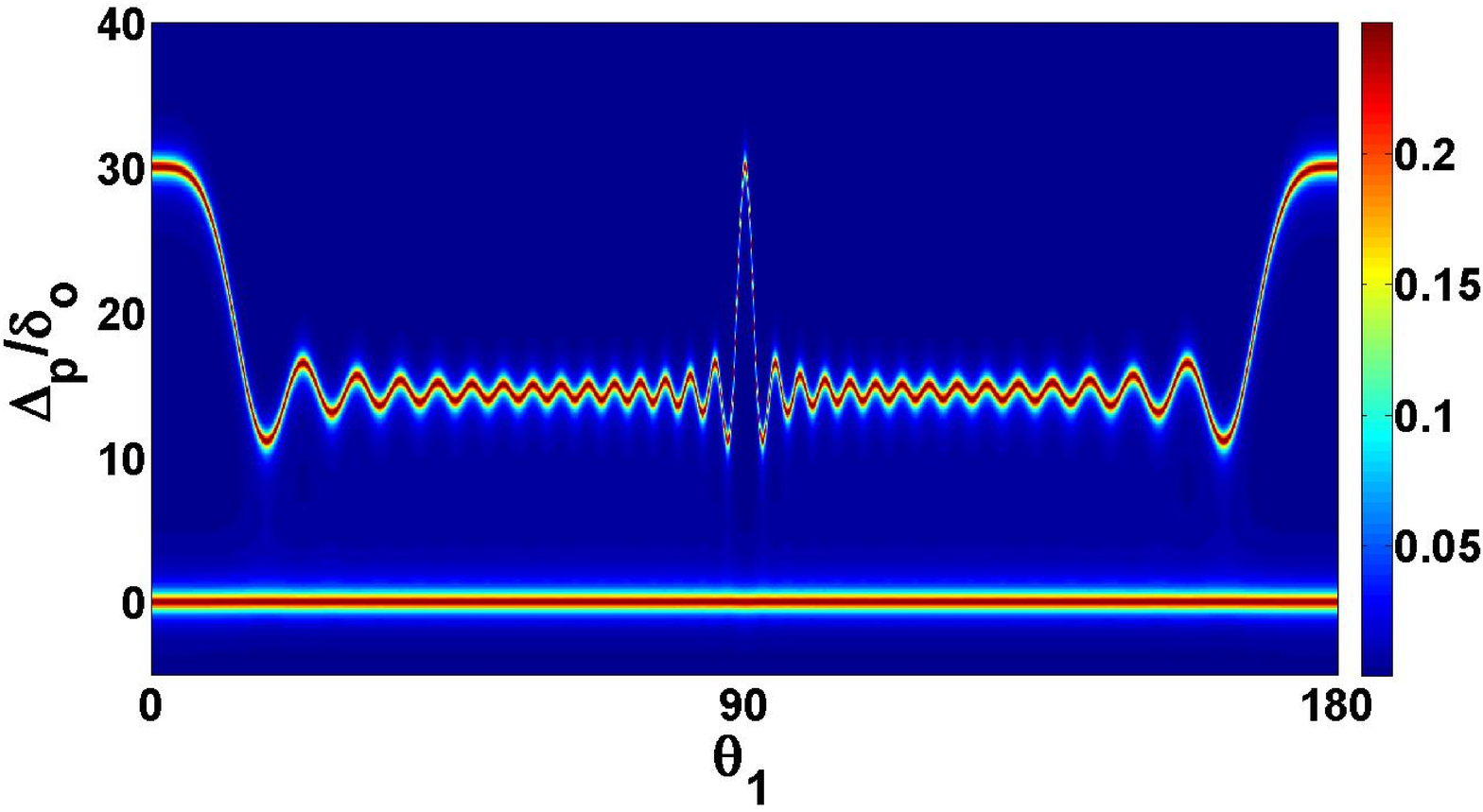} }
\subfloat[Part 1][]{\includegraphics[width=9cm, height=6.5cm]{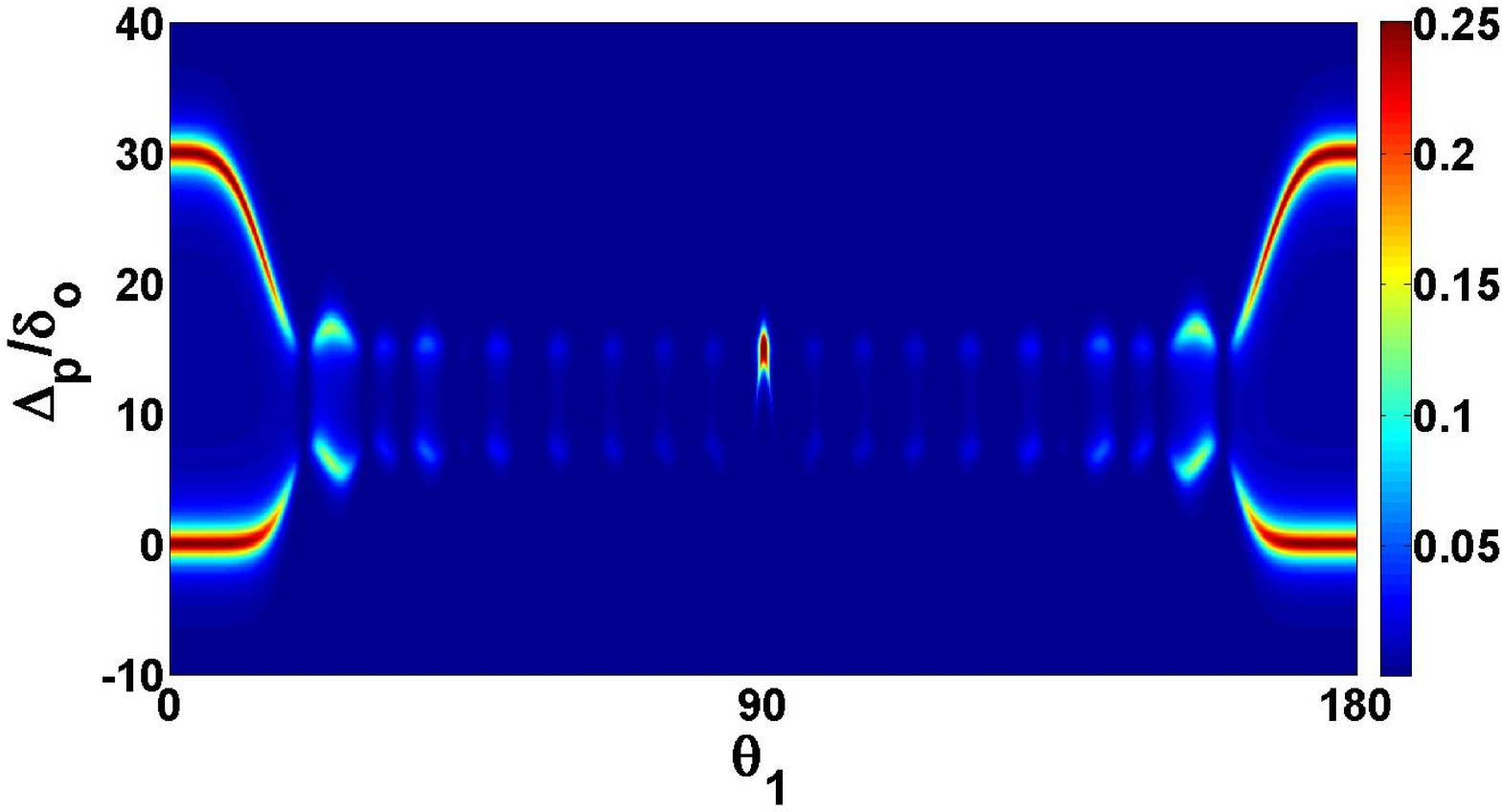} }\\ 
\subfloat[Part 1][]{\includegraphics[width=9cm, height=6.5cm]{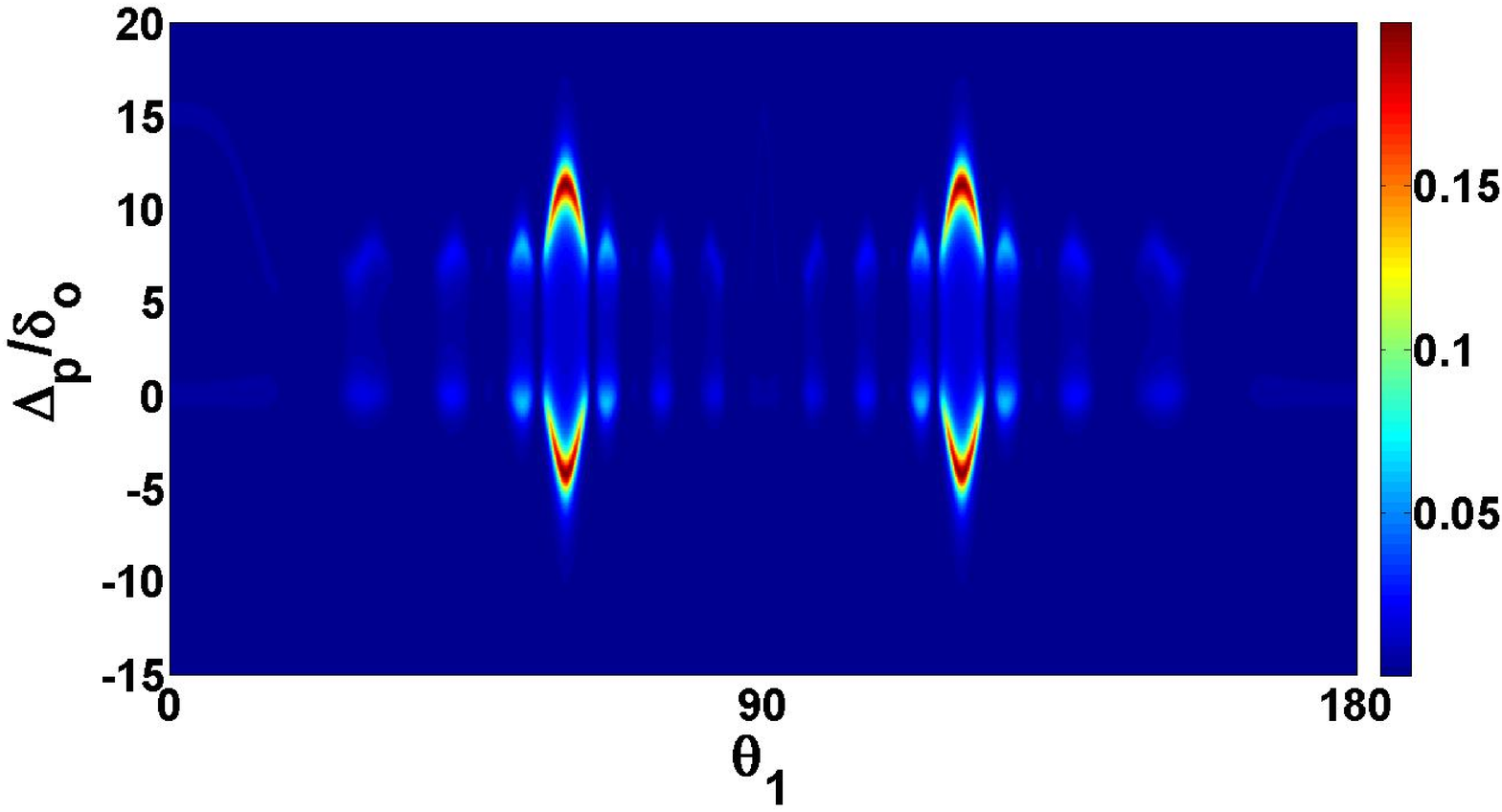} }
\subfloat[Part 1][]{\includegraphics[width=9cm, height=6.5cm]{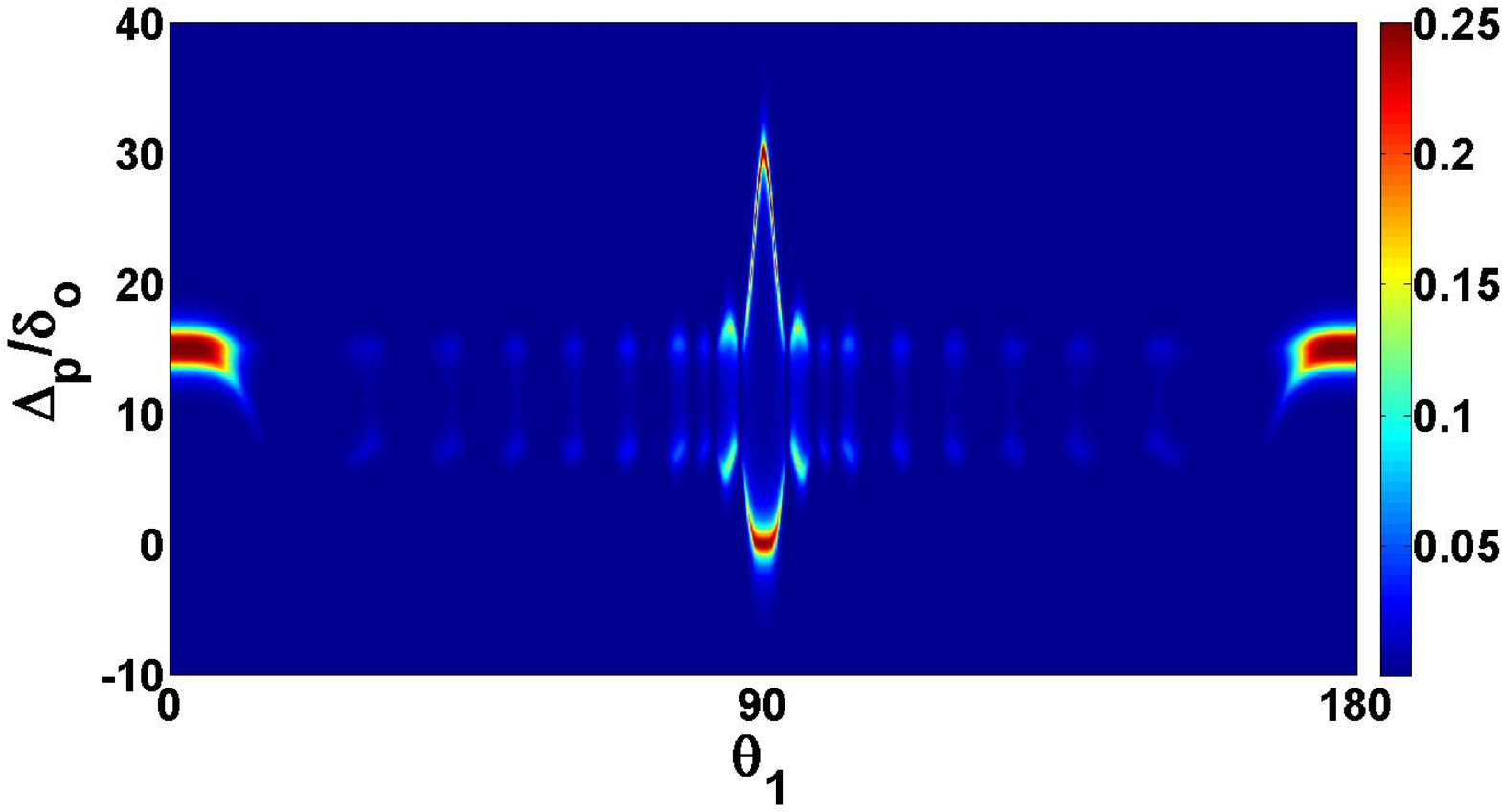}}
\caption{(Color Online) Variation of the photon number (\ref{inMI}) (color axis) with detuning $\frac{\Delta_{p}}{\delta_{0}}$ and the angles $\theta_{0}$ and $\theta_{1}$(in degrees) for N=M=30, K=15, when the atoms are in MI state, for double standing mode case. In (a)$\theta_{0}$=$\theta_{1}$, $\kappa$=$0.5\delta_{0}$. (b) shows the variation when $\theta_{0}$=$0^{\circ}$, $\theta_{1}$ is varied and $\kappa =\delta$. Similar cases are shown in (c) with $\theta_{0}$=$60^{\circ}$ and in (d) when $\theta_{0}$= $90^{\circ}$} \label{M21}
\end{figure}

This relation is also reflected in the plots of resulting transmission at certain demonstrative values of $\theta_{0},\theta_{1}$ as plotted in Fig. \ref{M21} and as explained below.

First we consider the case when both $\theta_{0}$  and $\theta_{1}$ are being varied from $0^{\circ}$ to $180^{\circ}$ 
always maintaining the relation $\theta_{0}=\theta_{1}$. The corresponding  $F(\theta_{0},\theta_{1},K)$ function shows a number of maxima along the line $\theta_{0} =\theta_{1}$ in Fig. \ref{TT1}(a). The photon number here is given by the expressions (\ref{tteq}, \ref{inMI}) where it was seen the normal modes will be zero and $2\mathcal{F}$. In Fig. \ref{M21}(a) we have plotted this variation in photon number along the color axis as a function of  $\theta_{1}$ and $\Delta_{p}/\delta_{0}$.Therefore at each $\theta_{1}$ one gets a maxima at a value $\frac{\Delta_{p}}{\delta_{0}}$ = 0 and at twice the value for a single mode case(Fig. \ref{M11}). 
 
However, in Fig. \ref{M21} (b), (c) and (d), we describe the case when $\theta_{0}$ is kept constant, while the other angle $\theta_{1}$ is constantly being varied. Corresponding plots show that the maximum number of photons scattered from one mode at an angle $\theta_{0}$ will be collected by $\hat{a}_{1}$ only when $\theta_{1}$ = $\pm \theta_{0}$ or $\pi \pm \theta_{0}$. When $\theta_{1}$ = $\theta_{0}$, the second mode is parallel to the first mode. When $\theta_{1}$ = -$ \theta_{0}$, angle of scattering is equal to the angle of reflection. This is also observed at $\pi \pm \theta_{0}$ \cite{mekhov2}. It is at this co-ordinate, the  $|F(\theta_{0},\theta_{1},K)|^2$ , mode splitting as well as the transmitted intensity will show maximum behavior. This is also seen from the $\theta_{0}$ = $\theta_{1}$ and $\theta_{0}$= $\pi - \theta_{1}$ lines in Fig. \ref{TT1}. For example in Fig. \ref{M21} (c),when  $\theta_{0}$ = $60^{\circ}$, the plot shows maximum mode splitting and intensity at  $\theta_{1}$ =$ 60^{\circ}$ and $120^{\circ}$.

One can also study the diffraction pattern of such system in the limit where the shift in the cavity frequency due to dispersion given in (\ref{tmode}) is neglected.
In that case the transmitted intensity will be directly proportional to the eigenvalue of $\hat{D}_{10}^{\dag}\hat{ D}_{10}$. This particular limit has been explored in \cite{mekhov2,mekhov6,mekhov8} for the two mode case and  shown to consist of two parts.
The first part is  due to classical diffraction and second part shows fluctuations from such classical pattern.
The above analysis in this work suggest an enrichment of these diffraction features to a considerable extent  once the frequency shift due to diffraction is taken into account. 
\subsubsection{Superfluid} 
Now we consider that the cold atomic condensate is in SF ground state.
The transmission through the SF can be obtained by taking the expectation value of
the photon number operator (\ref{tmode}) in a SF state. This gives 
\beq\langle \Psi | \hat{a}_{1}^{\dag} \hat{a}_{1} | \Psi \rangle = \frac{1}{M^{N}} \sum_{\langle n_{j} \rangle}\frac{N!}{n_{1}!n_{2}!...n_{M}!}\frac{\delta_{1}^{2} [F_s(\theta_{0},\theta_{1},K,n_{j})]^{2}|\eta_{0}|^{2}} {([\Delta_{p} - (f_{n_{j}} + \mathcal{F}_{n_{j}})]^2 +  \kappa^2)([\Delta_{p} - (f_{n_{j}} - \mathcal{F}_{n_{j}})]^2 +  \kappa^2)}
\label{inSF} \eeq
where,
\bea f_{n_{j}} & = & \frac{\delta_{1}}{2}(F_s(\theta_{1},K,n_{j}) + F_s(\theta_{0},K,n_{j}))   \nonumber \\
\mathcal{F}_{n_{j}} & = & \sqrt{\frac{\delta_{1}^{2} {(F_s(\theta_{1},K,n_{j}) - F_s(\theta_{0},K,n_{j}))}^{2}}{4} +   \delta_{1}^{2}[F_s(\theta_{0},\theta_{1},K,n_{j})]^{2}} \nonumber \eea 
are respectively the eigenvalues of $\hat{\omega}_{m}$ and $\hat{\Omega}_{m}$ operators acting on a particular
Fock state. These are in terms of $F_s(\theta_{0},K,n_{j})$ functions which were first described in (\ref{D00sf}). $F_s(\theta_{0},\theta_{1},K,n_{j})$ is given by 
\beq  F_s(\theta_{0},\theta_{1},K,n_{j}) = \sum_{j=1:K} cos(j\pi cos\theta_{0})cos(j\pi cos\theta_{1})n_{j} \label{fs} \eeq 


\begin{figure}[H]
\centering
\subfloat[Part 1][]{\includegraphics[width=9cm, height=6.5cm]{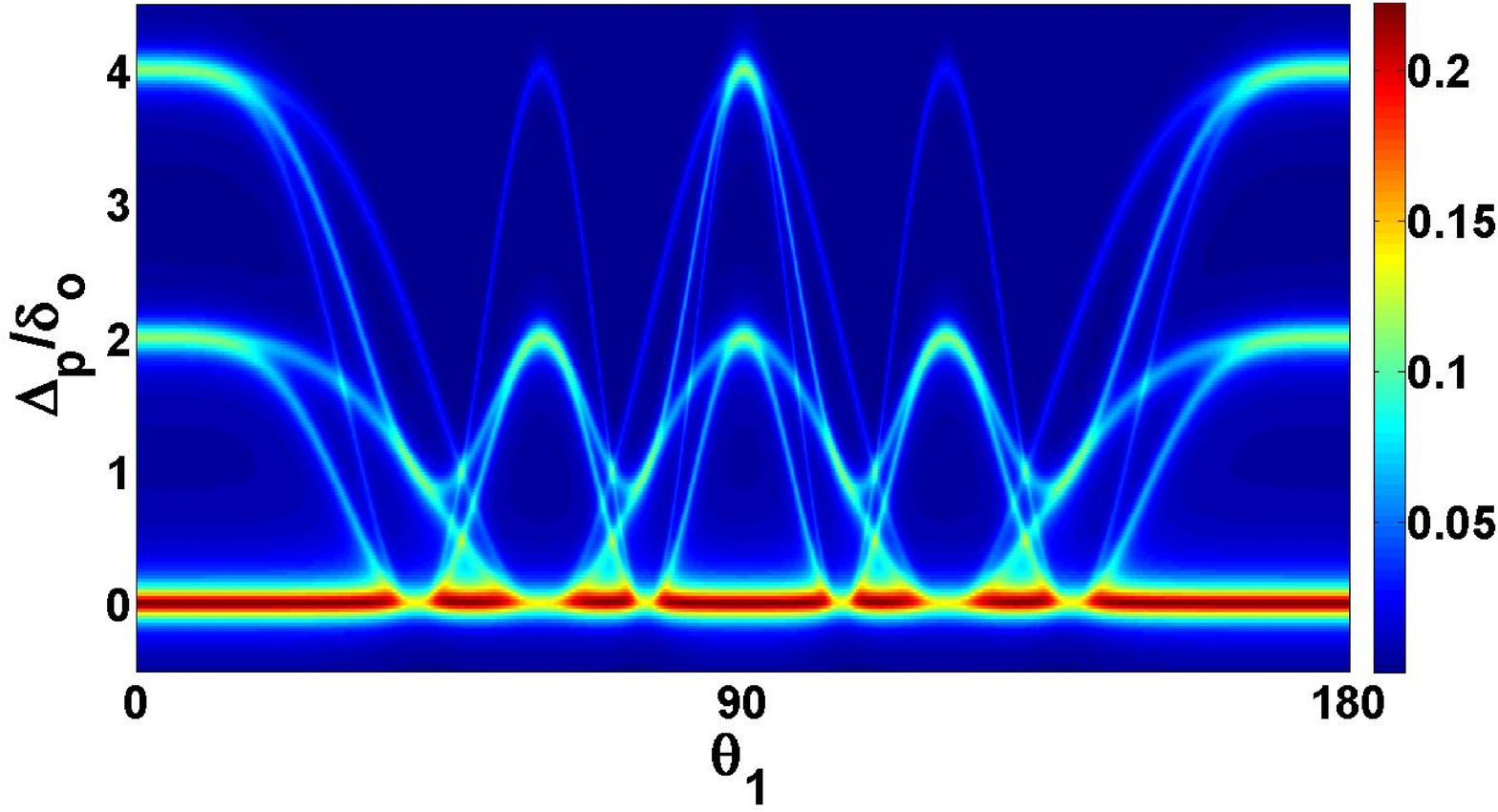}}
\subfloat[Part 1][]{\includegraphics[width=9cm, height=6.5cm]{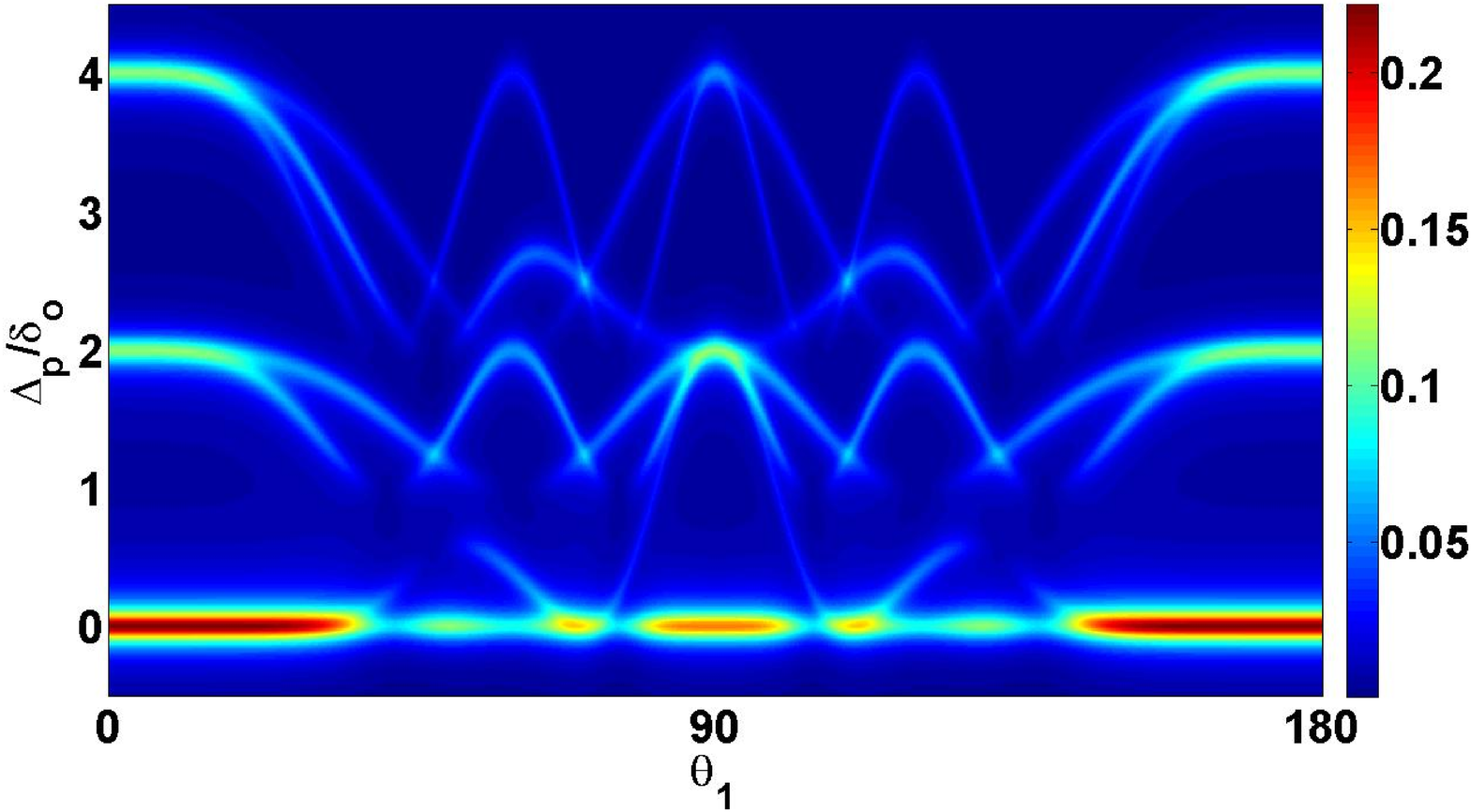}}\\
\subfloat[Part 1][]{\includegraphics[width=9cm, height=6.5cm]{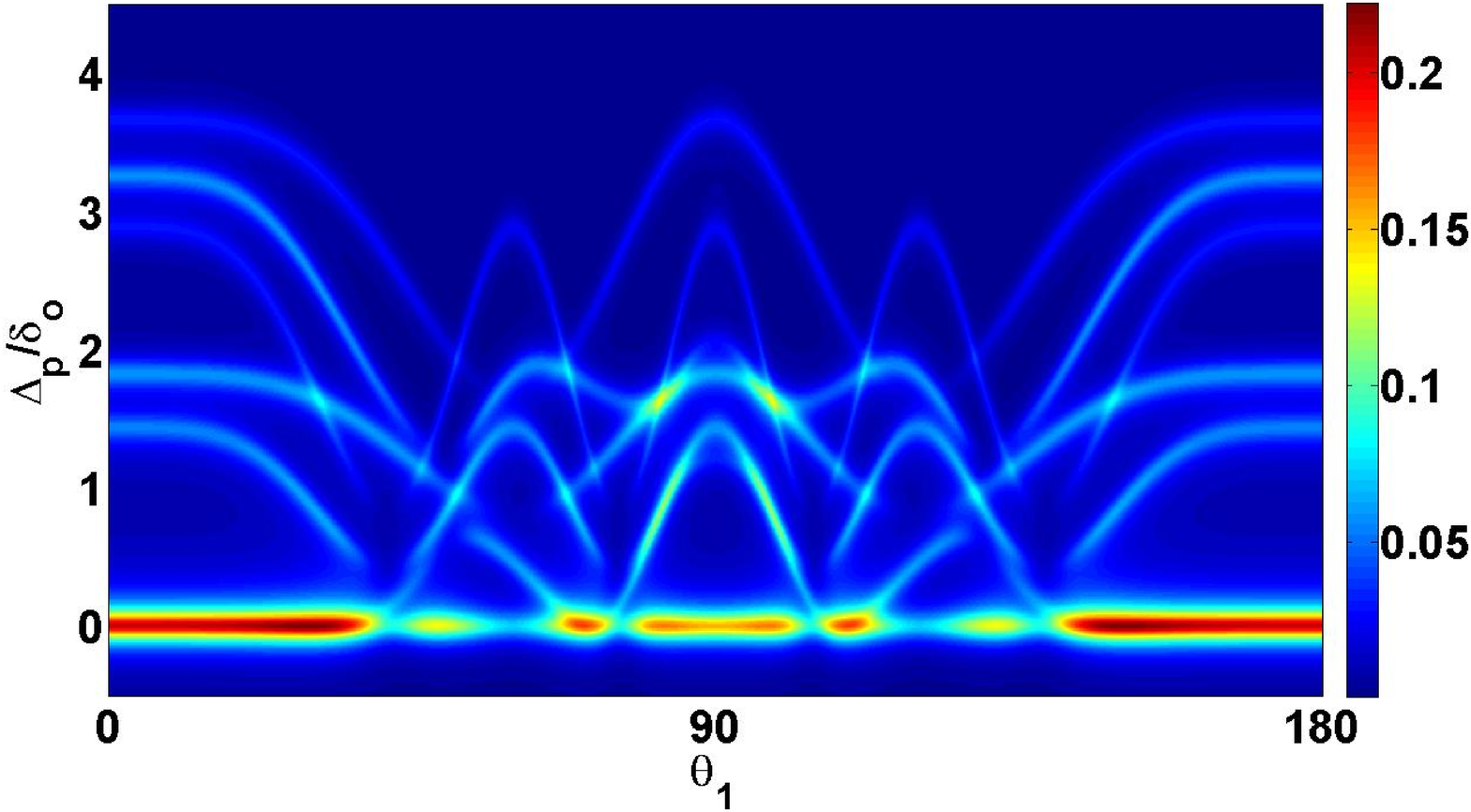}}
\subfloat[Part 1][]{\includegraphics[width=9cm, height=6.5cm]{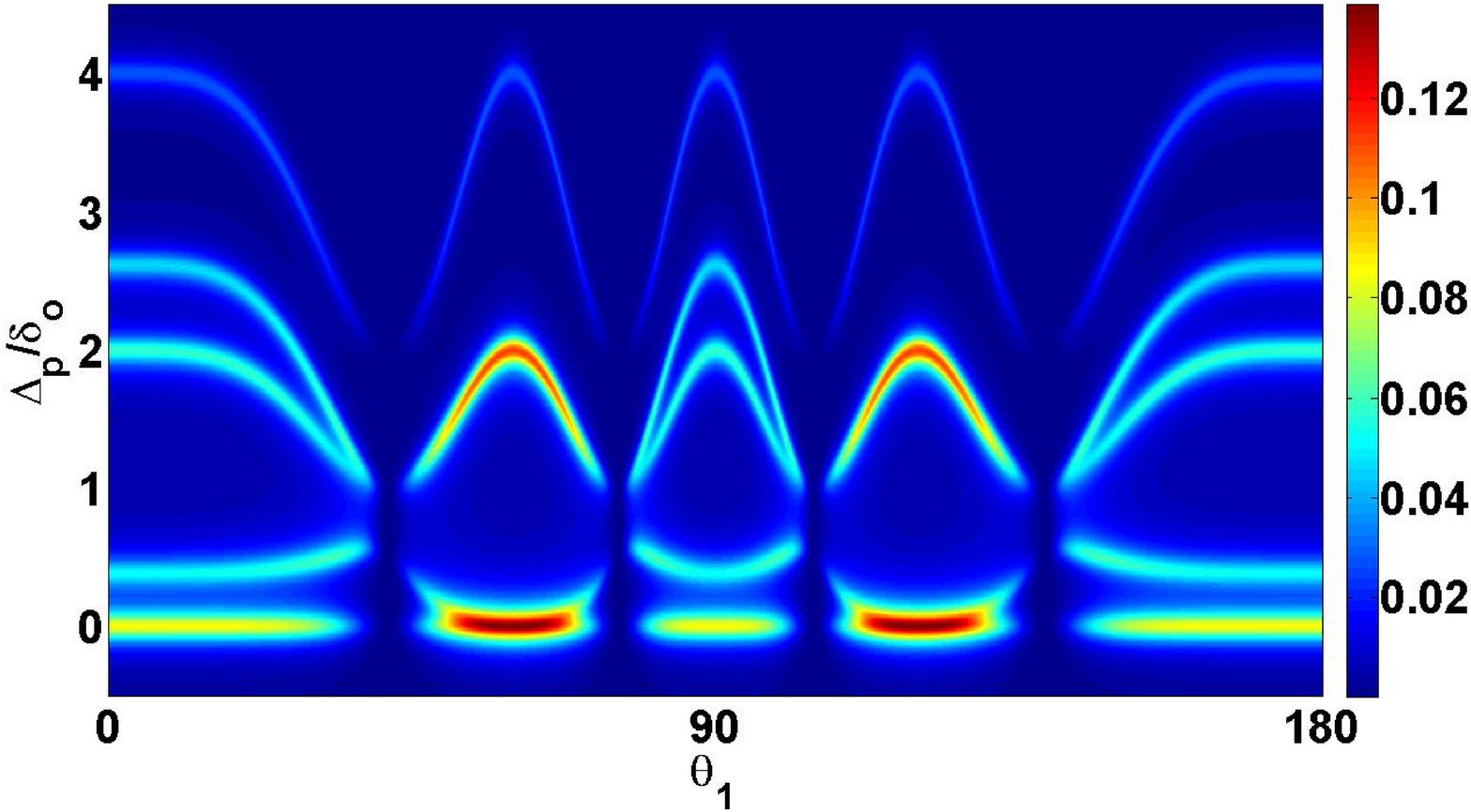}}
\caption{(Color Online)Variation of the photon number (\ref{inSF}) (color axis) with detuning $\frac{\Delta_{p}}{\delta_{0}}$ and angles $\theta_{0}$ and $\theta_{1}$(in degrees) for N=K=2, M=3, $\kappa$=$0.1\delta_{0}$ when the atoms are in SF state for double standing mode case; in (a)$\theta_{0}$=$\theta_{1}$ (b) $\theta_{0}$= $0^{\circ}$  (c)$\theta_{0}$= $30^{\circ}$ (d)$\theta_{0}$= $60^{\circ}$ }
\label{SF21}
\end{figure}
In Fig. \ref{SF21} (a), the cavity modes are oriented at the same angle with the lattice axis ie. $\theta_{0} = \theta_{1}$, and are together varied from $0^{\circ}$ to $180^{\circ}$. Here $F_s(\theta_{0}, K, n_{j})=F_s(\theta_{1}, K, n_{j}) = F_s(\theta_{0},\theta_{1}, K, n_{j})$ and thus corresponds to the case described in Eq.(\ref{tteq}). In this figure, the photon number has been plotted against the angle $\theta_{1}$ and $\Delta_{p}/\delta_{0}$. The plot exhibits that at each angle there is maxima in the photon number when the value of the dispersive shift is either zero or twice its corresponding value for the single mode case. 

\begin{figure}[H]
\centering
\subfloat[Part 1][]{\includegraphics[width=6cm, height=5cm]{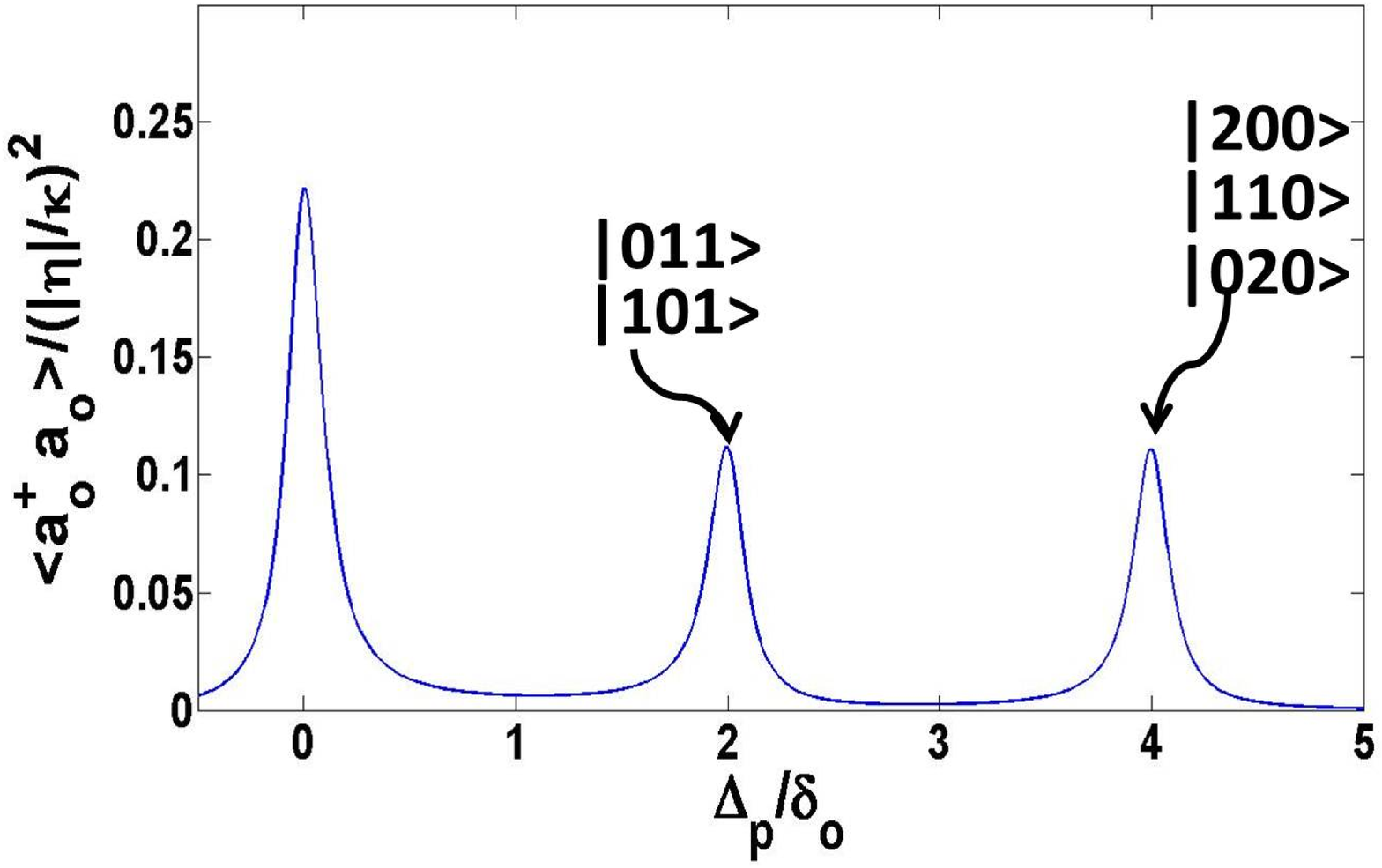}}
\subfloat[Part 1][]{\includegraphics[width=6cm, height=5cm]{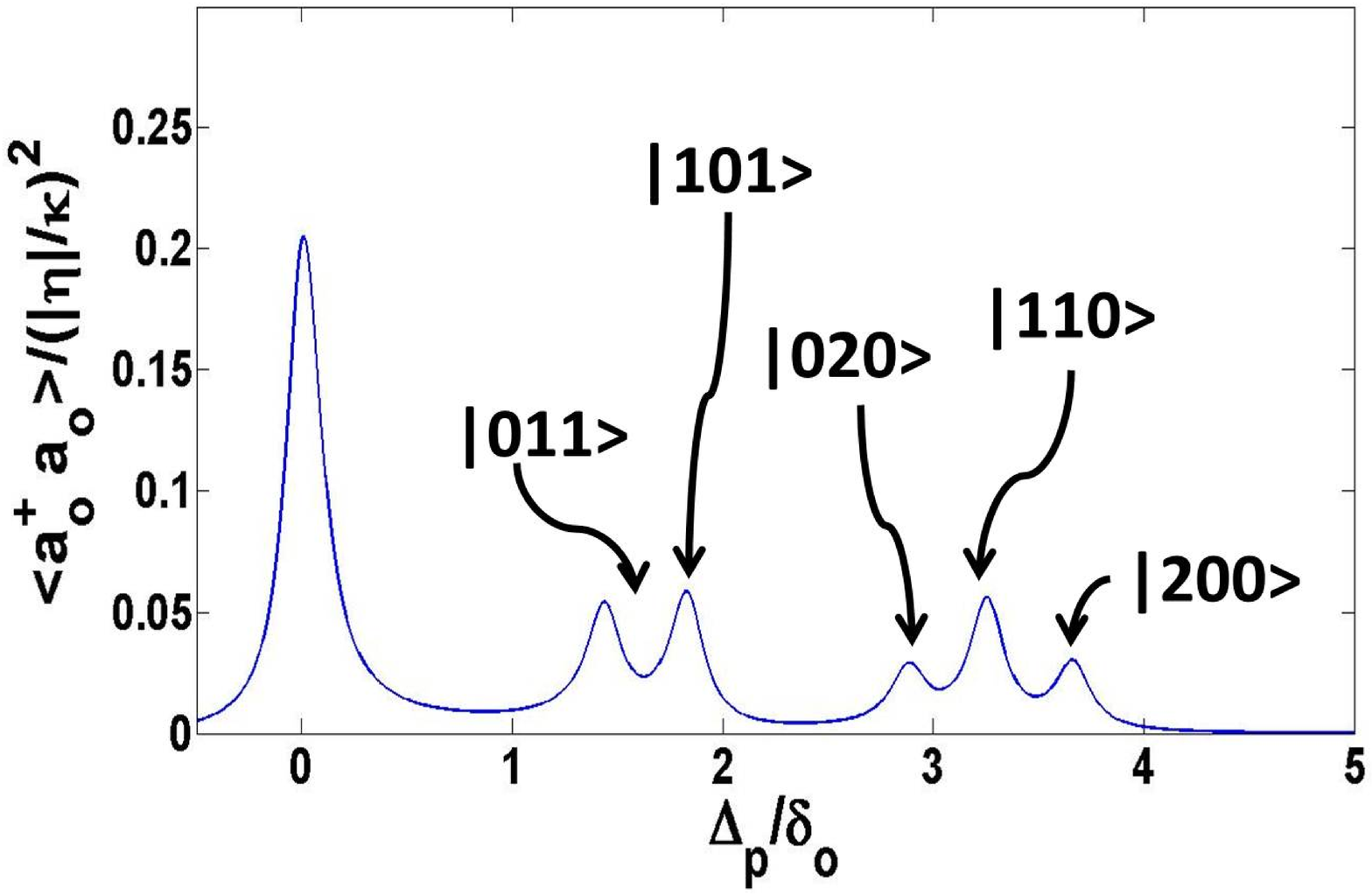}}
\subfloat[Part 1][]{ \includegraphics[width=6cm, height=5cm]{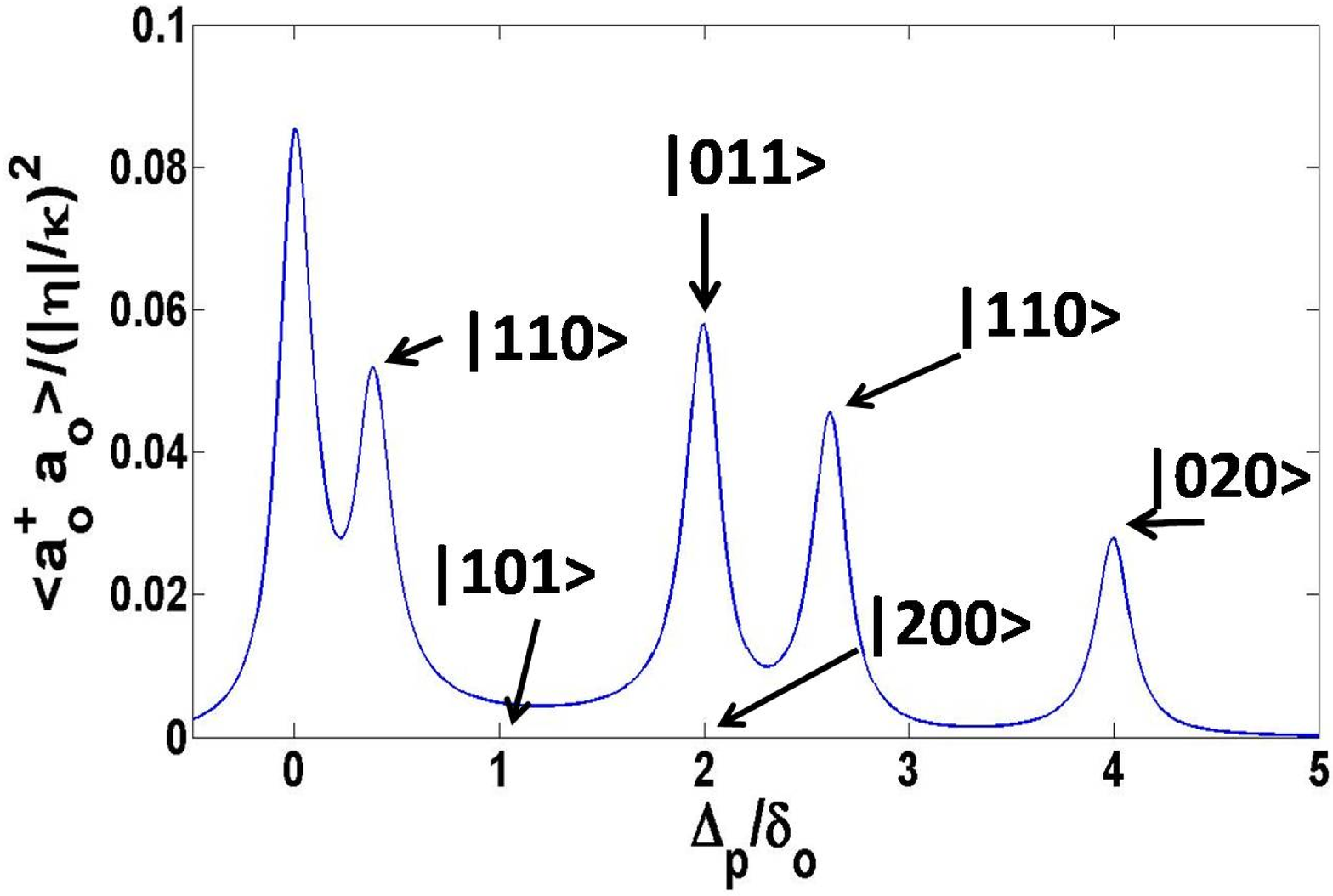} }\\
\subfloat[Part 1][]{\includegraphics[width=18cm, height=10cm]{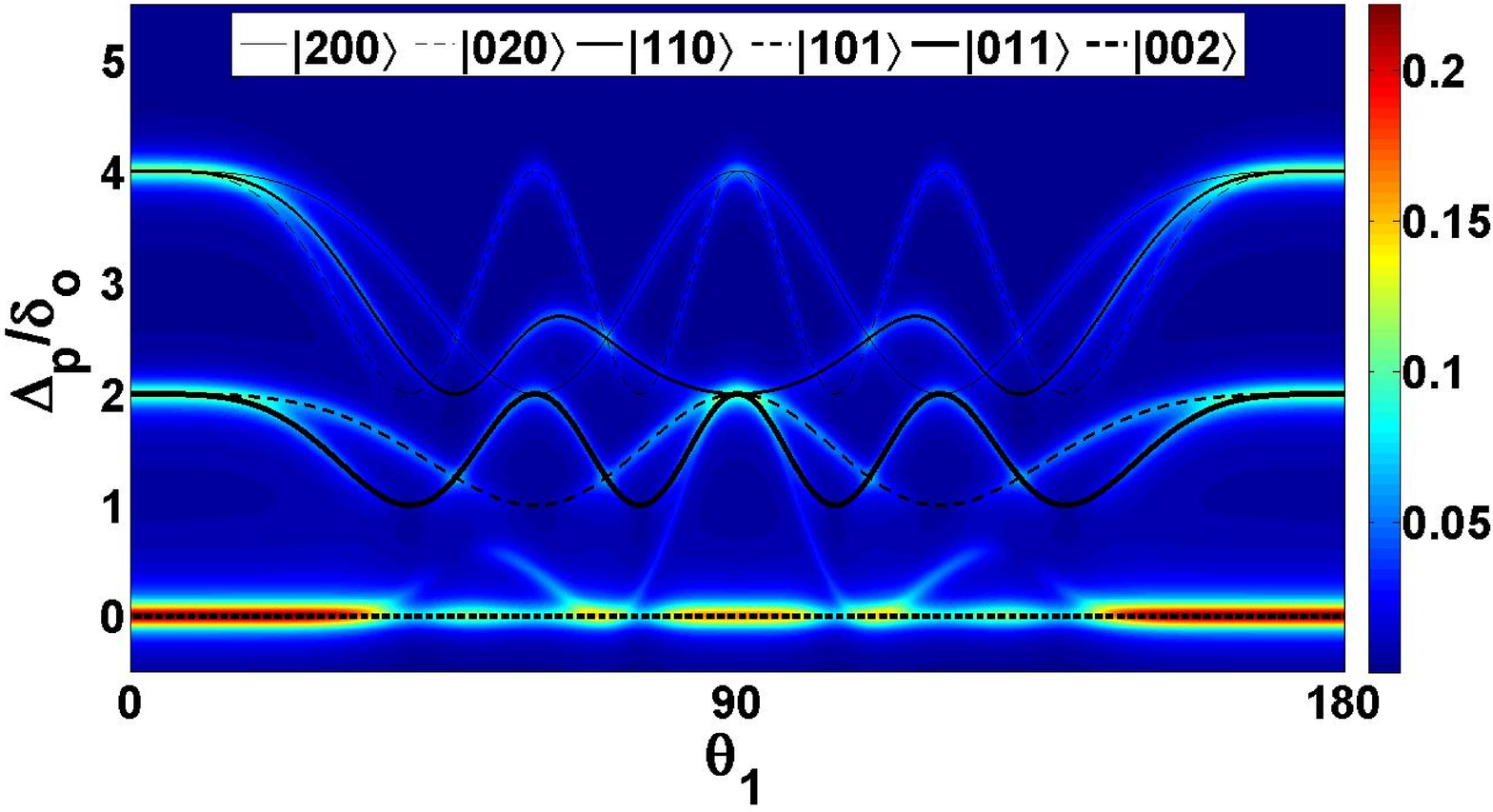}}
\caption{(Color Online) For N=K=2, M=3 . Plots (a)-(c)are the two dimensional plots for photon number with respect to $\Delta_p / \delta_0$ for  $\theta_0 = 0^\circ$ and different values of $\theta_1$. (a)$\theta_{1} = 0^{\circ}$. (b)$\theta_{1} = 30^{\circ}$ (c) $ \theta_{1} = 60^{\circ}$. Particularly here we have Fock states $|2,0,0 \rangle$ and $|1,0,1 \rangle $ that correspond to zero transmitted 
intensity located at finite value of $\frac{\Delta_{p}}{\delta_{0}}$ for the higher normal  mode frequency. 
(d) In this plot the  black lines   shows how  $\frac{f_{n_{j}} + \mathcal{F}_{n_{j}}}{\delta_{0}}$, the higher normal mode frequency for each Fock state varies as a function of $\theta_{1}$ when 
$\theta_{0}=0$. This has been superposed with the figure $\ref{SF21}(b)$.}
\label{SF2SW}
\end{figure}

Fig.\ref{SF21} (b), (c) and (d) describes the variation of the photon number ( transmitted intensity)  with the angle $\theta_{1}$ and $\Delta_{p}/\delta_{0}$, when  $\theta_{0}$ is kept constant. 
This makes $ F_s(\theta_{0}, K, n_{j})$, $F_s(\theta_{1},K,n_{j})$ and $F_s(\theta_{0},\theta_{1},K, n_{j})$ change separately for individual Fock states. To understand the features of the 
transmitted intensity better, we again consider the case when 2 atoms are placed in 3 sites among which 2 are illuminated. We will have 6 Fock states and now each Fock state gives intensity peaks for two values of $\Delta_{p}/\delta_{0}$ corresponding to the two normal modes $f_{n_{j}} \pm \mathcal{F}_{n_{j}}$.

Fig.\ref{SF21}(b), shows the intensity distribution when $\theta_{0} = 0^{\circ}$. Here, we observe that when $\theta_{1}$=$0^{\circ}$, the 6 Fock states distribute themselves into groups having 
$3,2,1$ states for the higher value of the two normal modes and each group has a distinct value for the dispersion shift depending on the occupancy. This is demonstrated more clearly with the help of 
two dimensional plots in Fig. \ref{SF2SW} (a). For each Fock state, the frequency shift corresponding to the higher of the two normal modes is twice of its value 
for the single mode case(Fig. \ref{SF00} (b)). Thus the difference between adjacent lorentzians has increased.
For the lower value of the normal modes, this frequency shift is zero when $\theta_{0}=\theta_{1}$ according to (\ref{tteq}). 
Thus the peak at $\frac{\Delta_{p}}{\delta_{0}} =0$ correspond to five Fock states from the lower branch. The state $| 0, 0, 2 \rangle$ 
where there is no atom on the illuminated sites corresponds to $\frac{\Delta_{p}}{\delta_{0}}=0$ as well as zero intensity. This happens for either of the normal modes.

In Fig. \ref{SF2SW} (d) we show how frequency shift at which the transmission peak occurs for each  Fock state varies with a change in $\theta_1$ while $\theta_0 = 0^{\circ}$. In the same figure
also for each Fock state  we show the corresponding angular variation of  the higher normal mode i.e.  $f_{n_{j}} + \mathcal{F}_{n_{j}}$ to show their interrelation with such transmission peaks. For $\theta_1= 30^{\circ}$, as  $F_s(\theta_{1},K,n_{j})$ will be different for each Fock state, the frequency shift for each Fock state will be separate. In the related two dimensional plot, given in Fig. \ref{SF2SW}(b), where we get six peaks thus distinctively mapping each Fock state for the higher normal mode. Again, the peak corresponding to $|0,0,2 \rangle$ where there is no atom on the illuminated sites is located 
at $\frac{\Delta_{p}}{\delta_{0}} =0$ and also has transmitted intensity $0$ for both the lower as well as higher normal mode frequency.In Fig. \ref{SF2SW}(b) the unmarked intensity peak in the left 
corresponds to the transmission peak of five other Fock states for the lower value of the normal mode frequency. Similarly in Fig. \ref{SF2SW}(c) we plot the transmitted intensity of all the  normal mode frequencies at $\theta_{1} = 60^{\circ}$ to show the grouping of the Fock states at a given intensity peak. 

\subsection{More general cases with two modes}
In the above analysis we set $\omega_{0}=\omega_{1}$. In a general case these two mode frequencies will be different and consequently 
various features associated with the mode splitting and the transmission spectrum described in the previous section will also change.  Particularly, for different mode frequencies $\omega$ the atom light coupling constant $g$ will be different since it is given by\cite{Knight}
\beq
g =\sqrt{\frac{d^2 \omega}{2 \hbar {\epsilon}_0 V}}   \label{diffmode} \eeq
Here  $d$ is the atomic dipole moment, $\epsilon_{0}$ is the free space permittivity, $V$  is the mode volume. 
According to the  expression ( \ref{diffmode})  the change of the mode profile or the cavity geometry  that change $V$, also leads to a change in $g$. 

For two different mode frequencies for which we denote the photon annihilation operators respectively as $\hat{a}_{0}$ and $\hat{a}_{1}$, the steady state solutions Eq. 
(\ref{heisen}) yields

\beq \hat{a}_{1}^{\dag} \hat{a}_{1} = \frac{\delta_{0}^{2} \hat{D}_{01} \hat{D}_{10} |\eta_{0}|^{2}} {([\Delta_{p} - (\hat{\omega}_{m} + \hat{\Omega}_{m})]^2 +  \kappa^2)([\Delta_{p} - (\hat{\omega}_{m} - \hat{\Omega}_{m})]^2 +  \kappa^2)} \label{tmodediff} \eeq 
Here again we are pumping the first mode, and, 
\bea \hat{\omega}_{m} & = & \frac{(\delta_{0}\hat{D}_{00} + \delta_{1}\hat{D}_{11})}{2}   \nonumber \\
\hat{\Omega}_{m} & = & \sqrt{\frac{(\delta_{0}\hat{D}_{00} - \delta_{1}\hat{D}_{11})^{2}}{4} +   \delta_{0}\delta_{1} \hat{D}_{01} \hat{D}_{10}} \label{omegadiff} \eea 

with \bea D_{00}  & = & cos^2(\frac{ 2\pi md}{\lambda_0} cos(\theta_0)) \nonumber \\
D_{11} & = & cos^2(\frac{ 2\pi md}{\lambda_1} cos(\theta_1)) \nonumber \\
D_{10}  & = & cos(\frac{ 2\pi md}{\lambda_1} cos(\theta_1))cos(\frac{ 2\pi md}{\lambda_0} cos(\theta_0)) \nonumber \\
D_{01}  & =  & cos(\frac{ 2\pi md}{\lambda_0} cos(\theta_0))cos(\frac{ 2\pi md}{\lambda_1} cos(\theta_1)) \eea 

For two different  mode frequencies the corresponding wavelengths will also be different. Hence  the number of illuminated sites corresponding to two different modes will be different from each other.
However, for a typical experimental case for ultra cold $^{87} Rb$ \cite{Bren1},  the atomic transition frequency $\omega_{a}= 3 \times 10^{5}$ GHz for $D_{2}$ line.  
The typical value of the cavity mode frequency is also in the optical range  and will be of the order of $10^{5}$ GHz. On the other hand the typical value of the cavity atom detuning 
parameter $|\Delta_{la}|= |\omega_{l} - \omega_{a}|$ in a typical experiment varies in the range $0-100$GHz \cite{Bren1, Colom1}. 
Thus the ratio $\frac{\omega_{l}}{ |\Delta_{la}|}$ is typically $ > 10^{3}$. 
This means that if the frequency of the two modes $\omega_{0}, \omega_{1}$
are slightly different from each other that will induce a large change in the corresponding ratio 
$\frac{\delta_{0}}{\delta_{1}} $ making it $\gg 1$.    

\begin{figure}[H]
\centering
\subfloat[Part 1][]{\includegraphics[width=18cm, height=10cm]{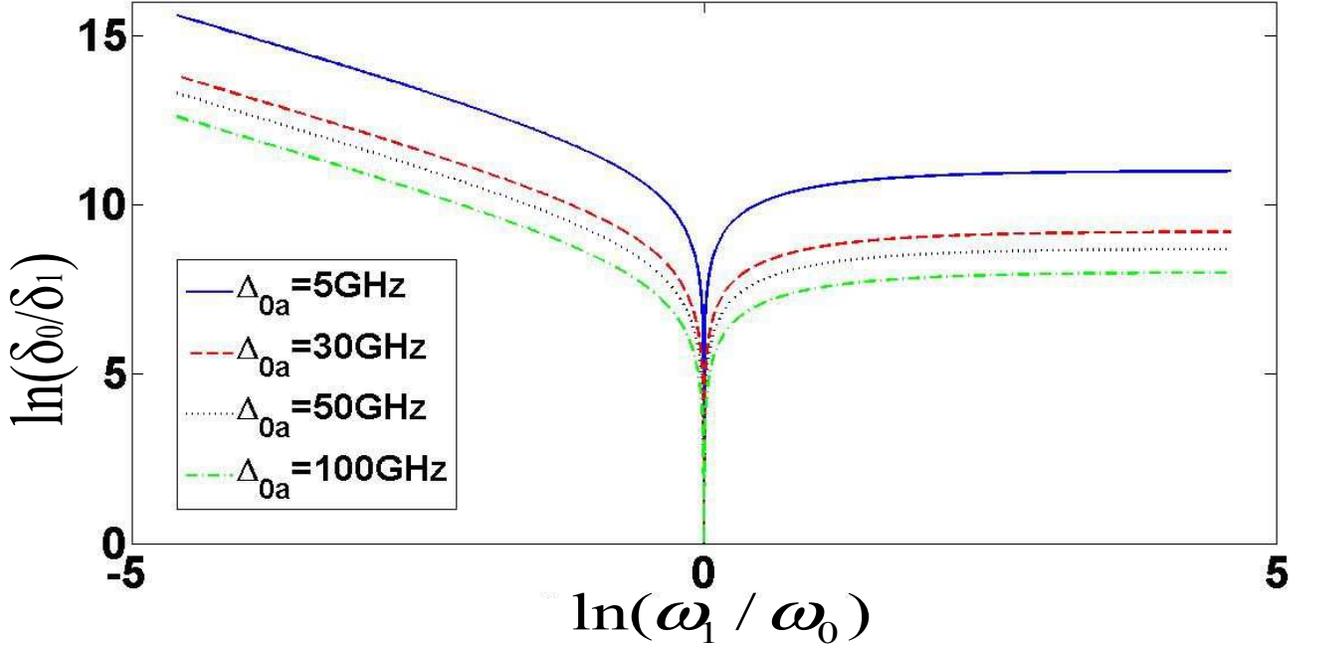}}\\
\subfloat[Part 1][]{\includegraphics[width=9cm, height=6.5cm]{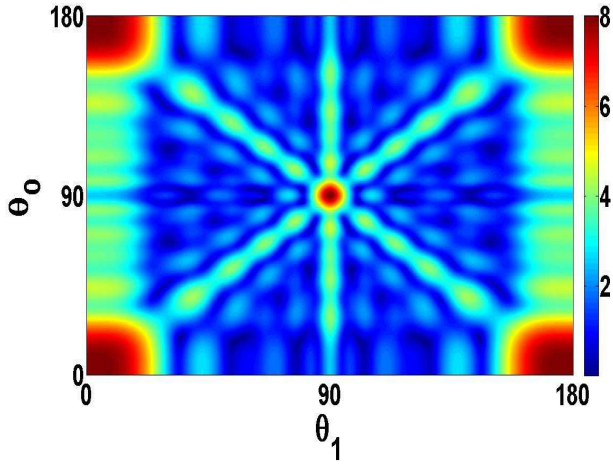}}
\subfloat[Part 1][]{\includegraphics[width=9cm, height=6.5cm]{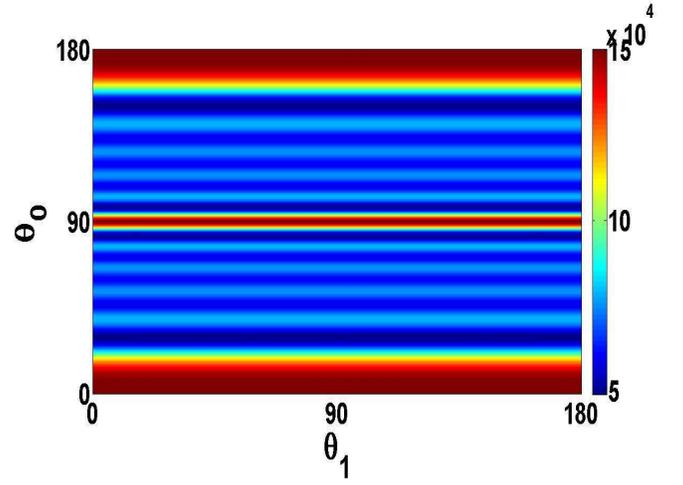}}
\caption{ (Color Online) In (a) Variation of $ln(\frac{\delta_{0}}{\delta_{1}})$ with $ln(\frac{\omega_{1}}{\omega_{0}})$ for $\Delta_{0a}= 5GHz$, $\Delta_{0a}= 30GHz$, $\Delta_{0a}= 50GHz$ and  $\Delta_{0a}= 100GHz$ . We see that in all these different values of $\Delta_{0a}$, the plots show that even for a small change in $\omega_{1}/ \omega_{0}$, the corresponding $\delta_{0}/\delta_{1}$ varies quite markedly. In (b) and (c), Mode splitting , given by the eigenvalue of $2\hat{\Omega}_m$, variation(color axis) in units of $\delta_{1}$ with angles $\theta_{0}$ and $\theta_{1}$,  N=5, K=5, M= 5. In both cases, $\theta_{0}$ and $\theta_{1}$ varies from $0^{\circ}$ to $180^{\circ}$. (b) shows the case when $\frac{\delta_{0}}{\delta_{1}}=0.5$ and $\frac{\omega_{1}}{\omega_{0}}=0.999992$, $\Delta_{0a}=5GHz$. (c) shows the mode splitting when $|\frac{\delta_{0}}{\delta_{1}}|=60000$ and $\frac{\omega_{1}}{\omega_{0}}=0.5$, $\Delta_{0a} = 5GHz$.}
\label{MSdelta}
\end{figure}

Fig. \ref{MSdelta} (a) depicts the above mentioned behavior through a log-log plot where  we plot the variation in $ln(\omega_{1}/\omega_{0})$,with $ln(\delta_{0}/\delta_{1})$ for different values of $\Delta_{0a}$. It shows a sharp dip at $\omega_0 = \omega_1$ since $\frac{\delta_{0}}{\delta_{1}}=1$ at this point.  Away from this point $\delta_1 \ll \delta_{0}$ as $\omega_{1}$ differs from $\omega_{0}$
even by a small fraction, because the ratio $\frac{\omega_{0}}{|\Delta_{0a}|}$ is of the order of $10^{3} - 10^{4}$, $|\Delta_{1a}| \gg |\Delta_{0a}|$.  Physically this means the atomic transition frequency cannot couple effectively with the mode frequency $\omega_1$ and hence $\theta_1$ cannot anymore serve as a tuning parameter. This can be seen in the Fig.\ref{MSdelta} (b)-(c) where we have plotted the mode splitting function for a Mott Insulator $(N=M=K=5)$. In Fig.\ref{MSdelta} (b) with $\omega_0$ and $\omega_1$ are almost equal, $\delta_0/\delta_1$ is 0.5 and modifies the mode splitting plot from the previously shown $\delta_0=\delta_1$ figure \ref{TT1} (b) . However  in Fig. \ref{MSdelta} (c) for a value of $\omega_1= 0.5 \omega_0$ ($\Delta_{0a}$ = 5GHz), $|\delta_0/\delta_1|$ is  60000 and we observe that the function becomes independent of $\theta_1$. 

In the literature some other variants of the  interaction between atoms and two cavity modes were also considered for two level \cite{Gou} and three level atoms with 
$\Lambda$ configuration \cite{Gerry}. However we have not considered such cases here. 

The above  analysis 
suggests that to achieve extra tuning parameter in 
the two-mode case, one should have two nearly degenerate modes. Unless there is some sort of degeneracy, two different modes in the same cavity are separated from each other by 
different harmonics and  in such a situation the transmission spectrum as well as the mode splitting is dependent on only one of the angles. 

It is also possible to have two different modes in the same cavity. If the frequencies of these two modes are different, then the corresponding analysis will be similar to the  one in the preceding section. However it is also possible to have degenerate modes with different polarization. In such cases if the interaction between light and atom is sensitive to the polarization degrees of freedom 
then the transmission spectrum will also be dependent on the polarization direction. Such situations, however in the absence of  a cavity was considered recently \cite{Burnett}. 
 We have not explicitly done this analysis. Other possible cases are where the mode functions will have a different spatial dependence as compared to the plane wave type considered here.
However interaction between such cavity modes and atoms will be an interesting case of study for ultra cold atoms in higher dimensional optical lattices.  
\section{Travelling waves}\label{TW}

\begin{figure}[H]
\begin{center}
\includegraphics[width=9cm, height=7cm]{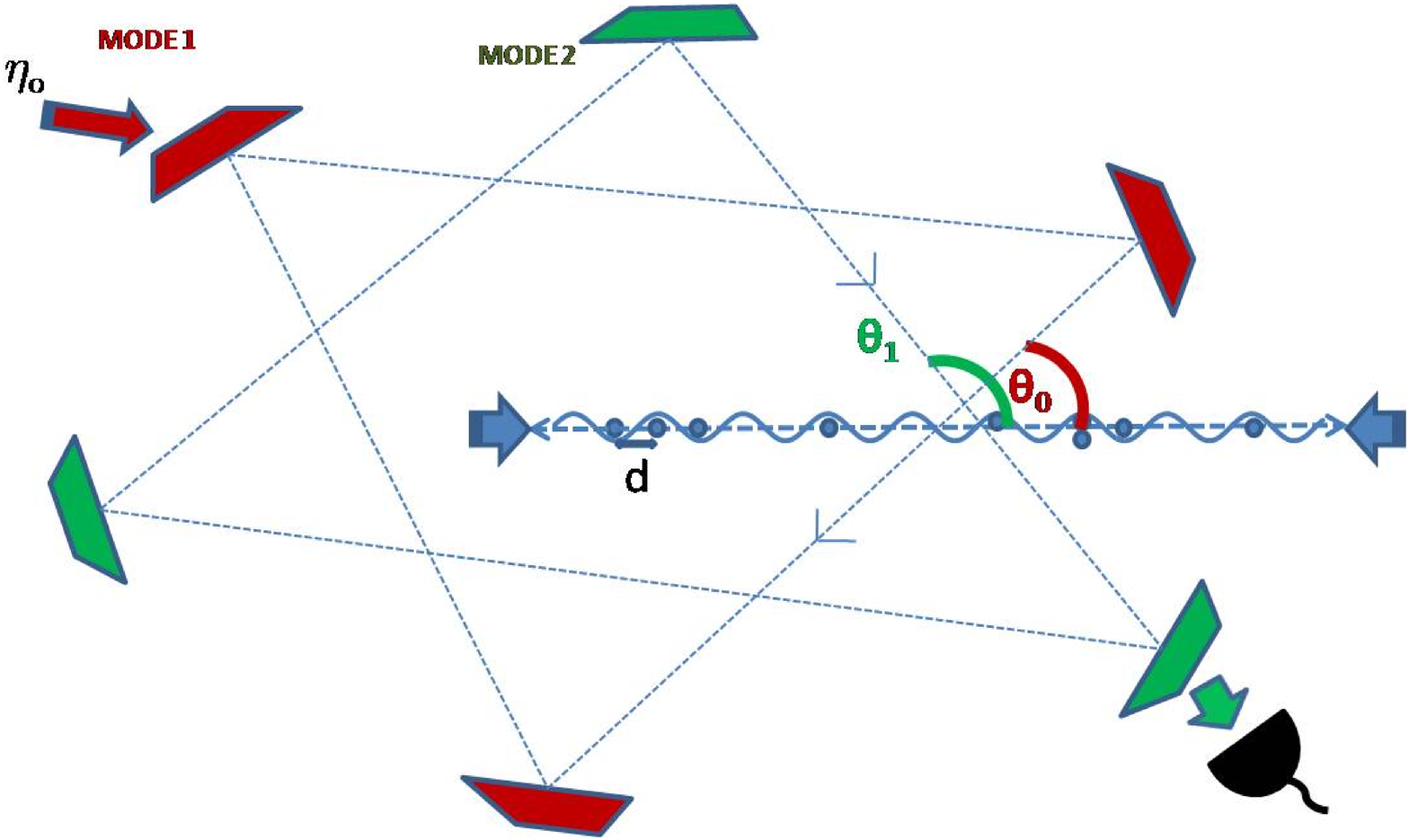}
\caption{(Color Online) Schematic diagram of the atom cavity system for travelling wave. The optical lattice is created from two counter propagating laser beams and has a site spacing $d$. The two ring wave cavity modes, MODE 1 and MODE 2 are at angles $\theta_{0}$ and $\theta_{1}$ respectively with the axis of the optical lattice. The MODE 1 is being pumped by a pump laser with amplitude $\eta_{0}$ while MODE 2 is not being pumped but is used to collect the scattered photons by a detector. In the single mode case, the detector is also fixed in MODE 1, and has not been shown in this figure. The ring cavities are set in a way that the waves are allowed to propagate only in one direction.}
\label{S2}
\end{center}
\end{figure}

 Fig. \ref{S2} depicts our model system where an optical lattice is shown to be illuminated by two ring cavities. We have considered that these cavities allow the waves to propagate only in one direction. Such cavities generate travelling wave modes \cite{ring,Bux, Nagorny}. These modes are described by \cite{mekhov1,ring}, $u({\bf{r}}_{j})_{TW} = \exp( i( \bs{k} \cdot \bs{r}_{j} + \phi))$ where $\phi$ is constant phase factor which has been set to zero. For such  waves, the operator  $\hat{D}_{00}$ becomes just
\beq
\hat{D}_{00} =\sum_{j=1:K} u_{l}^{*} u_{l} \hat{n_{j}} =  \sum_{j=1:K}\hat{n}_{j} 
\eeq
as $u_{l}^{*} u_{l} =1$. Thus the eigenvalue  of $\hat{D}_{00}$  for a given Fock state will be $nK$ which is just the number of atoms in the illuminated sites.  Thus the dispersive shift in single mode case will not  depend on the angle $\theta$.  
However this is not the case when two cavity modes are excited.
 As we have seen for the case of standing wave, the mode splitting which in turn influences the transmission 
through such cavity is closely related to the relative angle between the two modes through the function
$F(\theta_{0},\theta_{1},K)$ . In the current case the mode splitting is also dependent on 
the relative angle between the cavity modes since only the eigenvalue of the operator $\hat{D}_{10}$ is angle dependent which is given by,
\beq
\hat{D}_{10} = \sum_{j=1:K} e^{i(j\pi (cos\theta_{1} - cos\theta_{0}))} \hat{n}_{j}
\\ \label{D10}
\eeq
\subsection{Mott Insulator}
Again, we first consider the cold atomic condensate in a MI state. The eigenvalue of  $\hat{\omega}_{m}$ and $\hat{\Omega}_{m}$ (\ref{omega}) when acting on the MI state (\ref{mott}) are
\bea g&= &\langle \Psi |\hat{ \omega}_{m} |\Psi \rangle = \frac{nK\delta_{1} + nK\delta_{1}}{2} = nK\delta_{1}  \nonumber \\
\mathcal{G}&=& \langle \Psi |\hat{ \Omega}_{m} |\Psi \rangle = \sqrt { (\frac{nK\delta_{1} - nK\delta_{1}}{2})^2+ |G(\theta_{0},\theta_{1}, K) n\delta_{1}|^2} = |G(\theta_{0},\theta_{1}, K)| n\delta_{1} \eea
Here $G(\theta_{0},\theta_{1}, K)$ is
\beq
G(\theta_{0},\theta_{1}, K) = \frac{sin(K\pi \frac{cos\theta_{0} - cos\theta_{1}}{2}) }{sin(\pi \frac{cos\theta_{0} - cos\theta_{1}}{2})}
\eeq 

This system is equivalent to two coupled linearized harmonic oscillators (\ref{harmonic}), but with same natural frequencies ie., $\omega_{1}=\omega_{2}= \omega_{\circ}$ and coupled by a perturbation $\zeta$. The normal modes for such a system is given by  $\omega_{\circ} \pm \zeta$. In the current problem, the normal modes are hence given by $g \pm \mathcal{G}$ and therefore the amount of mode splitting is $2\mathcal{G}$.

The photon number (\ref{tmode}) is, 
\beq\langle \Psi_{MI} | \hat{a}_{1}^{\dag} \hat{a}_{1} |\Psi_{MI} \rangle = \frac{|\eta_{0}\mathcal{G}|^{2}} {([\Delta_{p} - (g + \mathcal{G})]^2 +  \kappa^2)([\Delta_{p} - (g - \mathcal{G})]^2 +  \kappa^2)} \label{inMITW}
\eeq 

\begin{figure}[H]
\centering
\subfloat[Part 1][]{\includegraphics[width=9cm, height=6.5cm]{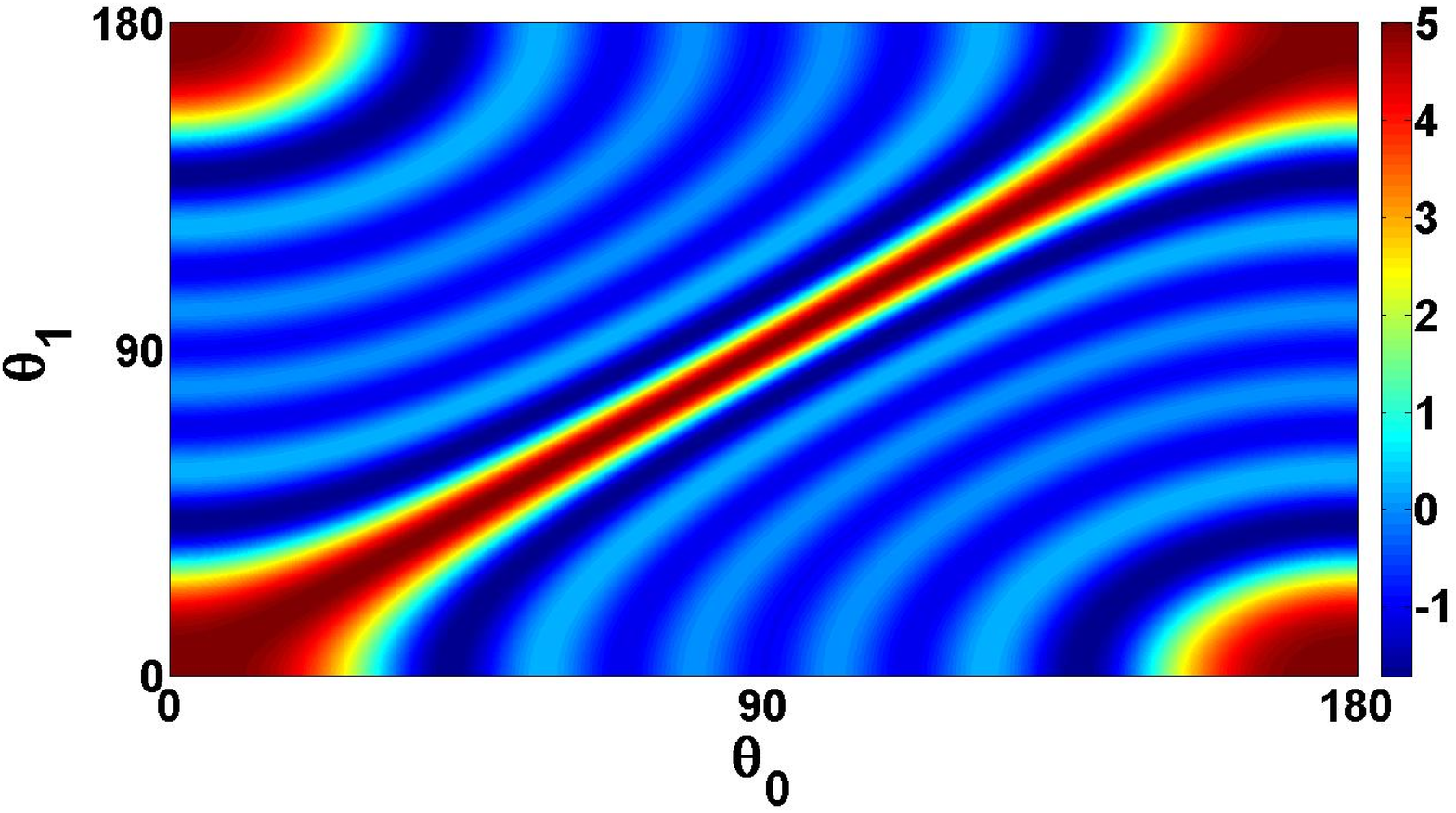}}
\subfloat[Part 1][]{\includegraphics[width=9cm, height=6.5cm]{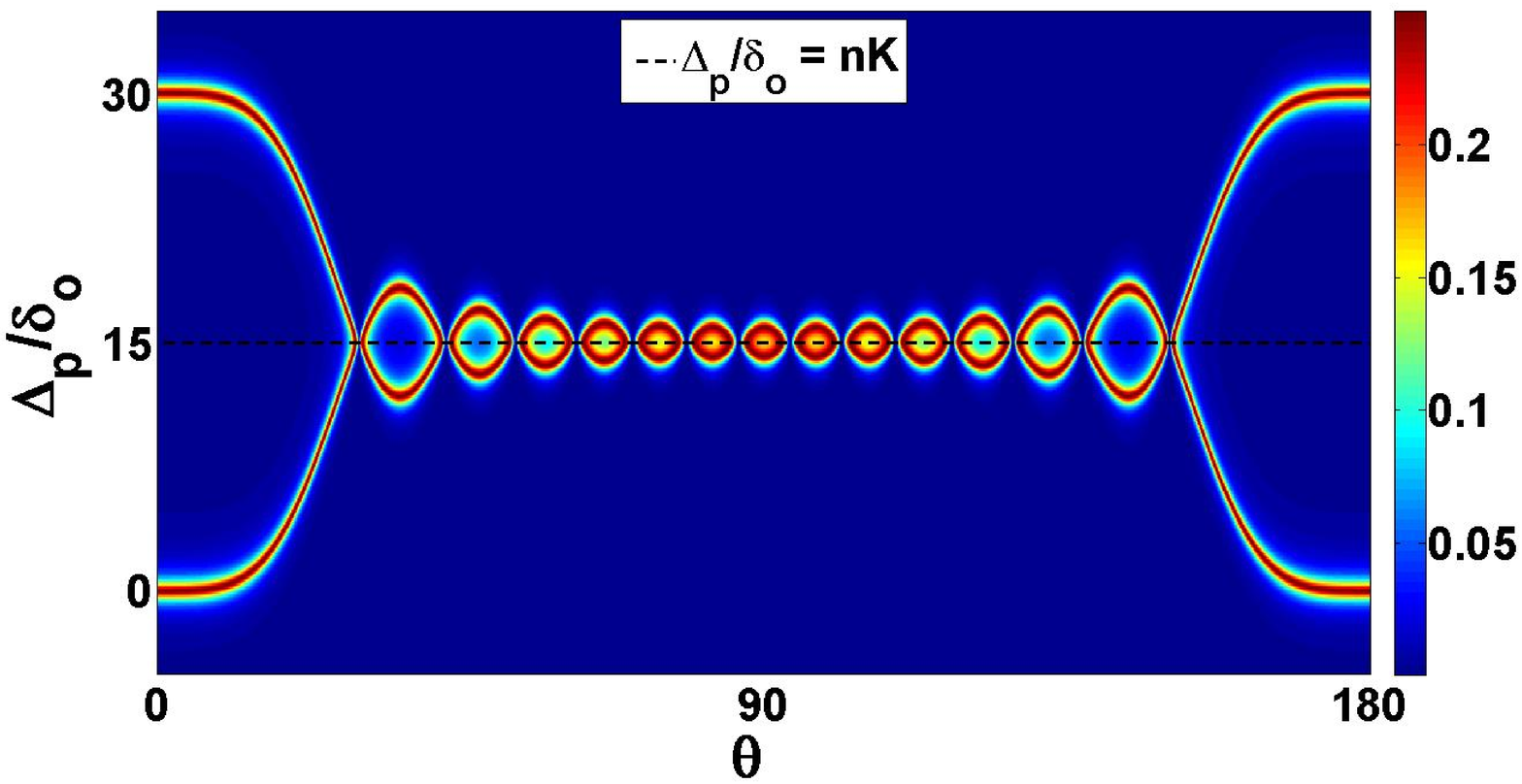}}\\
\subfloat[Part 1][]{\includegraphics[width=9cm, height=6.5cm]{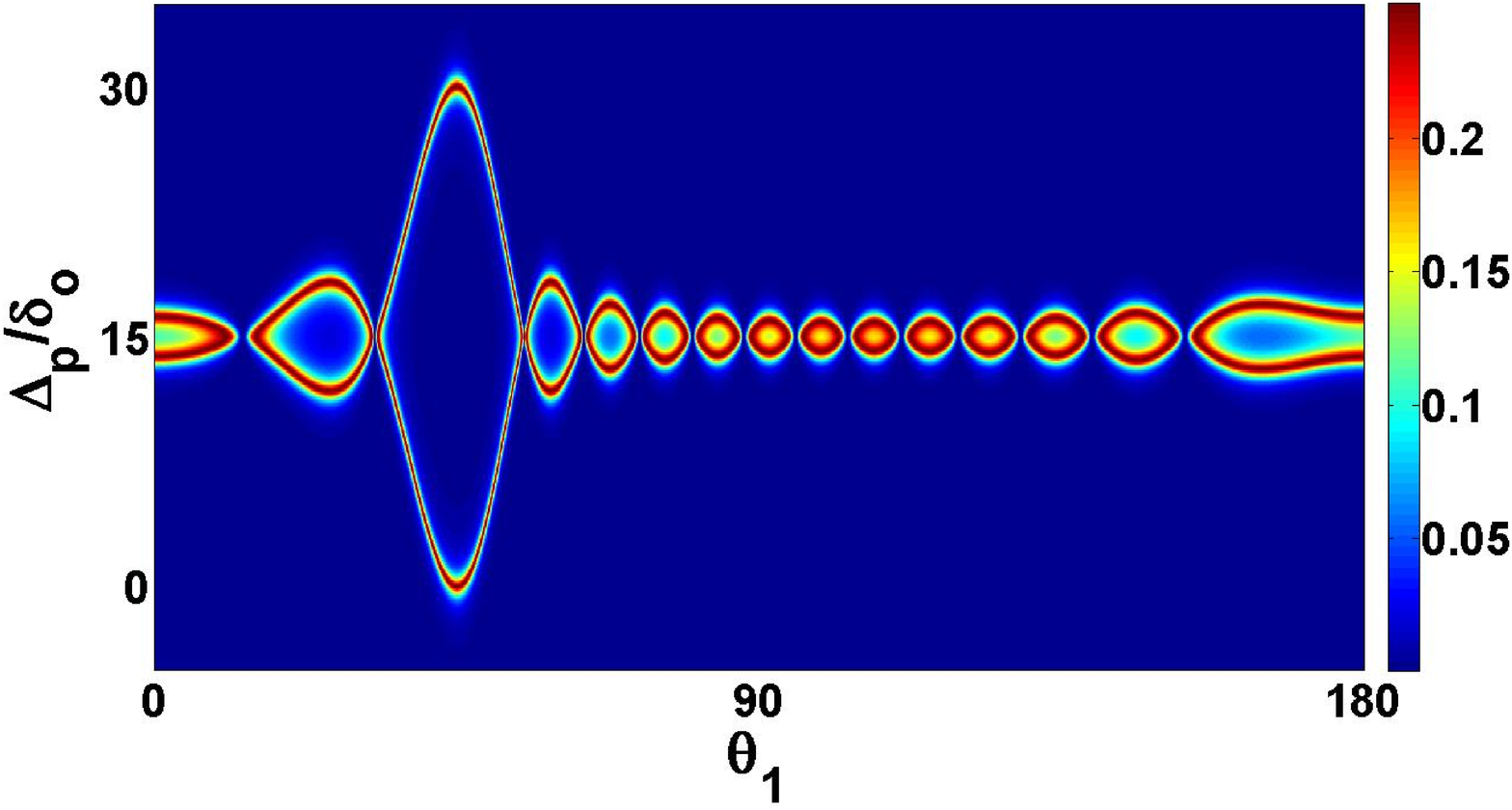}}
\subfloat[Part 1][]{\includegraphics[width=9cm, height=6.5cm]{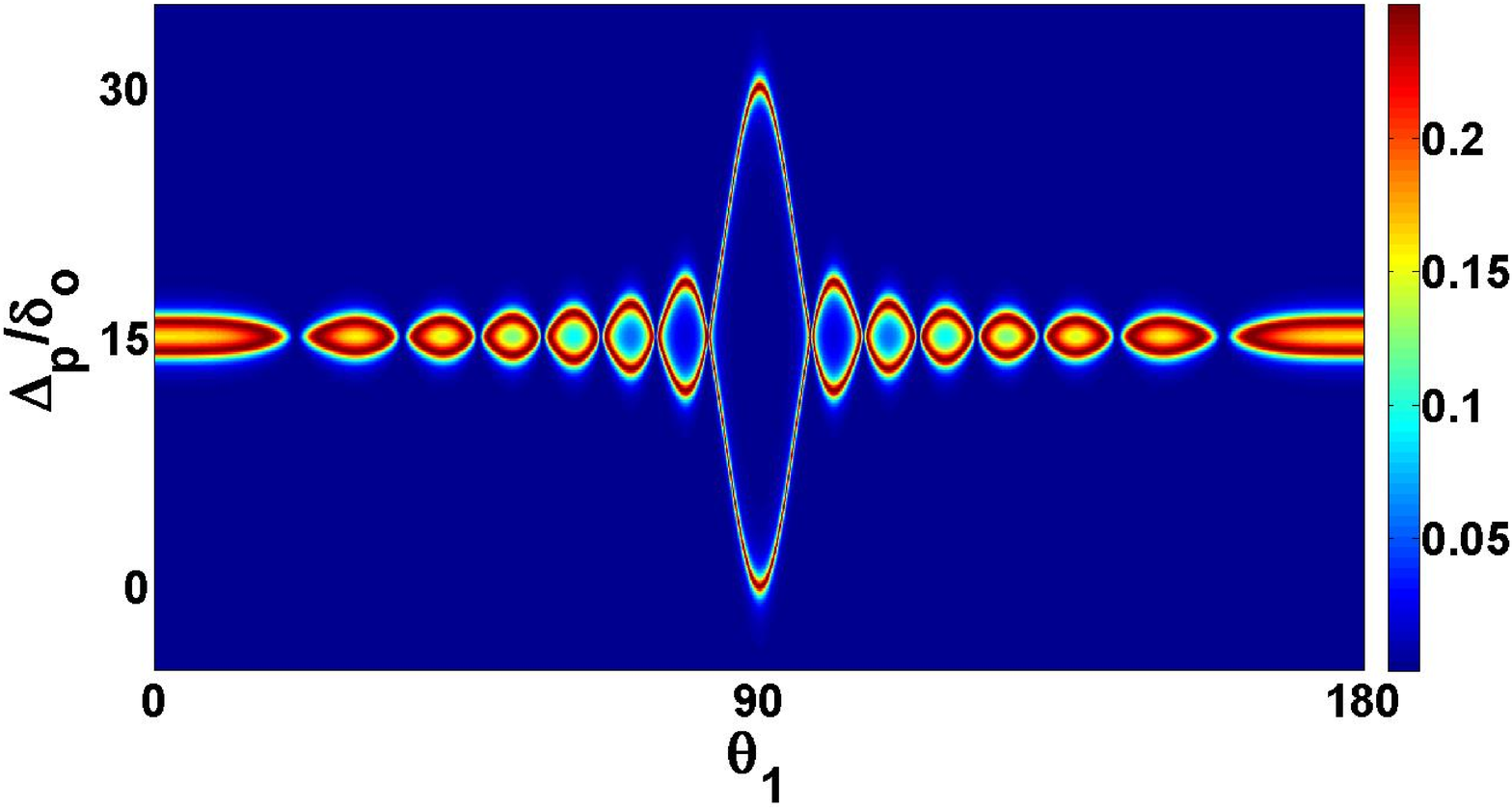}}
\caption{ (Color Online)(a)Variation in $G(\theta_{0},\theta_{1}, K=5)$(color axis) with $\theta_{0}$ and $\theta_{1}$(in degrees). (b)Variation of the photon number(color axis) with detuning $\frac{\Delta_{p}}{\delta_0}$ and $\theta_{1}$(in degrees), N=M=30, K=15,  $\kappa$ =0.5$\delta_{0}$ when the atoms are in MI state, single travelling mode case. $\theta_{0}$=$ 0^{\circ}$ (c)$\theta_{0}$= $45^{\circ}$ (d)$\theta_{0}$= $90^{\circ}$ }
\label{TT3}
\end{figure}

Fig.  \ref{TT3} (a) depicts the variation of function $G(\theta_{0},\theta_{1},K=5)$ with $\theta_{1}$ and $\theta_{0}$. For a particular value of $\theta_{0}$ and $\theta_{1}$, this function takes the maxima value when the argument of the function, ie., (${cos\theta_{0} - cos\theta_{1}}$) will become zero, ie., when $\theta_{1}= \pm \theta_{0}$.This can be seen from the $\theta_{0}=\theta_{1}$ line. 

Fig. \ref{TT3} (b)-(d) depicts the variation of intensity with $\Delta_{p}/\delta_{0}$ and $\theta_{1}$ for a fixed value of $\theta_{0}$. The plots show two symmetrically placed transmission peaks, whose separation is again proportional to $G(\theta_{0},\theta_{1},K)$ and therefore will also show a maxima when $\theta_{0}= \pm \theta_{1}$.  Physically $\theta_{1}=\theta_{0}$ corresponds to the case when both the ring cavities are oriented at the same angle, while $\theta_{0}= -\theta_{1}$, corresponds to the case when scattering is at the angle of reflection. However, in Fig. \ref{TT3} (b), when $\theta_{0}$=$0^{\circ}$, we observe an additional maxima at $\theta_{1}=180^{\circ}$ because, at this value, function $G(\theta_{0},\theta_{1},K) = \frac{sin(K\pi)}{sin{\pi}}$ also shows a maximum behavior (Fig. \ref{TT3}(a)). 
Also it is clearly seen that in all these plots, both the normal modes symmetrically vary around the average value $i.e.,$ $nK$. This average value is shown by a dotted black line in the Fig. \ref{TT3} (b). As clearly seen, it is independent of the angles between the lattice axis and the cavity modes, and is only dependent on the total number of atoms present in the illuminated sites.

\subsection{Superfluid}
\begin{figure}[H]
\centering
\subfloat[Part 1][]{\includegraphics[width=9cm, height=6.5cm]{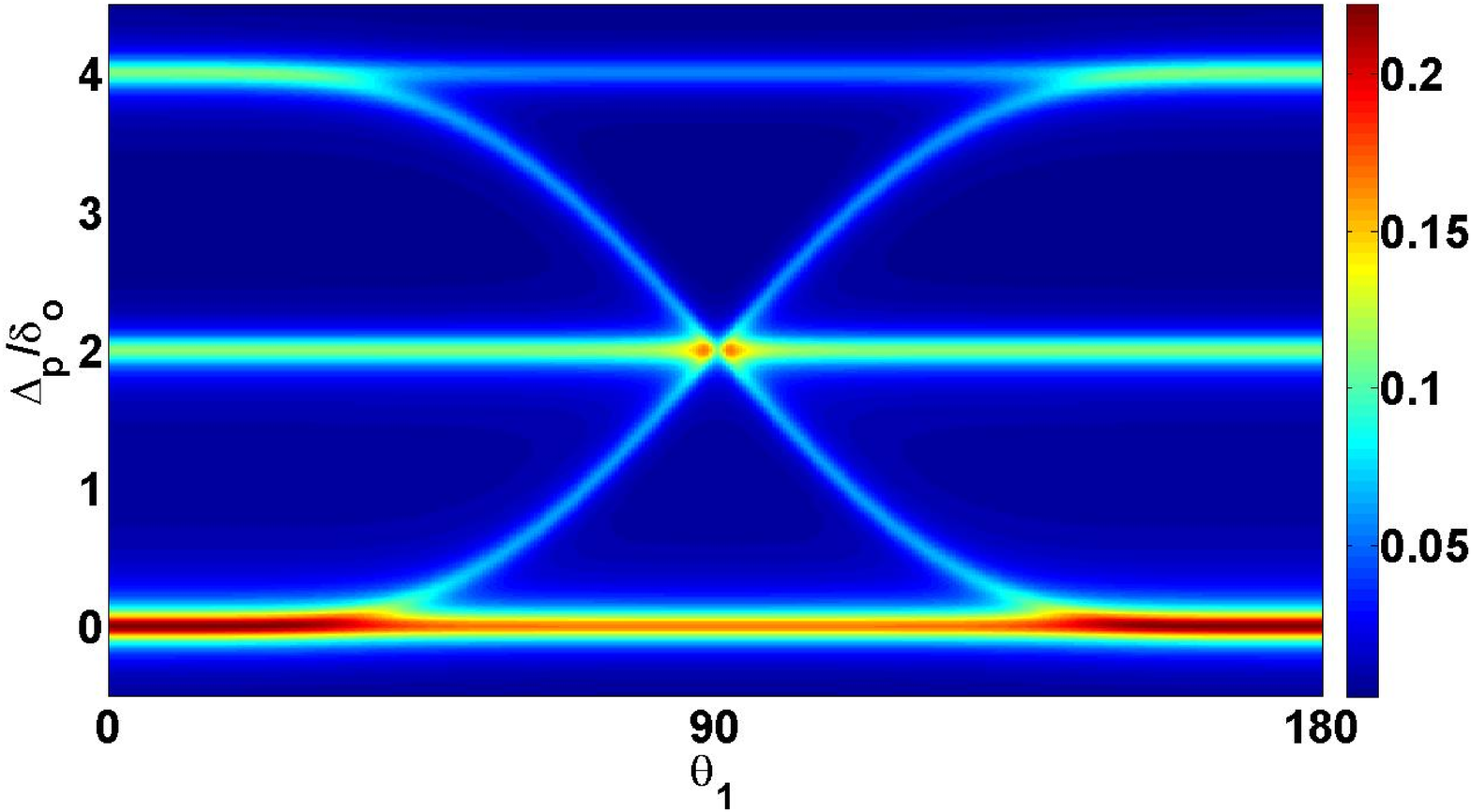}}
\subfloat[Part 1][]{\includegraphics[width=9cm, height=6.5cm]{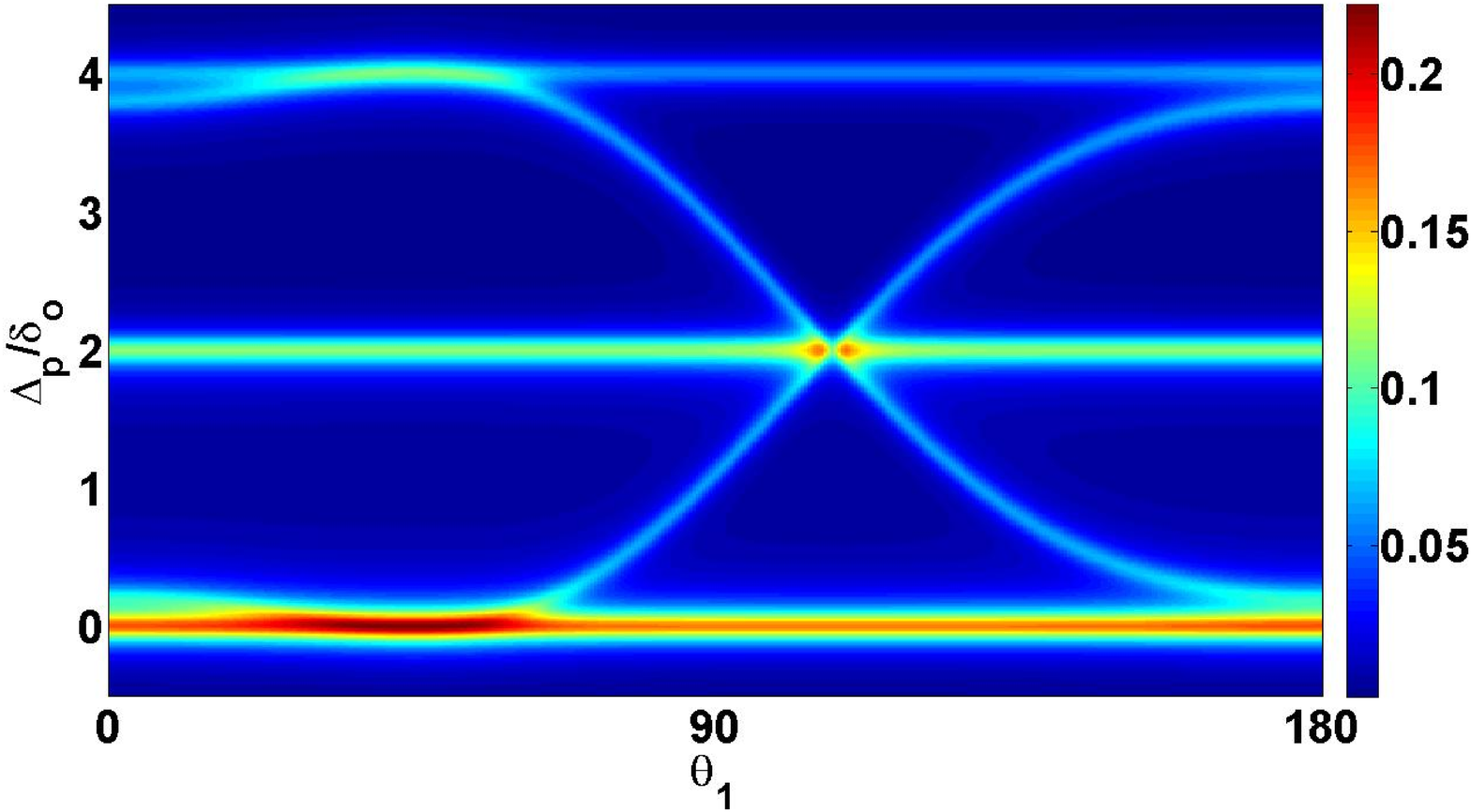}}\\
\subfloat[Part 1][]{ \includegraphics[width=9cm, height=6.5cm]{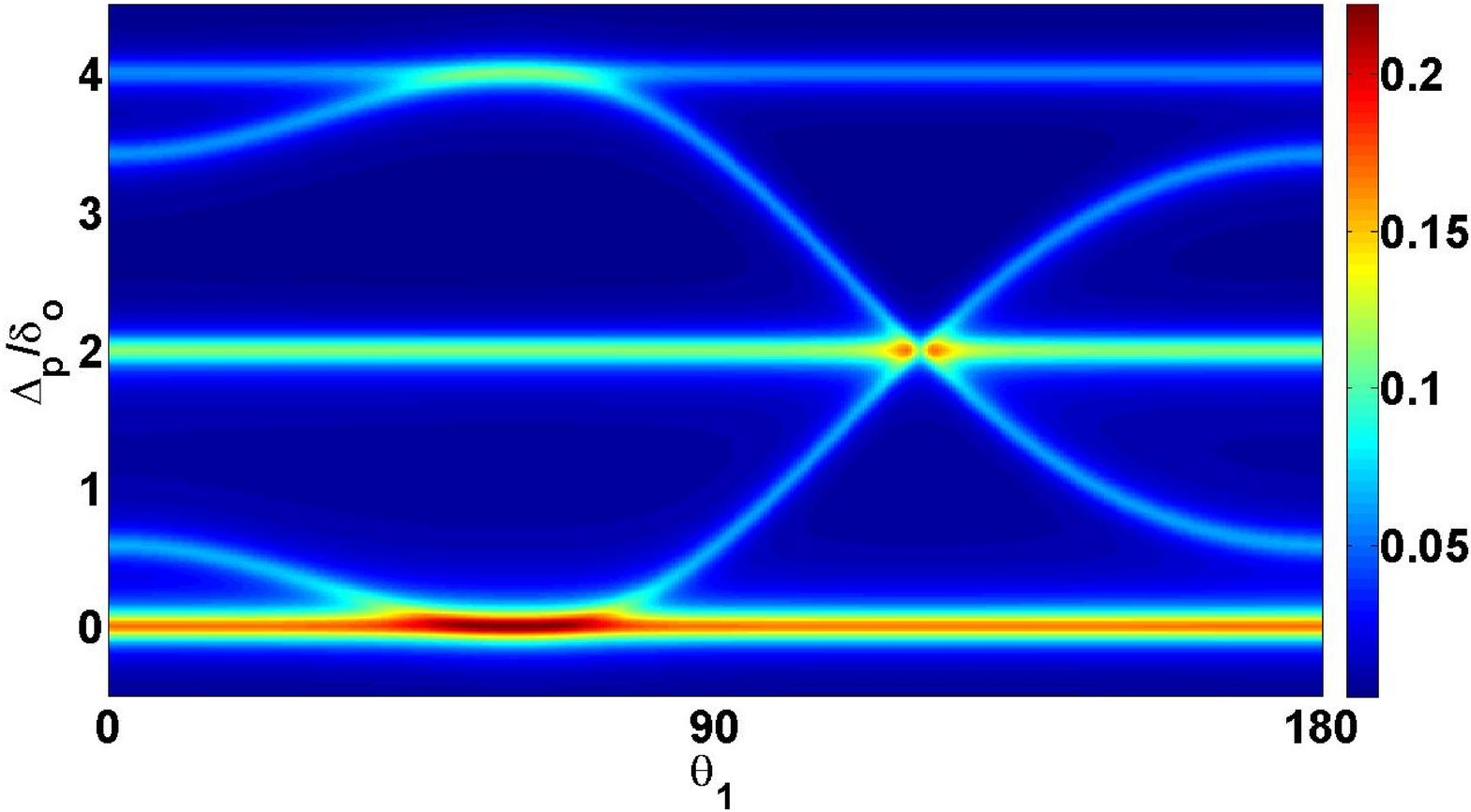} }
\subfloat[Part 1][]{\includegraphics[width=9cm, height=6.5cm]{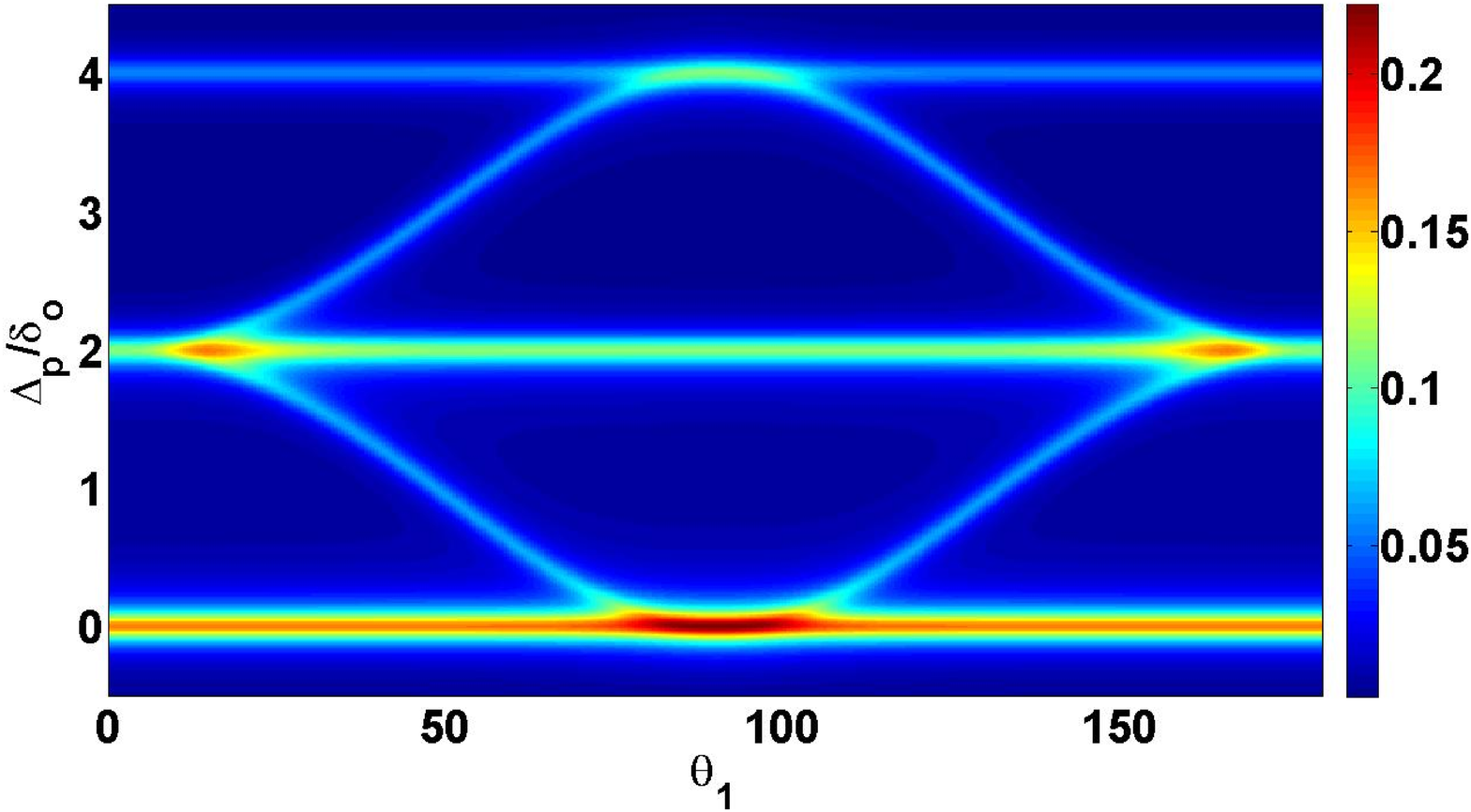}}
\caption{(Color Online) Variation of the photon number(Color Axis) with detuning $\frac{\Delta_{p}}{\delta_{0}}$ and  $\theta_{1}$(in degrees), N=K=2,M=3, $\kappa$ =0.1$\delta_{0}$ when the atoms are in SF state, two mode case for travelling wave cavity. In (a)$\theta_{0}$= $0^{\circ}$ (b)$\theta_{0}$= $45^{\circ}$ (c)$\theta_{0}$= $60^{\circ}$ (d)$\theta_{0}$= $90^{\circ}$ }
\label{SF2TW00}
\end{figure}

In this case also, the mode splitting only depends on the eigenvalue of the operator $\hat{D}_{10}$, but the eigenvalues are different  for different Fock states. 
Fig.\ref{SF2TW00} depicts the same for travelling mode case.

\begin{figure}[H]
\centering
\subfloat[Part 1][]{\includegraphics[width=6cm, height=5cm]{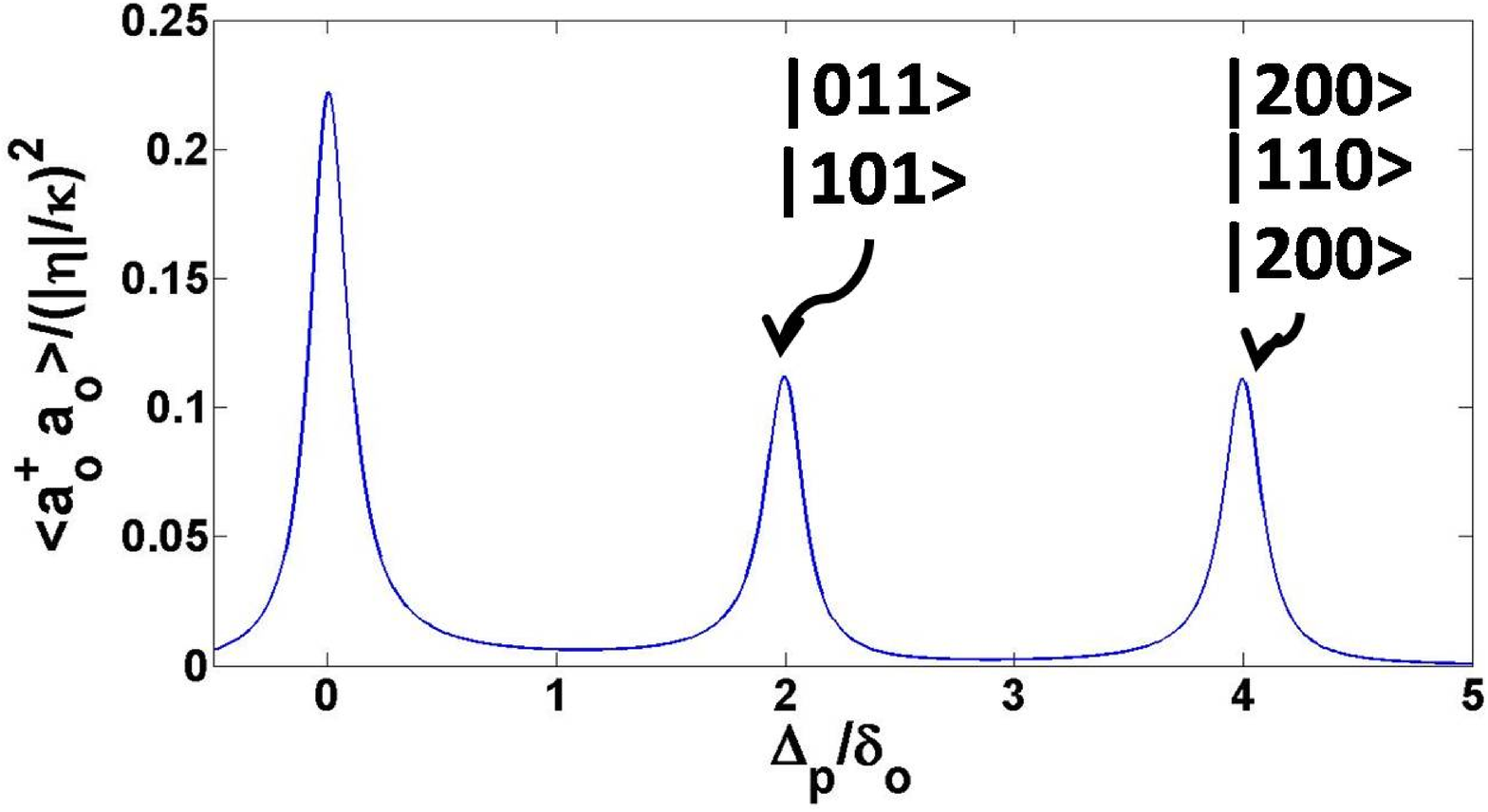}}
\subfloat[Part 1][]{\includegraphics[width=6cm, height=5cm]{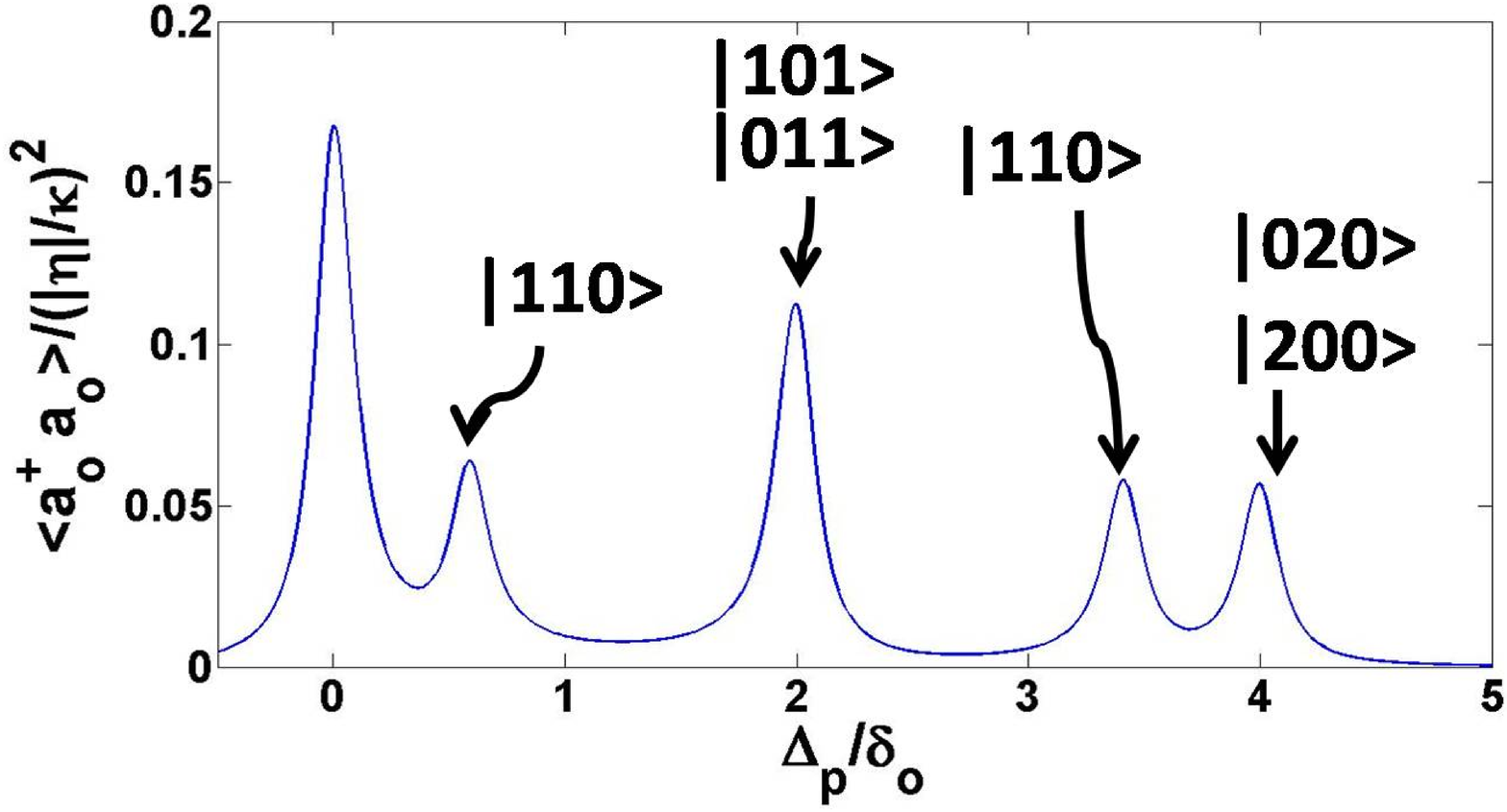}}
\subfloat[Part 1][]{ \includegraphics[width=6cm, height=5cm]{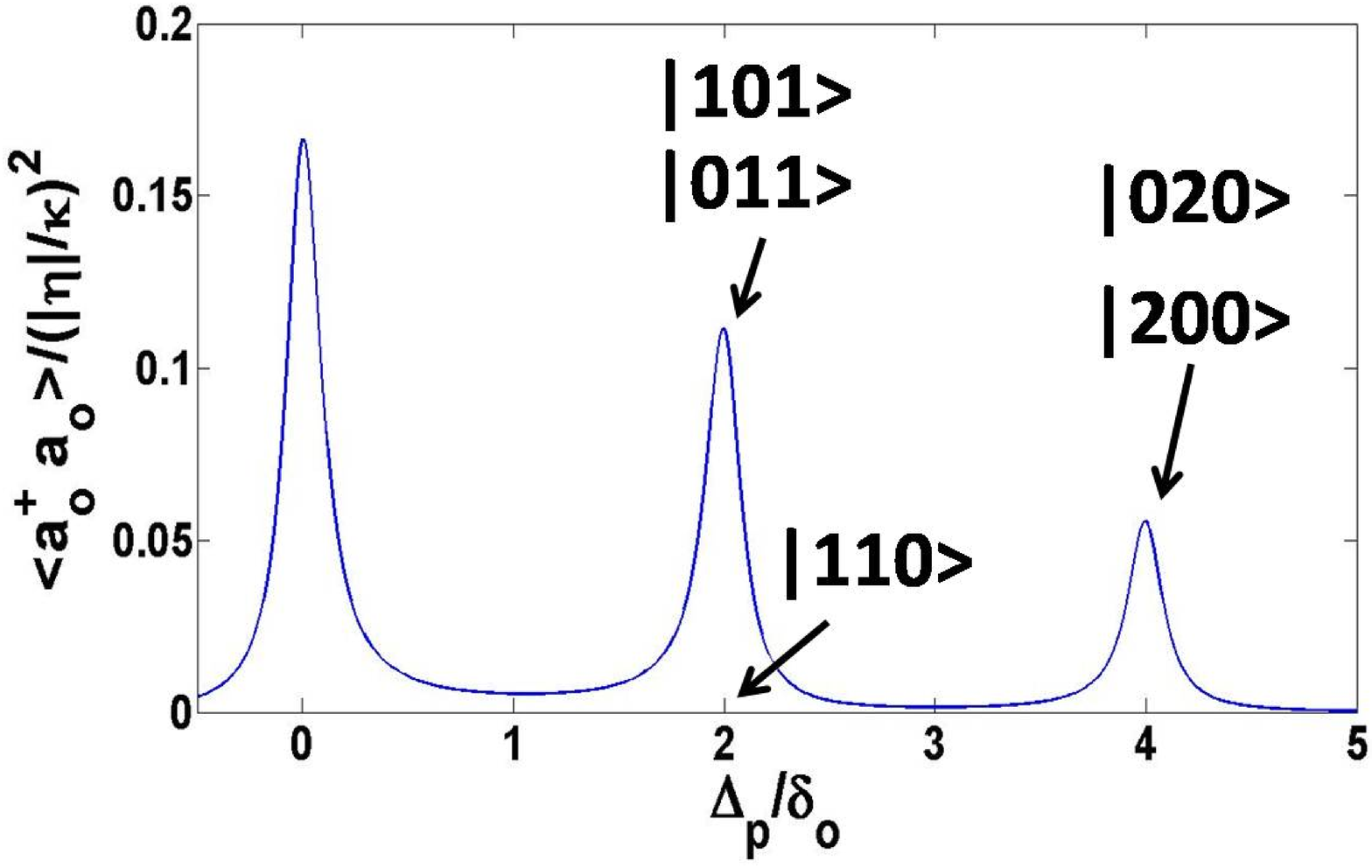} }\\
\subfloat[Part 1][]{\includegraphics[width=18cm, height=10cm]{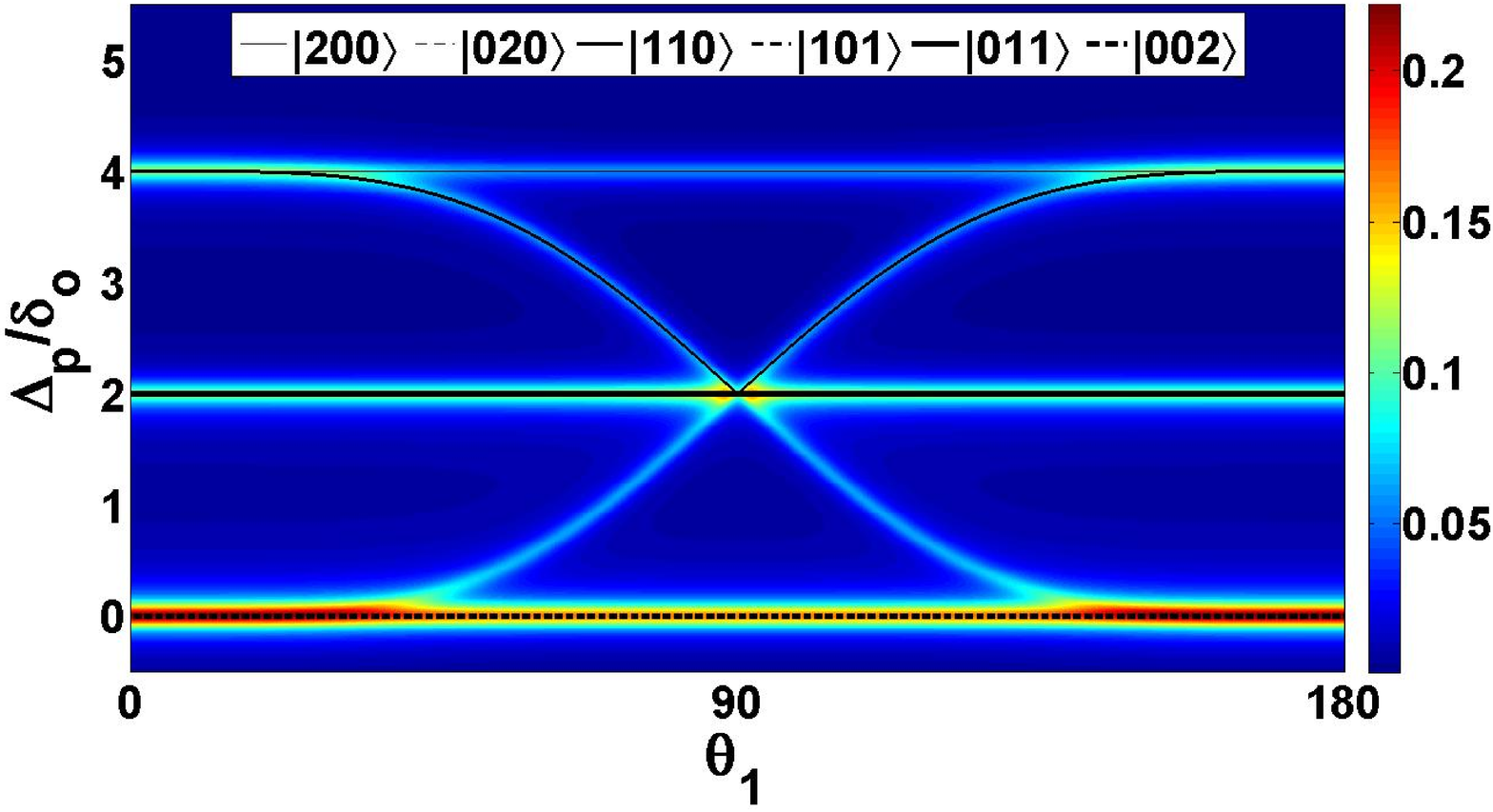}}
\caption{(Color Online)As before for N=K=2, M=3 we have six Fock states. Shown above are the two dimensional plots for photon number with respect to $\Delta_p / \delta_0$ for  $\theta_0 = 0^\circ$ and different values of $\theta_1$. (a) $\theta_{1} = 0^{\circ}$. (b)When $\theta_{1} = 60^{\circ}$ we observe five peaks. (c) $ \theta_{1} = 90^{\circ}$ and we again observe three peaks corresponding to different Fock states. The figure (d) shows how the individual Fock states(black lines) corresponding to different peaks change with $\theta_{1}$. This has been superposed with the figure $\ref{SF2TW00}(a)$.  For this two mode case, although each Fock state can show maxima at two values of $\Delta_p/\delta_0$, however here we have only shown $g_s + {\mathcal{G}_s}$ for each Fock state. }
\label{SF2TW}
\end{figure}

The photon number for this case will be given by,
\bea\langle \Psi | a_{1}^{\dag} a_{1} | \Psi \rangle& =& \frac{1}{M^{N}} \sum_{\langle n_{j} \rangle}\frac{N!}{n_{1}!n_{2}!...n_{M}!} \frac{(\mathcal{G}_s)^{2}|\eta_{0}|^{2}} {([\Delta_{p} - (g_s+ \mathcal{G}_s)]^2 +  \kappa^2)([\Delta_{p} - (g_s - \mathcal{G}_s)]^2 +  \kappa^2)}
\label{inSFTW} \eea
where $g_s= (\sum_{j=1:K}{n_j})\delta_{1}$ and $\mathcal{G}_s=G(\theta_{0},\theta_{1}, K,n_{j}) \delta_{1}$ where, $G(\theta_{0},\theta_{1}, K,n_{j})$ is the eigenvalue of $\hat{D}_{10}$ operator on a Fock state with $n_{j}$ particles on the $j$-th site. It may be again noted that in a SF state 
$n_{j}$ varies with the site index $j$ for a given Fock state.
The transmission spectra is shown in the Fig.(\ref{SF2TW00}) for certain demonstrative values of $\theta_{0},\theta_{1}$. 

For each Fock state, the transmission is expected to show two 
peaks at two values of $\frac{\Delta_{p}}{\delta_{0}}$ respectively given by $g_{s} \pm \mathcal{G}_{s}$ due to the mode splitting. 
In Fig. \ref{SF2TW} (d)  we superpose, the higher of these two normal modes , namely  $g_s + \mathcal{G}_s$ for different Fock states ( black lines)  on Fig. \ref{SF2TW00}(a). From the plot we note that all the Fock states do not show variation with $\theta_{1}$. Only the $|1,1,0 \rangle$ state shows an angle dependent shift. The Fock state $|0,0,2 \rangle$ do not show any frequency shift, while the other Fock states shift it by a constant value. This can be clearly seen from the Fig. \ref{SF2TW}(a), in which $\theta_{0}=\theta_{1}=0^{\circ}$. In this case both $g_s$ and $\mathcal{G}_s$ for each Fock state is $= q \delta_1$, where $q$ are the number of atoms in illuminated sites. Thus the Fock states group into sets of 1,2,3 for the higher normal mode 
similar to the case of standing wave modes. Now as $\theta_1$ is varied, the frequency shift corresponding to Fock state $|1,1,0\rangle$ shows variation ( see 
Fig. \ref{SF2TW}(b)).  However, at $\theta_{1}= 90^{\circ}$, its contribution to central peak at $\Delta_p/\delta_0$ = 2 is zero as $\mathcal{G}_s$ for this particular Fock state becomes zero, thus the intensity for this state becomes zero(Fig. \ref{SF2TW} (c)).

Thus we see that the shift in the frequency of the cavity mode depends not only on the local atomic configuration of a particular Fock state in a superfluid, but also on the type of quantization of 
the cavity modes. Hence we note that the change in boundary condition of the cavity mode, changes the nature of quantum diffraction through such cavity.

\section{Conclusion}\label{conclusion}
In our work, we have analyzed cold atomic condensates formed by bosonic atoms in an optical lattice at ultra cold temperatures. It has been suggested that such system when illuminated by cavity modes, can imprint their characteristics on the transmitted intensity. We have studied the off resonant scattering from such correlated systems by varying the angles that the cavity modes make with the optical lattice and thus obtained the transmission spectrum as a function of the detunings and the dispersive shifts.

The main result of our work reveals the pattern in the  shifts of the cavity mode frequency as the relative angle between the cavity mode and the optical lattice is changed. 
As we have pointed out in section \ref{SMI} that a change in the dispersion shift implies the effective change of the refractive index. Thus our finding implies even for a given quantum phase,
as the relative angle between the mode propagation vector and the optical lattice changes, the cavity induced dispersion shift or the effective refractive index of the medium also
changes. This highlights the uniqueness of such quantum phase of matter as medium of optical dispersion.

For the  single mode case discussed in section \ref{smode},
in MI phase, we have seen that the transmitted intensity depends on the number of atoms in the illuminated sites, since the presence of an atom shifts the cavity resonance and this shift is directly proportional to the number of illuminated atoms. The SF phase is however a superposition of many Fock states and set of Fock states group correspond to same shift. However changing the angle, these group of Fock states change thus providing more information about the system.

As discussed in next section \ref{dmode} when  two cavity modes are considered, the system shows mode splitting between the cavity modes coupled by the atomic ensemble. This was clearly visible in the MI case. In the SF state, at some specific angles of illumination, the Fock states of SF distinctly map to different frequency shift. Thus giving the Fock state structure of the system. However, it was noticed that such a system can only be achieved through high finesse cavities, as such characteristic features in the plots for the SF phase become blurred for an increase in $\frac{\kappa}{\delta_{0}}$ values. Some generalizations of this two mode case were also discussed. 

Such system when illuminated by ring cavities show different features of intensity transmission as shown in the section \ref{TW} that describes the situation where the cavity modes 
are travelling waves. Thus the nature of diffraction pattern of light scattered from such ultra cold atoms in a cavity is also dependent on the nature of the quantization of  
 the cavity mode.  It may be mentioned that such dependence on the mode of quantization of light is also observed in the complementary study where the diffraction properties of the atoms 
by quantized electromagnetic wave was studied \cite{Meystre2}.

Thus our analysis shows that the variation of the relative angle between the cavity mode and the optical lattice 
can resolve the Fock space structure of a quantum many body state of ultra cold atoms by varying the effective number of illuminated sites.
It has been pointed out in experiment described in ref. \cite{Colom1} that it is possible to study the correlated many body states of few ultra cold atoms in such cavity within the currently available technology. A few body correlated system of 
ultra cold fermions was also experimentally achieved recently \cite{Fermion}. In current work also, for example in Fig. \ref{SF00} and Fig. \ref{SF21} it has been shown that in the limit of small cavity decay rate $\kappa$ and for few number of particles in the 
illuminated sites, in a superfluid phase or more correctly in a few body analogue of a superfluid state it is possible to identify 
the extent of superposition of Fock states in different parameter regime. Such identification is potentially helpful in various types 
of many body quantum state preparation. 
\section{Acknowledgement}	
One of us (JL) thanks Prof. H. Ritsch for helpful discussion.  


\begin{thebibliography}{99}
\bibitem{Jaksch}D. Jaksch {\it et al.}, Phys. Rev. Lett. {\bf 81}, 3108 (1998).
\bibitem{RMP}I. Bloch, J. Dalibard and  W. Zwerger, Rev. Mod. Phys. {\bf 80}, 885 (2008).
\bibitem{Greiner1}M. Greiner {\it et al.}, Nature {\bf 415}, 39 (2002).
\bibitem{Cirac} J. J. Garcza-Ripoll, J. I. Cirac, Phil. Trans. R. Soc. Lond. A {\bf 361}, 1537 (2003) .
\bibitem{Raimond} J.M. Raimond, M. Brune, S. Haroche, Rev. Mod. Phys. {\bf 73},565 (2001).
\bibitem{Morice}O. Morice, Y. Castin and J. Dalibard. Phys. Rev. A {\bf 51}, 3896 (1995).
\bibitem{Meystre}M. G. Moore,  O. Zobay and P. Meystre, Phys. Rev. A {\bf 60}, 1491 (1999). 
\bibitem{mekhov1} I.  B. Mekhov, C. Maschler and H. Ritsch,  Nat.  Phys {\bf 3}, 319 (2007). 
\bibitem{Bren1} F. Brennecke  {\it et al.}, Nature {\bf 450}, 268 (2007). 
\bibitem{Colom1} Y. Colombe, T. Steinmetz {\it et al.}, Nature {\bf 450}, 272 (2007). 
\bibitem{Zimmerman1} S. Slama {\it et al.}  Phys. Rev. Lett. {\bf 98},  053603 (2007). 
\bibitem{Bren2} F. Brennecke, S. Ritter,T. Donner, T. Esslinger, Science {\bf 322}, 235 (2008). 
\bibitem{Gupta} S. Gupta, K. L. Moore, K. Murch, D. M. Stamper-Kurn, Phys. Rev. Lett {\bf 99}, 213601 (2007).
\bibitem{Miyake}H. Miyake {\it et al.}, Phys. Rev. Lett. {\bf 107}, 175302 (2011). 
\bibitem{Weit}C. Weitenberg {\it et al.}, Phys. Rev. Lett. {\bf 106}, 215301 (2011).
\bibitem{mekhov2} I. B. Mekhov,  C. Maschler and H. Ritsch, Phys. Rev. A {\bf 76}, 053618 (2007).
\bibitem{mekhov3}I. B. Mekhov, H. Ritsch, Phys. Rev. Lett {\bf 102}, 020403 (2009).
\bibitem{mekhov4} C. Maschler, I.B. Mekhov, H. Ritsch Eur. Phys. J. D {\bf 46}, 545 (2008).
\bibitem{mekhov5}I. B. Mekhov, H. Ritsch, Phys. Rev. A {\bf 80}, 013604 (2009).
\bibitem{mekhov6}I. B. Mekhov, H. Ritsch, Laser Physics, {\bf 19}, No. 4, 610 (2009).
\bibitem{mekhov7}B.Wunsch  {\it et al.}, Phys. Rev. Lett. {\bf 107}, 073201 (2011). 
\bibitem{mekhov8}I. B. Mekhov, H. Ritsch, Laser Physics, {\bf 20}, No. 3, 694 (2010).
\bibitem{Larson}J. Larson, B. Damski, G. Morigi and M. Lewenstein, Phys. Rev. Lett, {\bf 100}, 050401 (2008).
\bibitem{Chen1}W. Chen {\it{et al.}}, Phys. Rev. A {\bf 80}, 011801 (2009).
\bibitem{Chen2}W. Chen, D. S. Goldbaum, M. Bhattacharya, P. Meystre, Phys. Rev. A {\bf 81}, 053833 (2010) 
\bibitem{Aranya}A. B. Bhattacherjee, Phys. Rev. A {\bf 80}, 043607 (2009); T. Kumar, A. B. Bhattacharjee and ManMohan, Phys. Rev. {\bf 81}, 013835 (2010). 
\bibitem{Gopal1}S. Gopalakrishnan, B. L. Lev and P. M. Goldbart, Nat. Phys. {\bf 5}, 845 (2009). 
\bibitem{Gopal2}S. Gopalakrishnan, B. L. Lev and P. M. Goldbart, Phys. Rev. A {\bf 82}, 043612 (2010)
\bibitem{Vidal} S.F. Vidal, G. Chiara, J. Larson, G. Morigi, Phys. Rev. A {\bf 81}, 043407 (2010).
\bibitem{Bren3} K. Baumann, C. Guerlin, F. Brennecke,T. Esslinger, Nature, {\bf 464} 1301 (2010).
\bibitem{Frahn} W. E.. Frahn, Riv. Nuovo Cim, {\bf 7}, 499  (1977).
\bibitem{chapter}H. Tanji-Sujuki {\it et al.}, Adv. At. Mol. Opt. Phys. {\bf 60}, 201 (2011).
\bibitem{Meystre2}D. Meiser, C. P. Search and P. Meystre, Phys. Rev. A {\bf 71}, 013404 (2005).
\bibitem{Born}M. Born and E. Wolf, Principles of Optics, Cambridge University Press(1999).
\bibitem{Ghatak}A. Ghatak and K. Thyagarajan, Optical Electronics, Cambridge University Press(1989).
\bibitem{Raman}C. V. Raman  and N. S. N. Nath, Proc. Indian Acad. Sci {\bf 4}, 222 (1936)
\bibitem{Brill}L. Brillouin, Annales des Physique {\bf 17}, 88 (1922). 

\bibitem{Knight} B. W. Shore and P. L. Knight, Jour. of Mod. Opt. {\bf 40}, 1195 (1993). 
\bibitem{Gou}  S. C. Gou, Phys. Rev. A {\bf 40}, 5116 (1989).
\bibitem{Gerry} C. C. Gerry and J. H. Eberly, Phys. Rev. A {\bf 42}, 6805 (1990).
\bibitem{Burnett}J. S. Douglas and K. Burnett, Phys. Rev. A { \bf 82}, 033434 (2010). 
\bibitem{ring} M Gangl and H. Ritsch, Phys. Rev. A  {\bf 61}, 043405 (2000).
\bibitem{Bux}S. Bux {\it et al.}, Phys. Rev. Lett. {\bf 106}, 203601 (2011).
\bibitem{Nagorny}B. Nagorny, T. Els${\ddot{a}}$sser, A. Hemmerich,  Phys. Rev. Lett. {\bf 91}, 153003 (2003).
\bibitem{Fermion}F. Serwane, G. Z$\ddot{u}$rn, T. Lompe, T. B. Ottenstein, A. N. Wenz and S. Jochim, Science, {\bf 332}, 336 (2011).
\end{thebibliography}
\end{document}